\documentclass{elsart}

\usepackage[square,comma]{natbib}
\usepackage{graphicx}
\usepackage{amssymb}

\begin{document}

%\begin{widetext}
\thispagestyle{empty}
\begin{Large}
\textbf{DEUTSCHES ELEKTRONEN-SYNCHROTRON}

\textbf{\large{in der HELMHOLTZ-GEMEINSCHAFT}\\}
\end{Large}

DESY 05-109

June 2005

\begin{eqnarray}
\nonumber &&\cr \nonumber && \cr \nonumber &&\cr
\end{eqnarray}
\begin{eqnarray}
\nonumber
\end{eqnarray}
\begin{center}
\begin{Large}
\textbf{Understanding transverse coherence properties of X-ray
beams in third generation Synchrotron Radiation sources}
\end{Large}
\begin{eqnarray}
\nonumber &&\cr \nonumber && \cr
\end{eqnarray}

\begin{large}
Gianluca Geloni, Evgeni Saldin, Evgeni Schneidmiller and Mikhail
Yurkov
\end{large}
\textsl{\\Deutsches Elektronen-Synchrotron DESY, Hamburg}
\begin{eqnarray}
\nonumber
\end{eqnarray}
\begin{eqnarray}
\nonumber
\end{eqnarray}
\begin{eqnarray}
\nonumber
\end{eqnarray}
ISSN 0418-9833
\begin{eqnarray}
\nonumber
\end{eqnarray}
\begin{large}
\textbf{NOTKESTRASSE 85 - 22607 HAMBURG}
\end{large}
\end{center}
%\end{widetext}
\clearpage
\newpage

\begin{frontmatter}



\title{Understanding transverse coherence properties of X-ray beams in third generation Synchrotron Radiation sources}


\author[DESY]{Gianluca Geloni}
\author[DESY]{Evgeni Saldin}
\author[DESY]{Evgeni Schneidmiller}
\author[DESY]{Mikhail Yurkov}

\address[DESY]{Deutsches Elektronen-Synchrotron (DESY), Hamburg,
Germany}

\begin{abstract}
This paper describes a theory of transverse coherence properties
of Undulator Radiation. Our study is of very practical relevance,
because it yields specific predictions of Undulator Radiation
cross-spectral density in various parts of the beamline. On the
contrary, usual estimations of coherence properties assume that
the undulator source is quasi-homogeneous, like thermal sources,
and rely on the application of van Cittert-Zernike (VCZ) theorem,
in its original or generalized form, for calculating transverse
coherence length in the far-field approximation. The VCZ theorem
is derived in the frame of Statistical Optics using a number of
restrictive assumptions: in particular, the quasi-homogeneous
assumption is demonstrated to be inaccurate in many practical
situations regarding undulator sources. We propose a technique to
calculate the cross-spectral density from undulator sources in the
most general case. Also, we find the region of applicability of
the quasi-homogeneous model and we present an analytical
expression for the cross-spectral density which is valid up to the
exit of the undulator. For the case of more general undulator
sources, simple formulas for the transverse coherence length,
interpolated from numerical calculations and suitable for beamline
design applications are found. Finally, using a simple vertical
slit, we show how transverse coherence properties of an X-ray beam
can be manipulated to obtain a larger coherent spot-size on a
sample. This invention was devised almost entirely on the basis of
theoretical ideas developed throughout this paper.
\end{abstract}

\begin{keyword}

X-ray beams \sep Undulator radiation \sep Transverse coherence
\sep Van Cittert-Zernike theorem \sep Emittance effects

\PACS 41.60.m \sep 41.60.Ap \sep 41.50 + h \sep 42.50.Ar

\end{keyword}

\end{frontmatter}

% main text

\clearpage
\section{\label{sec:intro} Introduction}

In recent years, continuous evolution of third generation light
sources has allowed dramatic increase of brilliance with respect
to older designs, which has triggered a number of new techniques
and experiments unthinkable before. Among the most exciting
properties of today third generation facilities is the high flux
of coherent X-rays provided. The availability of intense coherent
X-ray beams has fostered the development of new coherence-based
techniques like fluctuation correlation dynamics, phase imaging,
coherent X-ray diffraction (CXD) and X-ray holography. In this
context, understanding the evolution of transverse coherence
properties of Synchrotron Radiation (SR) along the beam line is of
fundamental importance.

In general, when dealing with this problem, one should account for
the fact that Synchrotron Radiation is a random statistical
process. Therefore, the evolution of transverse coherence
properties should be treated in terms of probabilistic statements:
the shot noise in the electron beam causes fluctuations of the
beam density which are random in time and space. As a result, the
radiation produced by such a beam has random amplitudes and
phases.

Statistical Optics \cite{GOOD, MAND, NEIL} affords convenient
tools to deal with fluctuating electromagnetic fields in an
appropriate way. Among the most important quantities needed to
describe coherent phenomena in the framework of Statistical Optics
is the correlation function of the electric field. In any
interference experiment one needs to know the system (second
order) correlation function of the signal at a certain time and
position with the signal at another time and position.
Alternatively, and equivalently, one can describe the same
experiment in frequency domain. In this case one is interested in
the correlation function of the Fourier transform of the time
domain signal at a certain frequency and position with the Fourier
transform of the time domain signal at another frequency and
position. The signal one is interested to study is, indeed, the
Fourier transform of the original signal in time domain. In SR
experiments the analysis in frequency domain is much more natural
than that in the time domain. In fact, up-to-date detectors are
limited to about $100$ ps time resolution and they are by no means
able to resolve a single X-ray pulse in time domain. They work,
instead, by counting the number of photons at a certain frequency
over an integration time longer than the radiation pulse.
Therefore, in this paper we will deal with signals in the
frequency domain and we will often refer to the "Fourier transform
of the electric field" simply as "the field".

For some particular experiment one may be interested in higher
order correlation functions (for instance, in the correlation
between the intensities) which, in general, must be calculated
separately. In the particular case when the field fluctuations can
be described as a Gaussian process, the field is often said to
obey Gaussian statistics. In this case, with the help of the
Moment Theorem \cite{GOOD} one can recover correlation functions
of any order from the knowledge of the second order one: this
constitutes a great simplification to the task of describing
coherence properties of light. A practical example of a field
obeying Gaussian statistics is constituted by the case of
polarized thermal light. This is more than a simple example: in
fact, Statistical Optics has largely developed in connection with
problems involving optical sources emitting thermal light like the
sun, other stars, or incandescent lamps. As a consequence,
Gaussian statistics is often taken for granted. Anyway, it is not
\textit{a priori} clear wether Synchrotron Radiation fields obey
it or not; our analysis will show that Synchrotron Radiation is
indeed a Gaussian random process. Therefore, as is also the case
for polarized thermal light and any other signal obeying Gaussian
statistics, when we deal with Synchrotron Radiation the basic
quantity to consider is the second order correlation function of
the field. Moreover, as already discussed, in Synchrotron
Radiation experiments it is natural to work in the space-frequency
domain, so that we will focus, in particular, on the second order
correlation function in the space-frequency domain.

Besides obeying Gaussian statistics, polarized thermal light has
two other specific properties allowing simplifications of the
theory: the first is stationarity\footnote{Here we do not
distinguish between different kind of stationarity because, under
the assumption of a Gaussian process, these concepts simply
coincide. } and the second is quasi-homogeneity. Exactly as the
property of Gaussian statistics, also stationarity and
quasi-homogeneity of the source are usually taken for granted in
Statistical Optics problems but, unlike it, they do not belong, in
general, to Synchrotron Radiation fields. In fact, as we will
show, Synchrotron radiation fields are intrinsically
non-stationary and not always quasi-homogeneous. Nevertheless, up
to now it has been a widespread practice to assume that undulator
sources are completely incoherent (i.e. homogeneous) and to apply
the well known van Cittert-Zernike (VCZ) theorem for calculating
the degree of transverse coherence in the far-field approximation
\cite{PETR}. Using the VCZ theorem, the electric field
cross-correlation function in the far field is usually calculated
(aside for a geometrical phase factor) as a Fourier transformation
of the intensity distribution of the source, customarily located
at the exit of the undulator.

Although the VCZ theorem only deals with completely incoherent
sources, there exists an analogous generalized version of it which
allows to extend the treat the case of quasi-homogeneous sources
as well. Actually there is no unambiguous choice of terminology in
literature regarding the scope of the VCZ theorem. For instance, a
very well-known textbook \cite{MAND} reports of a
"Zernike-propagation" equation dealing with any distance from the
source. Also, sometimes \cite{GOOD}, the generalized VCZ theorem
is referred to as Schell's theorem (and also used in some paper
\cite{CHAN}). In this paper we will refer to the VCZ theorem and
its generalized version only in the limit for a large distance
from the source and for, respectively, homogeneous and
quasi-homogeneous sources. However, irrespectively of different
denominations, the fundamental fact holds, that once a
cross-correlation function is known on a given source plane it can
be propagated through the beamline at any distance from the
source. It should be noted that, from this viewpoint, a source
simply denotes an initial plane down the beamline from which the
cross-correlation function is propagated further. Then, the
position of the source down the beamline is suggested only on the
ground of opportunity. On the contrary, when dealing with the VCZ
theorem, the source must be (quasi)-homogeneous which explains the
customary location at the exit of the undulator.

In some cases, the VCZ theorem or its generalized version may
provide a convenient method for calculating the degree of
transverse coherence in various parts of the beamline once the
transverse coherence properties of the photon beam are specified
at the exit of the undulator, that is at the source plane. In most
SR applications though, such treatment is questionable. First, the
source (even at the exit of the undulator) may not be
quasi-homogeneous. Second, even for specific sets of problem
parameters where the quasi-homogeneous model is accurate, the
specification of the far-field zone depends not only on the
electron beam sizes, but also on the electron beam divergencies
(in both direction) and on the intrinsic divergence of the
radiation connected with the undulator device. At the time being,
widespread and a-critical use of the VCZ theorem and its
generalization shows that there is no understanding of transverse
coherence properties of X-ray beams in third generation
Synchrotron Radiation sources.

If, on the one hand, the definition of the far-zone and the
possible non quasi-homogeneity of SR sources constitute serious
problems in the description of the coherence properties of
Synchrotron light, on the other hand the intrinsic
non-stationarity of the SR process does not play a very important
role. In particular, as we will show, assumption of a minimal
undulator bandwidth much larger than the characteristic inverse
bunch duration (which is always verified in practice) allows to
separate the correlation function in space-frequency domain  in
the product of two functions. The first function is a spectral
correlation describing correlation in frequency. The second
function describes correlation in space and is well-known also in
the case of stationary processes as the cross-spectral density of
the process. Then, the cross-spectral density can be studied
independently at any given frequency giving information on the
spatial correlation of the field. Subsequently, the knowledge of
the spectral correlation function brings back the full expression
for the space-frequency correlation.

In this paper we aim at the development of a theory of transverse
coherence capable of providing very specific predictions, relevant
to practice, regarding the cross-spectral density of undulator
radiation at various positions along the beam-line. A fully
general study of undulator sources is not a trivial one.
Difficulties arise when one tries to include simultaneously the
effect of intrinsic divergence of the radiation due to the
presence of the undulator, of electron beam size and electron beam
divergence into the insertion device. The full problem, including
all effects, poses an unsolvable analytical challenge, and
numerical calculations are to be preferred. Generally, the
cross-spectral density of the undulator radiation is controlled by
nine physical parameters which model both the electron beam and
the undulator: the horizontal and vertical geometrical emittances
of the electron beam $\epsilon_{x,y}$, the horizontal and vertical
minimal betatron functions $\beta_{x,y}^o$, the observation
distance down the beamline $z_o$, the observation frequency
$\omega$, the undulator resonant frequency $\omega_o$, the
undulator length $L_w$ and the length of the undulator period
$\lambda_w$.

We will make a consistent use of dimensional analysis. Dimensional
analysis of any problem, performed prior to analytical or
numerical investigations, not only reduces the number of
independent terms, but also allows one to classify the grouping of
dimensional variables in a way that is most suitable for
subsequent study. The algorithm for calculating the cross-spectral
density can be formulated as a relation between dimensionless
quantities. After appropriate normalization, the radiation
cross-spectral density from an undulator device is described by
six dimensionless quantities: the normalized emittances
$\hat{\epsilon}_{x,y} = \omega_o \epsilon_{x,y}/c$, the normalized
betatron functions $\hat{\beta}_{x,y} = \beta_{x,y}^o/L_w$, the
normalized observation distance $\hat{z}_o = z_o/L_w$ and the
normalized detuning parameter $\hat{C} = 2 \pi N_w (\omega -
\omega_o)/\omega_o$, where $N_w = L_w/\lambda_w$ is the number of
undulator periods.

At some point in this work we will find it convenient to pose
$\hat{C} = 0$. In other words we will assume that parameters are
tuned at perfect resonance. It is relevant to note that even under
this simplifying assumptions, conditions for the undulator source
to be quasi-homogeneous still include four parameters
$\hat{\epsilon}_{x,y}$ and $\hat{\beta}_{x,y}$. For storage rings
that are in operation or planned in the $\mathrm{\AA}$ngstrom
wavelength range, the parameter variation of $\hat{\epsilon}_{x}
\sim 10 -10^3 $, $\hat{\epsilon}_{y} \sim 10^{-1} -10$,
$\hat{\beta}_{x,y} \sim 10^{-1}-10 $ are possible: these include
many practical situations in which the assumption of
quasi-homogeneous sources and, therefore the (generalized) VCZ
theorem, is not accurate.

In this paper we will first deal with the most general case of
non-homogeneous sources. In fact, from a practical viewpoint, it
is important to determine the cross-spectral density as a function
of $\hat{\epsilon}_{x,y}$, $\hat{\beta}_{x,y}$, $\hat{C}$ and
$\hat{z}_o$. Once a general expression for the cross-spectral
density is found, it can be used as a basis for numerical
calculations. A second goal of this work is to find the region of
applicability of the quasi-homogeneous source model (i.e. of the
generalized VCZ theorem) which will arise automatically from the
dimensional analysis of the problem. Finally, we will derive
analytical expressions for the cross-spectral density at
$\hat{C}=0$ in various parts of the beamline.

Results may also be obtained using numerical techniques alone,
starting from the Lienard-Wiechert expressions for the
electromagnetic field and applying the definition of the field
correlation function without any analytical manipulation. Yet,
computer codes can calculate properties for a given set of
parameters, but can hardly improve physical understanding, which
is particularly important in the stage of planning experiments:
understanding of correct approximations and their region of
applicability with the help of a consistent use of dimensional
analysis can simplify many tasks a lot, including practical and
non-trivial ones. Moreover, at the time being, no code capable to
deal with transverse coherence problems has been developed at all.

It should be noted that some theoretical attempt to follow this
path has been proposed in \cite{TAKA}. Among the results of that
paper is the fact that van Cittert-Zernike theorem could not be
applied unless the electron beam divergence is much smaller than
the diffraction angle, which is never verified in practice in the
horizontal plane. We will show that this conclusion is incorrect.

We organize our work as follows. After this Introduction, in
Section \ref{sec:theory} we present a second-order theory of
coherence for fields generated by Synchrotron Radiation sources.
In Section \ref{sec:evol} we give a derivation of the
cross-spectral density for undulator-based sources in reduced
units. Subsequently, we analyze the evolution of the
cross-spectral density function through the beamline in the limit
for $\hat{C}=0$. A particular case of quasi-homogeneous sources
and its applicability region is treated under several simplifying
assumptions in Section \ref{sec:quasi}. Effects of the vertical
emittance on the cross-spectral density are discussed in detail in
Section \ref{sec:twod}, while a treatment of some non
quasi-homogeneous source is given in the following Section
\ref{sec:nonh}. Obtained results include approximate design
formula capable of describing in very simple terms the evolution
of the coherence length along the beamline in many situation of
practical interest. A good physical insight is useful to identify
possible applications of given phenomena. In particular in Section
\ref{sec:spot} we selected one practical application to exploit
the power of our approach. We show that, by means of a simple
vertical slit, it is possible to manipulate transverse coherence
properties of an X-ray beam to obtain a convenient coherent
spot-size on the sample. This invention was devised almost
entirely on the basis of theoretical ideas of rather complex and
abstract nature which have been described in this paper. Finally,
in Section \ref{sec:concl}, we come to conclusions.

\section{\label{sec:theory} Second-order coherence theory of fields generated by Synchrotron Radiation sources}

\subsection{\label{sub:def} Thermal light and Synchrotron Radiation: some concepts and definitions}

A great majority of optical sources emits thermal light. Such is
the case of the sun and the other stars, as well as of
incandescent lamps. This kind of radiation consists of a large
number of independent contributions (radiating atoms) and is
characterized by random amplitudes and phases in space and time.
The electromagnetic fields can be then conveniently described in
terms of Statistical Optics, a branch of Physics that has been
intensively developed during the last few decades. Today one can
take advantage of a lot of existing experience and theoretical
basis for the descriptions of fluctuating electromagnetic fields
\cite{GOOD, MAND}.

\begin{figure}
\begin{center}
\includegraphics*[width=140mm]{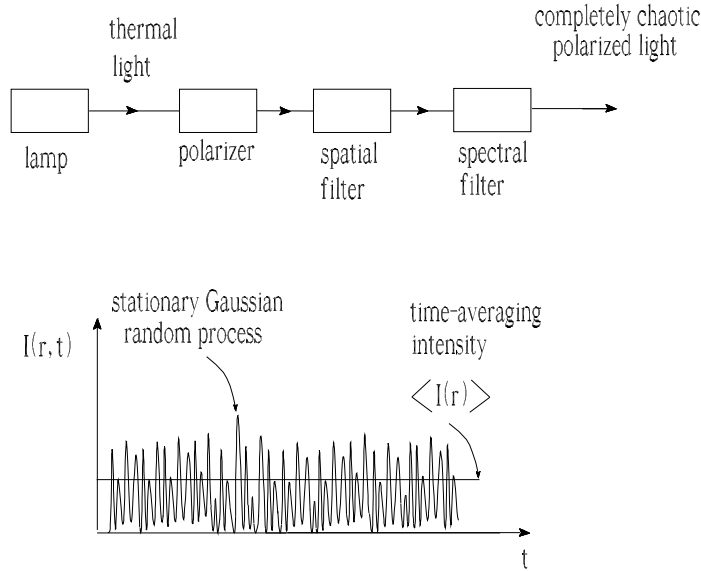}% Here is how to import EPS art
\caption{\label{thermal} Light intensity from an incandescent lamp
driven by a constant electric current. A statistically stationary
wave has an average that does not vary with time.}
\end{center}
\end{figure}
Consider the light emitted by a thermal source passing through a
polarization analyzer (see Fig. \ref{thermal}). Properties of
polarized thermal light are well-known in Statistical Optics, and
are referred to as properties of completely chaotic, polarized
light \cite{GOOD, MAND}. Thermal light is a statistical random
process and statements about such process are probabilistic
statements. Statistical processes are handled using the concept of
statistical ensemble, drawn from Statistical Mechanics, and
statistical averages are performed indeed, over many ensembles, or
realizations, or outcomes of the statistical process under study.

Polarized thermal light is a very particular kind of random
process in that it is Gaussian, stationary and ergodic. Let us
discuss these characteristics in more detail.

The properties of Gaussian random processes are well-known in
Statistical Optics. For instance, the real and imaginary part of
the complex amplitudes of the electric field from a polarized
thermal source have Gaussian distribution, while the instantaneous
radiation power fluctuates in accordance with the negative
exponential distribution. Gaussian statistics alone, guarantees
that higher-order correlation functions can be expressed in terms
of second-order correlation functions. Moreover, it can be shown
\cite{GOOD} that a linearly filtered Gaussian process is also a
Gaussian random process. As a result, the presence of a spectral
filter (monochromator) and a spatial filter as in the system
depicted in Fig. \ref{thermal} do not change the statistics of the
signal, because they simply act as linear filters.

Stationarity is a subtle concept. There are different kinds of
stationarity. Strict-stationarity means that all ensemble averages
are independent on time. Wide-sense stationarity means that the
signal average is independent on time and that the second order
correlation function in time depends only on the difference of the
observation times. However, for Gaussian processes strict and
wide-sense stationarity coincide \cite{GOOD,MAND}. As a
consequence of the definition of stationarity, necessary condition
for a certain process to be stationary is that the signal last
forever. Yet, if a signal lasts much longer than its coherence
time $\tau_c$ (which fixes the short-scale duration of the field
fluctuations) and it is observed for a time much shorter than its
duration $\sigma_T$, but much longer than its coherence time it
can be reasonably considered as everlasting and it has a chance to
be stationary as well, as in the case of thermal light.

Ergodicity is a stronger requirement than stationarity.
Qualitatively, we may state that if, for a given random process
all ensemble averages can be substituted by time averages, the
process under study is said to be ergodic: all the statistical
properties of the process can be derived from one single
realization. A process must be strictly stationary in order to be
ergodic. There exist stationary processes which are not ergodic.
One may consider, for instance, the  random constant process: this
is trivially strictly stationary, but not ergodic because a single
(constant) realization of the process does not allow one to
characterize the process from a statistical viewpoint. However,
this is a pathologic case when both the coherence time $\tau_c$
and the duration time of the signal $\sigma_T$ are infinite. On
the contrary, a stationary process like the radiation from an
incandescent lamp driven by a constant current has, virtually,
infinite duration. In this case different ensembles are simply
different observations, for given time intervals, of the same,
statistically identical phenomenon: then, the concept of ensemble
average and time average are equivalent and the process is also
ergodic.

Statistical Optics was developed starting with signals
characterized by  Gaussian statistics, stationarity and
ergodicity. Let us consider any Synchrotron radiation source. Like
thermal light, also Synchrotron Radiation is a random process. In
fact, relativistic electrons in a storage ring emit Synchrotron
Radiation passing through bending magnets or undulators. The
electron beam shot noise causes fluctuations of the beam density
which are random in time and space from bunch to bunch. As a
result, the radiation produced has random amplitudes and phases.
As already declared in the Introduction we will demonstrate that
the SR field obeys Gaussian statistics. In contrast with thermal
light though, Synchrotron Radiation is intrinsically
non-stationary (and, therefore, non-ergodic) because even if its
short pulse duration cannot be resolved by detectors working in
the time domain, it can nonetheless be resolved by detectors
working in frequency domain. For this reason, in what follows the
averaging brackets $\langle ... \rangle$ will always indicate the
ensemble average over bunches.
\begin{figure}
\begin{center}
\includegraphics*[width=140mm]{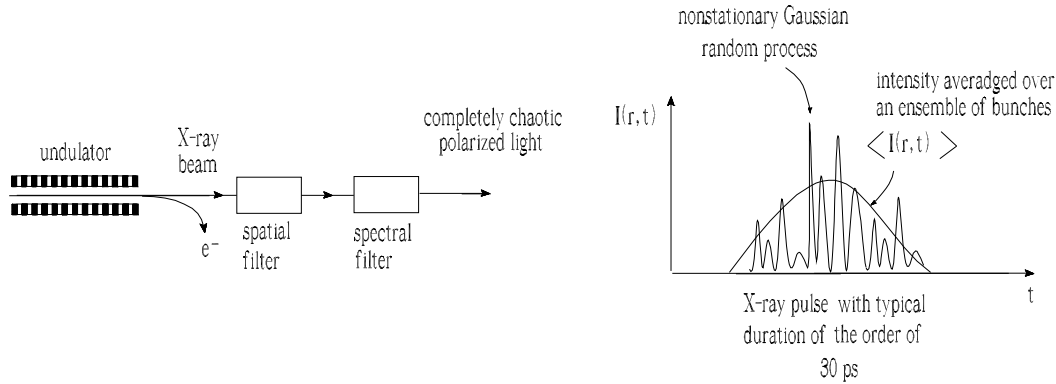}% Here is how to import EPS art
\caption{\label{SR} The intensity of an X-ray beam from a
Synchrotron Radiation source. A statistically non-stationary wave
has a time-varying intensity averaged over an ensemble of
bunches.}
\end{center}
\end{figure}
In spite of differences with respect to the simpler case of
thermal light, as we will see in this paper, also Synchrotron
Radiation fields can be described in terms of Statistical Optics.
Fig. \ref{SR} shows the geometry of the experiment under
consideration. The problem is to describe the statistical
properties of Synchrotron Radiation at the detector installed
after the spatial and spectral filters. Radiation at the detector
consists of a carrier modulation of frequency $\omega$ subjected
to random amplitude and phase modulation. The Fourier
decomposition of the radiation contains frequencies spread about
the monochromator bandwidth $\Delta \omega_\mathrm{m}$: it is not
possible, in practice, to resolve the oscillations of the
radiation fields which occur at the frequency of the carrier
modulation. It is therefore appropriate, for comparison with
experimental results, to average the theoretical results over a
cycle of oscillations of the carrier modulation.

\begin{figure}
\begin{center}
\includegraphics*[width=140mm]{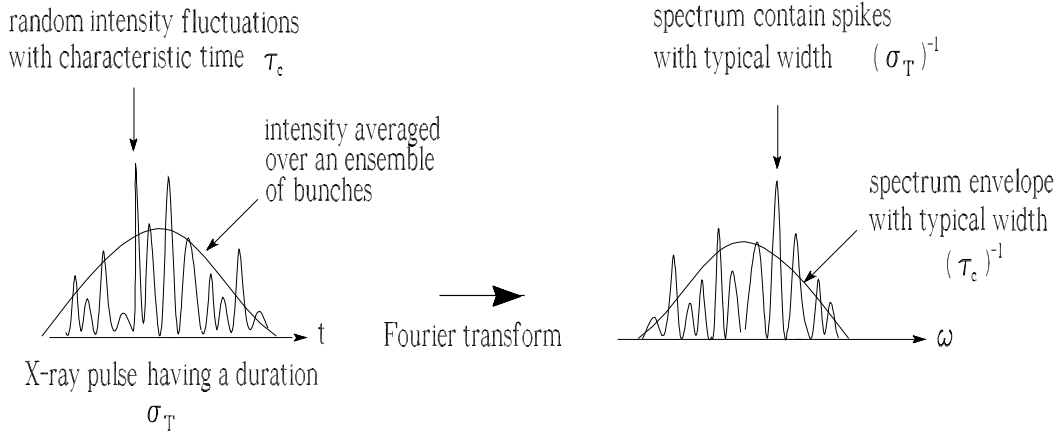}% Here is how to import EPS art
\caption{\label{avecarr} Reciprocal width relations of Fourier
transform pairs.}
\end{center}
\end{figure}
Fig. \ref{avecarr} gives a qualitative illustration of the type of
fluctuations that occur in cycle-averaged Synchrotron Radiation
beam intensity. Within some characteristic time, a given random
function appears to be smooth, but when observed at larger scales
the same random function exhibits "rough" variations. The time
scale of random fluctuations is the coherence time $\tau_c$. When
$\tau_c \ll \sigma_T$ %, $\sigma_T$ being the time scale of the
%electron beam duration,
the radiation beyond the monochromator is
partially coherent. This case is shown in Fig. \ref{avecarr}:
there, we can estimate $\tau_c \simeq \Delta
\omega_\mathrm{m}^{-1}$. If the radiation beyond the monochromator
is partially coherent, a spiky spectrum is to be expected. The
nature of the spikes is easily described in terms of Fourier
transform theory. We can expect that the typical width of the
spectrum envelope should be of order of $\Delta \omega/\omega \sim
(\tau_c \omega)^{-1}$. Also, the spectrum of the radiation from a
bunch with typical duration $\sigma_T$ at the source plane should
contain spikes with characteristic width $\Delta \omega/\omega
\simeq (\omega \sigma_T)^{-1}$, as a consequence of the reciprocal
width relations of Fourier transform pairs (see, again, Fig.
\ref{avecarr}).

\subsection{\label{sub:seco} Second-order correlations in
space-frequency domain}

We start our discussion in the most generic way possible,
considering a fixed polarization component of the Fourier
transform at frequency $\omega$ of the electric field produced at
location $(z_o, \vec{r}_{\bot o})$, in some cartesian coordinate
system, by a given collection of sources. We will denote it with
${\bar{E}}_\bot(z_o, \vec{r}_{\bot o}, \omega)$ and it will be
linked to the time domain field ${{E}}_\bot(z_o, \vec{r}_{\bot o},
t)$ through the Fourier transform

\begin{equation}
\bar{{E}}_\bot(\omega) = \int_{-\infty}^{\infty} dt {{E}}_\bot(t)
e^{i \omega t}~, \label{ftran}
\end{equation}
so that
\begin{equation}
{{E}}_\bot(t) = {1\over{2\pi}} \int_{-\infty}^{\infty} d\omega
\bar{{E}}_\bot(\omega) e^{-i \omega t}~.
 \label{fanti}
\end{equation}
This very general collection of sources includes the case of an
ultra relativistic electron beam going through a certain magnetic
system and in particular an undulator, which is our case of
interest. In this case $z_o$ is simply the observation distance
along the optical axis of the system and $\vec{r}_{\bot o}$ are
the transverse coordinates of the observer on the observation
plane. The contribution of the $k$-th electron to the field
Fourier transform at the observation point depends on the
transverse offset $({l}_{x k},{l}_{y k})$ and deflection angles
$(\eta_{x k},\eta_{y k})$ that the electron has at the entrance of
the system with respect to the optical axis. Moreover, an arrival
time $t_k$ at the system entrance has the effect of multiplying
the field Fourier transform by a phase factor $\exp{(i\omega
t_k)}$ (that is, in time domain the electric field is retarded by
a time $t_k$). At this point we do not need to specify explicitly
the dependence on offset and deflection. The total field Fourier
transform can be written as

\begin{equation}
{\bar{E}}_\bot(z_o,\vec{r}_{\bot o},\omega)=\sum_{k=1}^{N}
\bar{E}_{s\bot}(\vec{\eta}_k,\vec{l}_k,z_o,\vec{r}_{\bot
o},\omega) \exp{(i\omega t_k)} ~, \label{total}
\end{equation}
where $\vec{\eta}_k,\vec{l}_k$ and $t_k$ are random variables and
$N$ is the number of electrons in the beam. It follows from Eq.
(\ref{total}) that the Fourier transform of the Synchrotron
Radiation pulse at a fixed frequency and a fixed point in space is
a sum of a great many independent contributions, one for each
electron, of the form
$\bar{E}_{s\bot}(\vec{\eta}_k,\vec{l}_k,z_o,\vec{r}_{\bot
o},\omega) \exp{(i\omega t_k)}$. For simplicity we make three
assumptions about the statistical properties of elementary phasors
composing the sum, which are generally satisfied in Synchrotron
Radiation problems of interest.

1) We assume that for a beam circulating in a storage ring random
variables $t_n$ are independent from $\vec{\eta}_n$ and
$\vec{l}_n$. This is always verified, because the random arrival
times of electrons, due to shot noise, do not depend on the
electrons offset and deflection with respect to the $z$-direction.
Eq. (\ref{total}) states that the $k$-th elementary contribution
to the total ${\bar{E}}_\bot$ can be written as a product of the
complex phasors $\exp{(i\omega t_k)}$, and $\bar{E}_{s\bot}$ that,
in its turn, can be written as a product of modulus and phase as
$\bar{E}_{s\bot}(\vec{\eta}_k,\vec{l}_k,z_o,\vec{r}_{\bot
o},\omega) = \mid \bar{E}_{s\bot k} \mid \exp{(i \phi_k)}$. Under
the assumption of statistical independence of $t_n$ from
$\vec{\eta}_n$ and $\vec{l}_n$  the complex phasors $\exp{(i\omega
t_k)}$, and $\bar{E}_{s\bot}$ are statistically independent of
each other and of all the other elementary phasors for different
values of $k$. The ensemble average of a given function $f$ of
random variables $\vec{\eta}_n$, $\vec{l}_n$ and $t_n$ is by
definition:

\begin{eqnarray}
\langle f(\vec{\eta}_n ,\vec{l}_n, t_n) \rangle =
\int_{-\infty}^{\infty} d \eta_{x n} \int_{-\infty}^{\infty} d
\eta_{y n}\int_{-\infty}^{\infty} d l_{x n}\int_{-\infty}^{\infty}
d l_{y n}\int_{-\infty}^{\infty} d t_n &&\cr \times f(\vec{\eta}_n
,\vec{l}_n, t_n)
P(\vec{\eta}_n,\vec{l}_n,t_n)~,\label{ensembledef}
\end{eqnarray}
where $P(\vec{\eta}_n,\vec{l}_n,t_n)$ is the probability density
distribution in the joint random variables $\vec{\eta}_n$,
$\vec{l}_n$, $t_n$. Independence of $t_n$ from $\vec{\eta}_n$ and
$\vec{l}_n$ allows us to write

\begin{equation}
P(\vec{\eta}_n,\vec{l}_n,t_n) =
F_{\eta_{x},l_{x}}(\eta_{xn},l_{xn})
F_{\eta_{y},l_{y}}(\eta_{yn},l_{yn})
 F_{t}(t_n)~,
\label{independence}
\end{equation}
where we also assumed that the distribution in the horizontal and
vertical planes are not correlated. Since electrons arrival times
are completely uncorrelated from transverse coordinates and
offsets, the shapes of $F_{\eta_{x},l_{x}}$, $F_{\eta_{y},l_{y}}$
and $F_{t}$ are the same for all electrons.

2) We assume that the random variables  $\mid \bar{E}_{s\bot k}
\mid$ (at fixed frequency $\omega$), are identically distributed
for all values of $k$, with a finite mean $\langle \mid
\bar{E}_{s\bot k} \mid \rangle$ and a finite second moment
$\langle \mid \bar{E}_{s\bot k} \mid^2 \rangle $. This is always
the case in practice because electrons are identical particles.

3) We assume that the electron bunch duration $\sigma_T$ is large
enough so that $\omega \sigma_T \gg 1$: under this assumption the
phases $\omega t_k$ can be regarded as uniformly distributed on
the interval $(0, 2\pi)$. The assumption $\omega \sigma_T \gg 1$
is justified by the fact that $\omega$ is the undulator resonant
frequency, which is high enough to guarantee that $\omega \sigma_T
\gg 1$ for any practical choice of  $\sigma_T$.

\begin{figure}
\begin{center}
\includegraphics*[width=140mm]{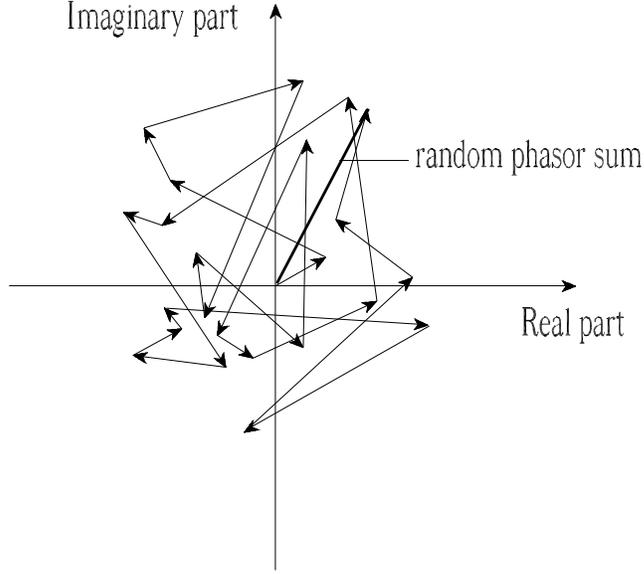}% Here is how to import EPS art
\caption{\label{phasor} Amplitude and phase of the resultant
vector (total complex amplitude) formed by a large number of
complex phasors having random length and random phase.}
\end{center}
\end{figure}
The formal summation of phasors with random lengths and phases is
illustrated in Fig. \ref{phasor}. Under the three previously
discussed assumptions we can use the central limit theorem to
conclude that the real and the imaginary part of $\bar{E}_\bot$
are distributed in accordance to a Gaussian law. Detailed proof of
this fact is given in Appendix A. As a result, Synchrotron
Radiation is a Gaussian random process and second-order field
correlation function is all we need in order to specify the field
statistical properties. In fact, as already remarked, higher-order
correlation functions can be expressed in terms of second-order
correlation functions.

In Synchrotron Radiation experiments with third generation light
sources detectors are limited to about $100$ ps time resolution
and are by no means able to resolve a single X-ray pulse in time
domain: they work, instead, by counting the number of photons at a
certain frequency over an integration time longer than the pulse.
Therefore, for Synchrotron Radiation related issues the frequency
domain is much more natural a choice than the time domain, and we
will deal with signals in the frequency domain throughout this
paper. The knowledge of the second-order field correlation
function in frequency domain

\begin{equation}
\Gamma_{\omega}(z_o,\vec{r}_{\bot o1},\vec{r}_{\bot
o2},\omega,\omega') = \left< {\bar{E}}_\bot(z_o,\vec{r}_{\bot
o1},\omega){\bar{E}}^*_\bot(z_o,\vec{r}_{\bot o2},\omega')
\right>~, \label{gamma}
\end{equation}
is all we need to completely characterize the signal from a
statistical viewpoint. For the sake of completeness it is
nonetheless interesting to remark that it is possible (and often
done, in Statistical Optics) to give equivalent descriptions of
the process in time domain as well. First, note that the time
domain process ${E}_\bot(t)$ is linked to
${\bar{{E}}}_\bot(\omega)$ by Fourier transform, and that a
linearly filtered Gaussian process is also a Gaussian process (see
\cite{GOOD} 3.6.2). As a result, ${E}_\bot(t)$ is a Gaussian
process as well. Second,  the operation of ensemble average is
linear with respect to Fourier transform integration. This
guarantees, that the knowledge of $\Gamma_{\omega}$ in frequency
domain is completely equivalent to the knowledge of the
second-order correlation function between
${E}_\bot(z_o,\vec{r}_{\bot o1},t_1)$ and
${E}_\bot(z_o,\vec{r}_{\bot o2},t_2)$. The latter is usually known
as mutual coherence function and was first introduced in
\cite{WOLF}:

\begin{equation}
\Gamma_{t}(z_o,\vec{r}_{\bot o1},\vec{r}_{\bot o2},t_1,t_2) =
\left< {{E}}_\bot(z_o,\vec{r}_{\bot
o1},t_1){{E}}^*_\bot(z_o,\vec{r}_{\bot o2},t_2) \right>~.
\label{gammatime}
\end{equation}
For the rest of this paper we will abandon almost entirely any
reference to the time domain and work consistently in frequency
domain with the help of Eq. (\ref{gamma}) because, as has already
been said, this is a natural choice for Synchrotron Radiation
applications. In particular, as has already been anticipated,
under non-restrictive assumptions on characteristic bandwidths of
the process, it is possible to break the correlation function
$\Gamma_{\omega} (\omega,\omega')$ in space-frequency domain in
the product of two factors, the spectral correlation function
${F}_\omega(\omega-\omega')$ , and the cross-spectral density of
the process  $G_{\omega}(z_o,\vec{r}_{\bot o1},\vec{r}_{\bot o2},
\omega)$ \cite{MAND}. The cross-spectral density can be studied
independently at any given frequency giving information on the
spatial correlation of the field. Subsequently, the knowledge of
the spectral correlation function brings back the full expression
for $\Gamma_\omega$.

Substituting Eq. (\ref{total}) in Eq. (\ref{gamma}) one has

\begin{eqnarray}
\Gamma_{\omega}(z_o,\vec{r}_{\bot o1},\vec{r}_{\bot o2}, \omega,
\omega') = \left<\sum_{m=1}^{N} \bar{E}_{s\bot
}(\vec{\eta}_m,\vec{l}_m,z_o,\vec{r}_{\bot o1}, \omega)
\right.&&\cr \times \left. \sum_{n=1}^{N}
\bar{E}^*_{s\bot}(\vec{\eta}_n,\vec{l}_n,z_o,\vec{r}_{\bot o2},
\omega') \exp{[i( \omega t_m- \omega' t_n)]} \right>~.
\label{gamma2}
\end{eqnarray}
Expanding Eq. (\ref{gamma2}) one has

\begin{eqnarray}
\Gamma_{\omega}(z_o,\vec{r}_{\bot o1},\vec{r}_{\bot o2}, \omega,
\omega')
 = \sum_{m=1}^{N} \Bigg\langle \bar{E}_{s\bot}
(\vec{\eta}_m,\vec{l}_m,z_o,\vec{r}_{\bot o1}, \omega) && \cr
\times \bar{E}^*_{s\bot}(\vec{\eta}_m,\vec{l}_m,z_o,\vec{r}_{\bot
o1}, \omega') \exp{[i( \omega- \omega') t_m]} \Bigg\rangle &&\cr +
\sum_{m\ne n} \Bigg\langle \bar{E}_{s\bot
}(\vec{\eta}_m,\vec{l}_m,z_o,\vec{r}_{\bot o1}, \omega) \exp{(i
 \omega t_m)}\Bigg\rangle&&\cr \times
\Bigg\langle\bar{E}^*_{s\bot}
(\vec{\eta}_n,\vec{l}_n,z_o,\vec{r}_{\bot o2}, \omega') \exp{(- i
 \omega' t_n)} \Bigg\rangle~. \label{gamma3}
\end{eqnarray}
With the help of Eq. (\ref{ensembledef}) and Eq.
(\ref{independence}), the ensemble average $\langle \exp{(i \omega
t_k)} \rangle_{t}$ can be written as the Fourier transform of the
bunch longitudinal profile function $F_{t}(t_k)$, that is

\begin{equation}
\langle \exp{(i \omega t_k)} \rangle_{t} = \int_{-\infty}^{\infty}
d t_k F_{t}(t_k) e^{i\omega t_k} = {F}_{\omega}(\omega)~.
\label{FTlong}
\end{equation}
Using Eq. (\ref{FTlong}), Eq. (\ref{gamma3}) can be written as

\begin{eqnarray}
\Gamma_{\omega}(z_o,\vec{r}_{\bot o1},\vec{r}_{\bot o2}, \omega,
\omega')
 = \sum_{m=1}^{N} {F}_{\omega}( \omega- \omega')  \Bigg
\langle \bar{E}_{s\bot} (\vec{\eta}_m,\vec{l}_m,z_o,\vec{r}_{\bot
o1}, \omega) && \cr \times
\bar{E}^*_{s\bot}(\vec{\eta}_m,\vec{l}_m,z_o,\vec{r}_{\bot o2},
\omega')  \Bigg\rangle_{\vec{\eta},\vec{l}} + \sum_{m\ne n}
{F}_{\omega}( \omega){F}_{\omega}(- \omega') && \cr \times
\Bigg\langle
\bar{E}_{s\bot}(\vec{\eta}_m,\vec{l}_m,z_o,\vec{r}_{\bot o1},
\omega) \Bigg\rangle_{\vec{\eta},\vec{l}}
\Bigg\langle\bar{E}^*_{s\bot}(\vec{\eta}_n,
\vec{l}_n,z_o,\vec{r}_{\bot o2}, \omega')
\Bigg\rangle_{\vec{\eta},\vec{l}} ~, \label{gamma4}
\end{eqnarray}
where ${F}_{\omega}^*( \omega')={F}_{\omega}(- \omega')$ because
${F}_{t}$ is a real function. When the radiation wavelengths of
interest are much shorter than the bunch length we can safely
neglect the second term on the right hand side of Eq.
(\ref{gamma4}) since the form factor product ${F}_{\omega}(
\omega) {F}_{\omega}(- \omega')$ goes rapidly to zero for
frequencies larger than the characteristic frequency associated
with the bunch length: think for instance, at a centimeter long
bunch compared with radiation in the Angstrom wavelength range. It
should be noted, however, that when the radiation wavelength of
interested is longer than the bunch length the second term in Eq.
(\ref{gamma4}) is dominant with respect to the first, because it
scales with the number of particles \textit{squared}: in this
case, analysis of the second term leads to a treatment of Coherent
Synchrotron Radiation phenomena (CSR). In this paper we will not
be concerned with CSR and we will neglect the second term in Eq.
(\ref{gamma4}), assuming that the radiation wavelength of interest
is shorter than the bunch length: then, it should be noted that
${F}_{\omega}( \omega- \omega')$ depends on the
\textit{difference} between $ \omega$ and $ \omega'$, and the
first term cannot be neglected. We can therefore write

\begin{eqnarray}
\Gamma_{\omega}(z_o,\vec{r}_{\bot o1},\vec{r}_{\bot o2}, \omega,
\omega')
 = \sum_{m=1}^{N} {F}_{\omega}( \omega- \omega')  &&\cr \times \Bigg
\langle \bar{E}_{s\bot} (\vec{\eta}_m,\vec{l}_m,z_o,\vec{r}_{\bot
o1}, \omega)
\bar{E}^*_{s\bot}(\vec{\eta}_m,\vec{l}_m,z_o,\vec{r}_{\bot o2},
\omega') \Bigg\rangle_{\vec{\eta},\vec{l}} && \cr = N
{F}_{\omega}( \omega- \omega')   \Bigg \langle \bar{E}_{s\bot}
(\vec{\eta},\vec{l},z_o,\vec{r}_{\bot o1}, \omega)
\bar{E}^*_{s\bot}(\vec{\eta},\vec{l},z_o,\vec{r}_{\bot o2},
\omega') \Bigg\rangle_{\vec{\eta},\vec{l}} ~. \label{gamma5}
\end{eqnarray}
As one can see from Eq. (\ref{gamma5}) each electron is correlated
just with itself: cross-correlation terms between different
electrons was, in fact, included in the second term on the right
hand side of Eq. (\ref{gamma4}), which has been dropped. It is
important to note that if the dependence of $\bar{E}_{s\bot}$ on $
\omega$ and $ \omega'$ is slow enough, so that $\bar{E}_{s\bot}$
does not vary appreciably on the characteristic scale of
${F}_{\omega}$, we can substitute
$\bar{E}^*_{s\bot}(\vec{\eta},\vec{l},z_o,\vec{r}_{\bot o2},
\omega')$ with
$\bar{E}^*_{s\bot}(\vec{\eta},\vec{l},z_o,\vec{r}_{\bot o2},
\omega)$ in Eq. (\ref{gamma5}).
\begin{figure}
\begin{center}
\includegraphics*[width=110mm]{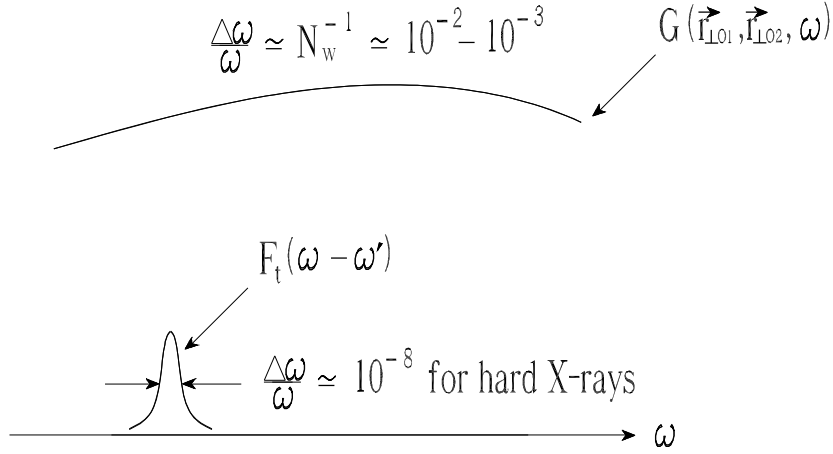}% Here is how to import EPS art
\caption{\label{rela} Schematic illustration of the relative
frequency dependence of the spectral correlation function
$F_\omega (\omega-\omega')$ and of the cross-spectral density
function (the cross-power spectrum) $G_{\omega}(z_o,\vec{r}_{\bot
o1},\vec{r}_{\bot o2}, \omega)$ of the Synchrotron Radiation at
points $\vec{r}_{\bot o1}$ and $\vec{r}_{\bot o2}$ at frequency
$\omega$.}
\end{center}
\end{figure}
The situation is depicted in Fig. \ref{rela}. On the one hand, the
characteristic scale of ${F}_{\omega}$ is given by $1/\sigma_T$,
where $\sigma_T$ is the characteristic bunch duration. On the
other hand, the bandwidth of single particle undulator radiation
at resonance is given by $\omega_o/N_w$, where $\omega_o$ is the
resonant frequency and $N_w$ is the number of undulator periods
(of order $10^2 - 10^3$). In the case of an electron beam the
undulator spectrum will exhibit a longer tail, as has been shown
in \cite{OURS}, which guarantees that $\omega_o/N_w$ is, indeed, a
minimum for the radiation bandwidth, and is the right quantity to
be compared with $1/\sigma_T$. As an example, for wavelengths of
order $1 \AA$, $N_w \sim 10^3$ and $\sigma_T \sim 30$ ps (see
\cite{PETR}), $\omega_o/N_w \sim 2\cdot 10^{16}$ Hz which is much
larger than $1/\sigma_T \sim 3\cdot 10^{10}$ Hz. As a result we
can simplify Eq. (\ref{gamma5}) to

\begin{eqnarray}
\Gamma_{\omega}(z_o,\vec{r}_{\bot o1},\vec{r}_{\bot o2}, \omega,
\omega')
 = N {F}_{\omega}( \omega- \omega') G_{\omega}(z_o,\vec{r}_{\bot o1},\vec{r}_{\bot o2}, \omega)
\label{gamma6prima}
\end{eqnarray}
where

\begin{equation}
G_{\omega}(z_o,\vec{r}_{\bot o1},\vec{r}_{\bot o2}, \omega)=
\Bigg\langle \bar{E}_{s\bot} (\vec{\eta},\vec{l},z_o,\vec{r}_{\bot
o1}, \omega)
\bar{E}^*_{s\bot}(\vec{\eta},\vec{l},z_o,\vec{r}_{\bot o2},
\omega) \Bigg\rangle_{\vec{\eta},\vec{l}} ~.\label{coore}
\end{equation}
Eq. (\ref{gamma6prima}) fully characterizes the system under study
from a statistical viewpoint. However, in practical situations,
the observation plane is behind a monochromator or, equivalently,
the detector itself is capable of analyzing the energy of the
photons. The presence of a monochromator simply modifies the right
hand side Eq. (\ref{gamma6prima}) for a factor
$T(\omega)T^*(\omega')$, where $T$ is the monochromator transfer
function:

\begin{eqnarray}
\Gamma_{\omega}(z_o,\vec{r}_{\bot o1},\vec{r}_{\bot o2}, \omega,
\omega')
 = N {F}_{\omega}( \omega- \omega') T(\omega) T^*(\omega')
 G_{\omega}(z_o,\vec{r}_{\bot o1},\vec{r}_{\bot o2}, \omega)~.
\label{gamma6}
\end{eqnarray}
Independently on the characteristics (and even on the presence) of
the monochromator, it should be noted that both in Eq.
(\ref{gamma6prima}) and Eq. (\ref{gamma6}), correlation in
frequency and space are expressed by two separate factors. In
particular, in both these equation, spatial correlation is
expressed by the cross-spectral density function
$G_{\omega}(z_o,\vec{r}_{\bot o1},\vec{r}_{\bot o2}, \omega)$. In
other words, we are able to deal separately with spatial and
spectral part of the correlation function in space-frequency
domain with the only non-restrictive assumption that $\omega_o/N_w
\gg 1/\sigma_T$. From now on we will be concerned with the
calculation of the correlation function
$G_{\omega}(z_o,\vec{r}_{\bot o1},\vec{r}_{\bot o2}, \omega)$,
independently on the shape of the remaining factors on the right
hand side of Eq. (\ref{gamma6}) which can have a simple or a
complicated structure, accounting for the characteristics of the
monochromator.

Before proceeding with the analysis of
$G_{\omega}(z_o,\vec{r}_{\bot o1},\vec{r}_{\bot o2}, \omega)$
though, let us spend some words on these remaining factors; the
presence of a monochromator introduces another bandwidth of
interest. If we indicate the bandwidth of the monochromator with
$\Delta \omega_\mathrm{m}$ and the central frequency of interest
at which the monochromator is tuned with $ \omega_o$ (typically,
the undulator resonant frequency), then $T$ is peaked around $
\omega_o$ and goes rapidly to zero as we move out of the range $(
\omega_o-\Delta \omega_\mathrm{m}/2, \omega_o+\Delta
\omega_\mathrm{m}/2)$. Now, if  the characteristic bandwidth  of
the monochromator, $\Delta \omega_\mathrm{m}$, is large enough so
that $T$ does not vary appreciably on the characteristic scale of
${F}_{\omega}$, i.e. $\Delta \omega_\mathrm{m} \gg 1/\sigma_T$,
then ${F}_{\omega}(\omega- \omega')$ is peaked at
$\omega=\omega'$. In this case the process resembles more and more
a stationary process, although it will be still intrinsically
non-stationary. Consider a signal observed for a time much shorter
than its duration, but much longer than its coherence time, and
such that the stationary model applies to it. Now imagine that we
extend the observation time to a duration  which is still much
shorter than the signal duration, but long enough that we need to
account for the intrinsical non-stationarity of the process due to
finite signal duration. In this case the stationary model does not
apply anymore strictly. To describe this situation, we can define
a property weaker than stationarity, but nonetheless very
interesting from a physical standpoint: quasi-stationarity. The
time domain correlation function (that is, the mutual coherence
function) can be written as

\begin{eqnarray}
\Gamma_t(z_o,\vec{r}_{\bot o1},\vec{r}_{\bot o2},t_1,t_2) =
\frac{N}{(2\pi)^2} \int_{-\infty}^{\infty} d\omega
\int_{-\infty}^{\infty} d\omega' {F}_{\omega}( \omega- \omega')
T(\omega) T^*(\omega') &&\cr \times  G_{\omega}(z_o,\vec{r}_{\bot
o1},\vec{r}_{\bot o2},\omega) \exp{(-i\omega t_1)} \exp{(i\omega'
t_2)} ~.\label{trasfgammabreak}
\end{eqnarray}
When $\Delta \omega_\mathrm{m} \gg 1/\sigma_T$, and with the help
of new variables $\Delta \omega = \omega - \omega'$ and $\omega$,
we can simplify Eq. (\ref{trasfgammabreak}) accounting for the
fact that ${F}_{\omega}(\omega- \omega')$ is strongly peaked
around $\Delta \omega=0$. In fact we can consider $T(\omega)
T^*(\omega') G_{\omega}(z_o,\vec{r}_{\bot o1},\vec{r}_{\bot
o2},\omega) \simeq |T({\omega})|^2 G_{\omega}(z_o,\vec{r}_{\bot
o1},\vec{r}_{\bot o2}, {\omega})$, so that we can integrate
separately in $\Delta \omega$ and ${\omega}$ to obtain

\begin{eqnarray}
\Gamma_t(z_o,\vec{r}_{\bot o1},\vec{r}_{\bot o2},t_1,t_2) =
\frac{N}{(2\pi)^2} \int_{-\infty}^{\infty} d\Delta \omega
~{F}_{\omega}(\Delta \omega) \exp{\left(-i \Delta{\omega}
{t_2}\right)} &&\cr \times \int_{-\infty}^{\infty} d{\omega}
~|T({\omega})|^2 G_{\omega}(z_o,\vec{r}_{\bot o1},\vec{r}_{\bot
o2},{\omega}) \exp{\left[-i {\omega} (t_1-t_2)\right]} &&\cr =
F_t\left({t_2}\right) G_t\left(z_o,\vec{r}_{\bot o1},\vec{r}_{\bot
o2},{t_1-t_2}\right)~.\label{trasfgammabreak2}
\end{eqnarray}
In other words, in the quasi-stationary case,
$\Gamma_t(z_o,\vec{r}_{\bot o1},\vec{r}_{\bot o2},t_1,t_2)$ is
split on the product of two factors, a "reduced mutual coherence
function", that is $G_t(z_o,\vec{r}_{\bot o1},\vec{r}_{\bot o2},
t_1-t_2)$, and an intensity profile, that is $F(t_2)$.

If we now assume $\Delta \omega_\mathrm{m} N_w / \omega_o \ll 1$
(that is usually true), $G_{\omega}(z_o,\vec{r}_{\bot
o1},\vec{r}_{\bot o2},\omega)$ $ = G_{\omega}(z_o,\vec{r}_{\bot
o1},\vec{r}_{\bot o2},\omega_o)$ is a constant function of
frequency within the monochromator line. In this case,
$G_{\omega}$ contains all the information about spatial
correlations between different point and is, in fact, the quantity
of central interest in our study, but it is independent on the
frequency $\omega$. As a result we have

\begin{equation}
\Gamma_t(z_o,\vec{r}_{\bot o1},\vec{r}_{\bot o2},t_1,t_2) = N
g_t(t_1-t_2) F_t\left({t_2}\right) G_\omega\left(z_o,\vec{r}_{\bot
o1},\vec{r}_{\bot o2},\omega_o\right)~, \label{CScase}
\end{equation}
which means that the mutual coherence function
$\Gamma_t(z_o,\vec{r}_{\bot o1},\vec{r}_{\bot o2},t_1,t_2)$ is
reducible, in the sense that it can be factorized as a product of
two factors, the first $N g_t(t_1-t_2) F_t({t_2})$, characterizing
the temporal coherence and the second $ G_\omega(z_o,\vec{r}_{\bot
o1},\vec{r}_{\bot o2},\omega_o)$ describing the spatial coherence
of the system\footnote{It is interesting, for the sake of
completeness, to discuss the relation between $G_\omega$ and the
mutual intensity function as usually defined in textbooks
\cite{GOOD, MAND} in \textit{quasimonochromatic} conditions. The
assumption $\Delta \omega_\mathrm{m} \gg 1/\sigma_T$ in the limit
$\sigma_T \longrightarrow \infty$ describes a stationary process.
Now letting $\Delta \omega_m \longrightarrow 0$ slowly enough so
that $\Delta \omega_\mathrm{m} \gg 1/\sigma_T$,  Eq.
(\ref{trasfgammabreak2}) remains valid while both $F_\omega$ and
$|T({\omega})|^2$ become approximated better and better by Dirac
$\delta$-functions, $\delta(\Delta\omega)$ and
$\delta(\omega-\omega_o)$, respectively. Then $\Gamma_t \sim
G_\omega \exp [-i \omega_o (t_1-t_2)]$. Aside for an unessential
factor, depending on the normalization of $F_\omega$ and
$|T({\omega})|^2$, this relation between $\Gamma_t$ and $G_\omega$
allows identification of the mutual intensity function with
$G_\omega$ as in \cite{GOOD,MAND}.}. This case is of practical
importance. In fact for $1 \AA$ radiation we typically have
$\Delta \omega_\mathrm{m} /\omega_o \simeq 10^{-4} \div 10^{-5}$
and $N_w \simeq 10^2 \div 10^3$, i.e. $\Delta \omega_\mathrm{m}
N_w / \omega_o \ll 1$. It should be noted that, although Eq.
(\ref{CScase}) describes the case when $\Delta \omega_\mathrm{m}
N_w / \omega_o \ll 1$ \textit{and} $\Delta \omega_\mathrm{m} \gg
1/\sigma_T$ , only the former assumption is important for the
mutual coherence function to be reducible. In fact, if the former
is satisfied but the latter is not, from Eq.
(\ref{trasfgammabreak}) one would simply have:

\begin{equation}
\Gamma_t(z_o,\vec{r}_{\bot o1},\vec{r}_{\bot o2},t_1,t_2) = N
\widetilde{g}_t(t_1,t_2) G_\omega\left(z_o,\vec{r}_{\bot
o1},\vec{r}_{\bot o2},\omega_o\right)~, \label{CScasegen}
\end{equation}
that is still reducible.

Eq. (\ref{trasfgammabreak2}) contains two important facts:

(a) The temporal correlation function, that is
$G_t(z_o,\vec{r}_{\bot o1},\vec{r}_{\bot o2},{t_1-t_2})$, and the
spectral density distribution of the source, that is
$H(z_o,\vec{r}_{\bot o1},\vec{r}_{\bot o2},\omega) = N
|T({\omega})|^2  G_{\omega}(z_o, \vec{r}_{\bot o1}, {\vec{r}_{\bot
o2}}, {\omega})$, form a Fourier pair.

(b) The intensity distribution of the radiation pulse $F_t({t_2})$
and the spectral correlation function ${F}_{\omega}(\Delta
\omega)$ form a Fourier pair.

The statement (a) can be regarded as an analogue, for
quasi-stationary sources, of the well-known Wiener-Khinchin
theorem, which applies to stationary sources and states that the
temporal correlation function and the spectral density are a
Fourier pair.  Since there is symmetry between time and frequency
domains, a "anti" Wiener-Khinchin theorem must also hold, and can
be obtained by the usual Wiener-Khinchin theorem by exchanging
frequencies and times. This is simply constituted by the statement
(b).

The assumption of quasi-stationarity is not vital for the
following of this work, since the cross-spectral density can be
studied in any case as a function of frequency. In this respect it
should be noted that, although in the large majority of the cases
monochromator characteristics are not good enough to allow
resolution of the non-stationary process, there are cases when it
is not allowed to treat the process as if it were
quasi-stationary. For instance, in \cite{NOST} a particular
monochromator is described with a relative resolution of $10^{-8}$
at wavelengths of about $1 \AA$, or $\omega_o \sim 2 \cdot
10^{19}$ Hz. Let us consider, as in \cite{NOST}, the case of
radiation pulses of $32$ ps duration. Under the already accepted
assumption $1/\sigma_T \ll \omega_o/N_w$, we can identify the
radiation pulse duration with $\sigma_T$. Then we have $\Delta
\omega_\mathrm{m} \sim 2 \cdot 10^{11}$ Hz  which is of order of
$2 \pi/\sigma_T \sim 2 \cdot 10^{11}$ Hz: this means that the
monochromator has the capability of resolving the non-stationary
processes in the frequency domain. On the contrary, also in this
case, the mutual coherence function is reducible, in the sense
specified before,  because $\Delta \omega_\mathrm{m} /\omega_o
\simeq 10^{-8}$ and $N_w \simeq 10^2 \div 10^3$, i.e. $\Delta
\omega_\mathrm{m} N_w / \omega_o \ll 1$. However, such accuracy
level is not usual in Synchrotron Radiation experiments.

To sum up, condition $\omega_o/N_w \gg 1/\sigma_T$ alone allows
separate treatment of transverse coherence properties at a given
frequency through the function $G_{\omega}(z_o,\vec{r}_{\bot
o1},\vec{r}_{\bot o2}, \omega)$. The condition for the
monochromator bandwidth $\Delta \omega_\mathrm{m} \gg 1/\sigma_T$
defines a quasi-stationary process. If the monochromator bandwidth
is such that $\Delta \omega_m > \omega_o/N_w$, the ratio
$N_w/(\omega_o \sigma_T)$ gives us a direct measure of the
accuracy of the stationary approximation which does not depend, in
this case, on the presence of the monochromator. When a
monochromator is present, with a bandwidth $\Delta \omega_m <
\omega_o/N_w$, it is the ratio $1/(\Delta \omega_\mathrm{m}
\sigma_T)$ which gives such a measure.  The condition $\Delta
\omega_\mathrm{m} \ll \omega_o/N_w$ ensures, instead, that the
mutual coherence function of the signal is reducible, in the sense
specified by Eq. (\ref{CScasegen}). In the following we will need
only the first of these conditions, $\omega_o/N_w \gg 1/\sigma_T$.
In fact, this is all we need in order to separate transverse and
temporal coherence effects, as shown in Eq. (\ref{gamma6}).

Once transverse and temporal coherence effects are separated one
can focus on the study of transverse coherence through the
function $G_{\omega}(z_o,\vec{r}_{\bot o1},\vec{r}_{\bot o2},
\omega)$.

There exists an important class of sources, called
quasi-homogeneous.  As we will see, quasi-homogeneity is  the
spatial analogue of quasi-stationarity.

\begin{figure}
\begin{center}
\includegraphics*[width=140mm]{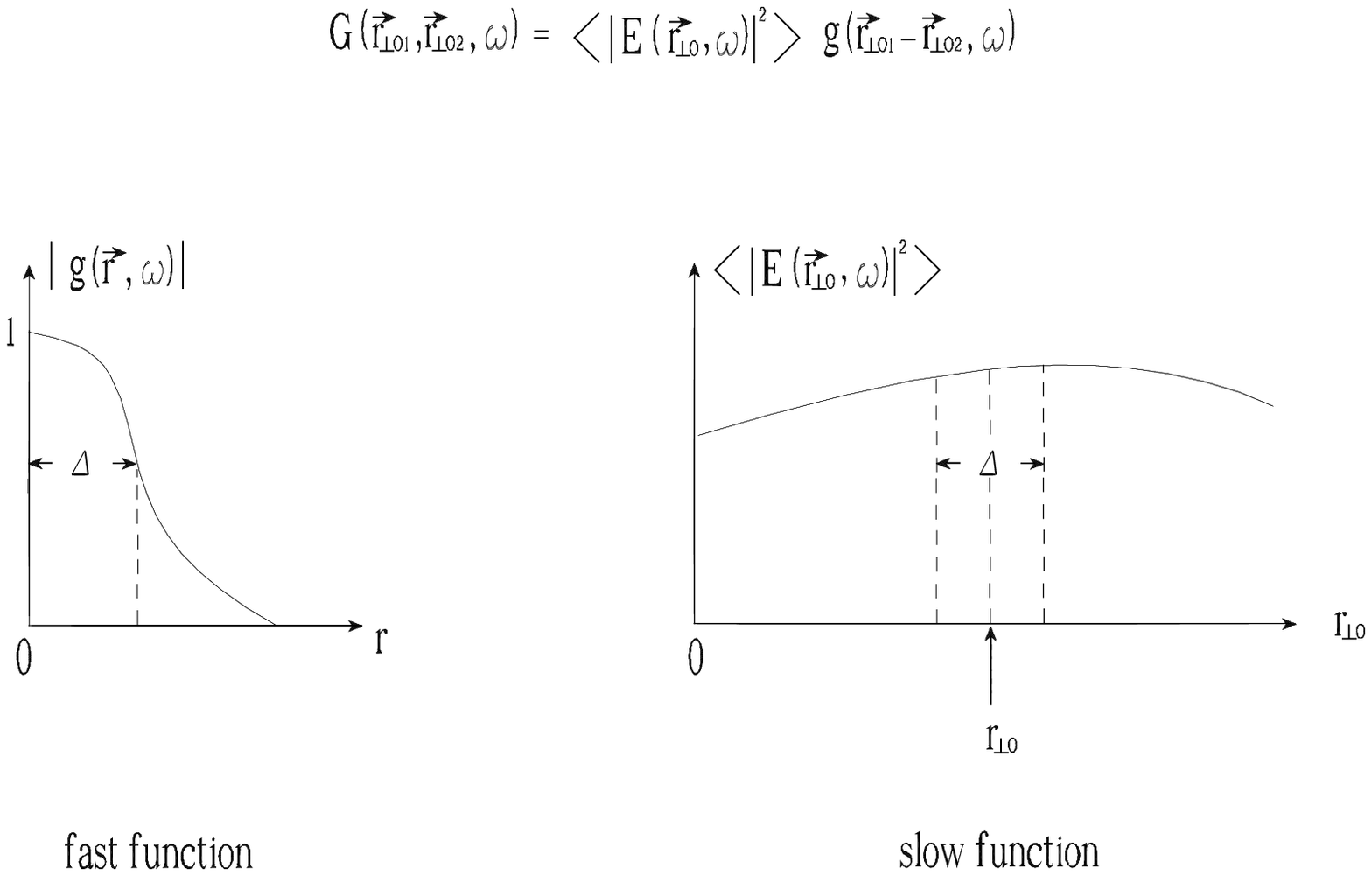}% Here is how to import EPS art
\caption{\label{QHOM} Illustrating the concept of
quasi-homogeneous source. Spectral density varies so slowly with
the position that it is approximatively constant over distances of
the order of the correlation length $\Delta$ across the source.}
\end{center}
\end{figure}
In general, quasi-homogeneous sources are a particular class of
Schell's model sources. Schell's model sources are defined by the
condition that their cross-spectral density at the source plane
(that is for a particular value of $z_o$) is of the form

\begin{eqnarray}
{G}_\omega({\vec{r}}_{\bot o1},{\vec{r}}_{\bot o2}, \omega) =
\left \langle \left| E({\vec{r}}_{\bot o1}, \omega) \right|^2
\right \rangle^{1/2} \left \langle \left| E({\vec{r}}_{\bot o2},
\omega) \right|^2 \right \rangle^{1/2} &&\cr \times
g({\vec{r}}_{\bot o2}-{\vec{r}}_{\bot o1}, \omega)
~,\label{introh1}
\end{eqnarray}
where $g({\vec{r}}_{\bot o2}-{\vec{r}}_{\bot o1}, \omega)$ is the
spectral degree of coherence (that is normalized to unity by
definition, i.e. $g(0,\omega) = 1$)\footnote{Sometimes, loosely
speaking, we will refer to $g$ as to "the cross-spectral density",
or to "the field correlation function" the difference being just a
normalization factor.}. Equivalently one may simply define
Schell's model sources using the condition that the spectral
degree of coherence depends on the positions across the source
only through the difference ${\vec{r}}_{\bot o2}-{\vec{r}}_{\bot
o1}$ (see \cite{MAND}, 5.2.2) from which Eq. (\ref{introh1})
follows.

\begin{figure}
\begin{center}
\includegraphics*[width=140mm]{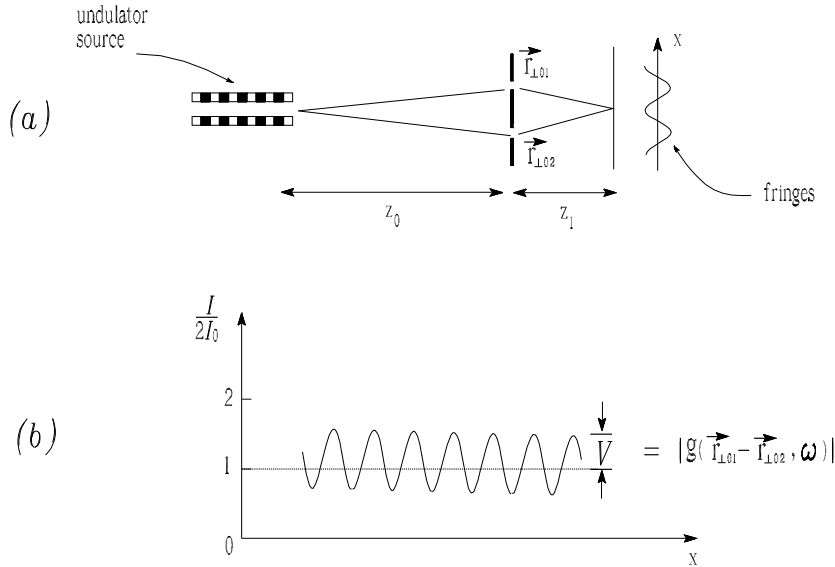}% Here is how to import EPS art
\caption{\label{ed12c} Measurement of the cross-spectral density
of a undulator source. (a) Young's double-pinhole interferometer
demonstrating the coherence properties of undulator radiation. The
radiation beyond the pinholes must be spectrally filtered by a
monochromator or detector (not shown in figure). (b) The fringe
visibility of the resultant interference pattern is equal to the
absolute value of the normalized cross-spectral density
$V=|g({\vec{r}}_{\bot o2}-{\vec{r}}_{\bot o1}, \omega)|$.}
\end{center}
\end{figure}
Quasi-homogeneous sources are Schell's sources obeying the
following extra-assumption: the spectral density
${G}_\omega(\vec{{r}}_{\bot o},\vec{{r}}_{\bot o}, \omega)$ at the
source plane, considered as a function of $\vec{{r}}_{\bot o}$,
varies so slowly with the position that it is approximatively
constant over distances across the source, which are of the order
of the correlation length $\Delta$ (that is the effective width of
$|{G}_\omega({\vec{r}}_{\bot o1},{\vec{r}}_{\bot o2}, \omega)|$,
see Fig. \ref{QHOM} for a qualitative illustration in one
dimension).

Since for quasi-homogeneous sources the spectral density is
assumed to vary slowly we are allowed to make the approximation

\begin{eqnarray}
{G}_\omega({\vec{r}}_{\bot o1},{\vec{r}}_{\bot o2}, \omega) =
I\left({\vec{r}}_{\bot o1}, \omega\right) g({\vec{r}}_{\bot
o2}-{\vec{r}}_{\bot o1}, \omega ) ~,\label{introh}
\end{eqnarray}
where

\begin{equation}
I\left({\vec{r}}_{\bot o1}, \omega\right) = \left \langle \left|
E\left({\vec{r}}_{\bot o1}, \omega\right) \right|^2
\right\rangle\label{intint}
\end{equation}
is the field intensity distribution. A schematic illustration of
the measurement of the cross-spectral density of a undulator
source is given in Fig. \ref{ed12c}.

We reported definitions and differences between Schell's sources
and quasi-homogeneous sources as treated in \cite{MAND} in order
to review some conventional language. However, in our paper we
will study and classify sources with the help of parameters from
dimensional analysis of the problem. In particular, in the
following we will introduce quantities  that model, in
dimensionless units, the electron beam dimension and divergence in
the $x$ and $y$ directions ($\hat{N}_{x,y}$ and $\hat{D}_{x,y}$,
respectively). When some of these parameters are much larger or
much smaller than unity, in certain particular combinations
discussed in the following part of this work, we will be able to
point out simplifications of analytical expressions. These
simplifications do not depend, in general, on the fact that the
source is quasi-homogeneous or not, but simply on the fact that
some of the above parameters are large or small. It will be
possible to describe some parameter combination in terms of
Schell's or quasi-homogeneous model, but this will not be the
case, in general. In order to link the physical properties of
certain kind of sources to a certain range of parameters we will
find it convenient to extend the concept of quasi-homogeneity to
the concept of "weak quasi-homogeneity". It is better to
familiarize with our new definition already here: a given
wavefront at fixed position $\hat{z}_o$ will be said to be weakly
quasi-homogeneous, by definition, when the \textit{modulus} of the
spectral degree of coherence $|g|$ depends on the position across
the source only through the difference ${\vec{r}}_{\bot
o2}-{\vec{r}}_{\bot o1}$, i.e. $|g|=w({\vec{r}}_{\bot
o2}-{\vec{r}}_{\bot o1},\omega)$. This is equivalent to generalize
Eq. (\ref{introh}) to

\begin{eqnarray}
|{G}_\omega({\vec{r}}_{\bot o1},{\vec{r}}_{\bot o2}, \omega)| =
I\left({\vec{r}}_{\bot o1}, \omega\right) w({\vec{r}}_{\bot
o2}-{\vec{r}}_{\bot o1}, \omega ) ~.\label{introh2}
\end{eqnarray}
As a remark to both the definitions of quasi-homogeneity and weak
quasi-homogeneity, it should be noted that they only involve
conditions on the cross-spectral densities through Eq.
(\ref{introh}) or Eq. (\ref{introh2}): one can apply these
definitions to any wavefront at any position ${z}_o$. This is
consistent with what has been remarked in the Introduction: the
choice of a source plane down the beamline is just a convention.
Therefore, our definition of weak quasi-homogeneity is completely
separated from the concept of source plane. It may seem, at first
glance, that the definition of "weak quasi-homogeneity" is somehow
artificial but it is, on the contrary, very convenient from a
practical viewpoint. In fact, in any coherent experiment, the
specimen is illuminated by coherent light from some kind of
aperture, or diaphragm. Think, for instance, to the usual process
of selection of transversely coherent light through a spatial
filter, where a diaphragm is placed downstream a pinhole.
Physically, when the modulus of the spectral degree of coherence
depends only on the coordinate difference ${\vec{r}}_{\bot
o2}-{\vec{r}}_{\bot o1}$ the coherence properties of the beam do
not depend on the position of the diaphragm with respect to the
transverse coordinate of the center of the pinhole, which we may
imagine on the $z$ axis: for instance, the coherence length will
depend only on ${\vec{r}}_{\bot o2}-{\vec{r}}_{\bot o1}$ and not
on the average position $({\vec{r}}_{\bot o2}+{\vec{r}}_{\bot
o1})/2$.  Mathematically, as has already been said, the definition
of weak quasi-homogeneity is linked with particular combination of
small and large parameters which will lead to simplifications of
equations and to analytical treatment of several interesting
cases.

After having introduced the definition of "weak quasi-homogeneity"
we should go back to the concept of usual quasi-homogeneity to
describe a vary particular feature of it: quasi-homogeneity can be
regarded as the spatial equivalent of quasi-stationarity, as
anticipated before. Exactly as the time domain has a reciprocal
description in terms of frequency, the space domain has a
reciprocal description in terms of transverse (two-dimensional)
wave vectors. However, since the frequency is fixed, the ratio
between the horizontal or vertical component of the wave vector
and the longitudinal wave number is representative of the
propagation angle of a plane wave at fixed frequency. Therefore
any given signal on a two-dimensional plane can be represented in
terms of superposition of plane waves with the same frequency and
different angles of propagation, which goes under the name of
angular spectrum. This defines an angular domain which is the
reciprocal of the space domain. Intuitively the angular spectrum
representation constitutes a picture of the effects of propagation
in the far zone whereas the near zone is described by the space
domain. In this Section, an analogous of the Wiener-Khinchin
theorem and its reciprocal form for quasi-stationary processes has
been discussed. Substitution of times with position vectors and
frequencies with angular vectors allows to derive similar
statements for the near (space domain) and far (angular domain)
zone \cite{MAND}. If we use $\vec{\theta} = \vec{r}_{\bot o}/z_o$
(and $\vec{\theta}_{1,2} = \vec{r}_{\bot o 1,2}/z_o$) as variables
to describe radiant intensity and cross-spectral density in the
far field, and if we identify points on the source plane with
$\vec{r}_{\bot}$ (and $\vec{r}_{\bot 1,2}$), we have:

(a') The cross-spectral density of the field at the source plane
$g({\vec{r}}_{\bot 2}-{\vec{r}}_{\bot 1})$ and the angular
distribution of the radiant intensity $I(\vec{\theta})$ are a
Fourier Pair.

(b') The cross-spectral density of the far field
$g({\vec{\theta}}_{2}-{\vec{\theta}}_{1})$ and the
source-intensity distribution $I(\vec{r}_{\bot})$  are, apart for
a simple geometrical phase factor, a Fourier Pair.

The statement (b') can be regarded as an analogue, for
quasi-homogeneous sources, of the far-zone form of the van
Cittert-Zernike theorem. The statement (a') instead, is due to the
symmetry between space and angle domains, and can be seen as an
"anti" VCZ theorem. This discussion underlines the link between
the VCZ theorem and the Wiener-Khinchin theorem.

\begin{figure}
\begin{center}
\includegraphics*[width=140mm]{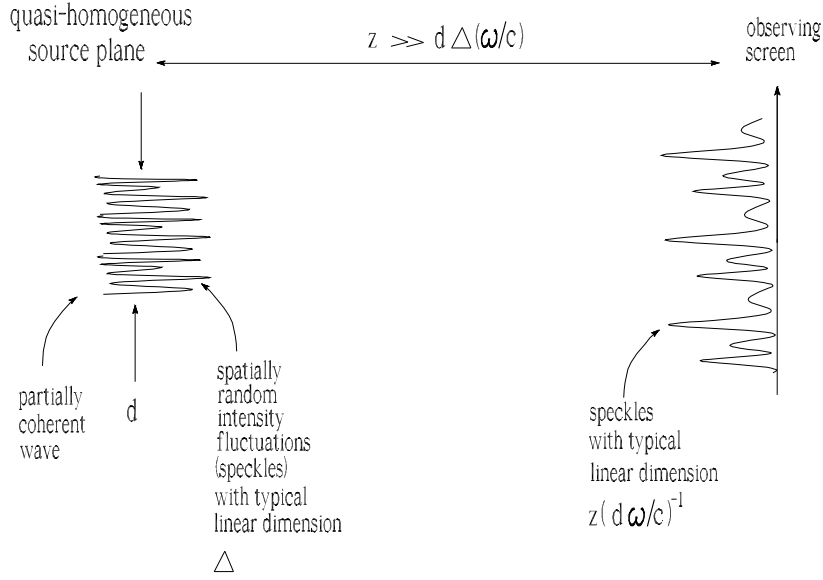}% Here is how to import EPS art
\caption{\label{trcoh} Geometry for propagation of transverse
coherence in the case of a quasi-homogeneous source.}
\end{center}
\end{figure}

Many undulator radiation sources are quasi-homogeneous sources in
the usual way. In the case of a quasi-homogeneous source of
typical linear dimension $d$, the angular spectrum at distance
$z_o \gg d \omega \Delta/c$ is expected to exhibit speckles with
typical linear dimension $z_o (d \omega/c)^{-1}$, as illustrated
in Fig. \ref{trcoh}, which shows an intuitive picture of the
propagation of transverse coherence.

They generate fields which are relatively simple to analyze
mathematically and still rich in physical features. Although both
VCZ and "anti" VCZ theorem are based on the assumption of usual
quasi-homogeneity we will see that these are often applicable, at
least in some sense, also in the case of "weakly
quasi-homogeneous" wavefronts.

\section{\label{sec:evol} Evolution of the cross-spectral density function through the undulator beamline}

\begin{figure}
\begin{center}
\includegraphics*[width=110mm]{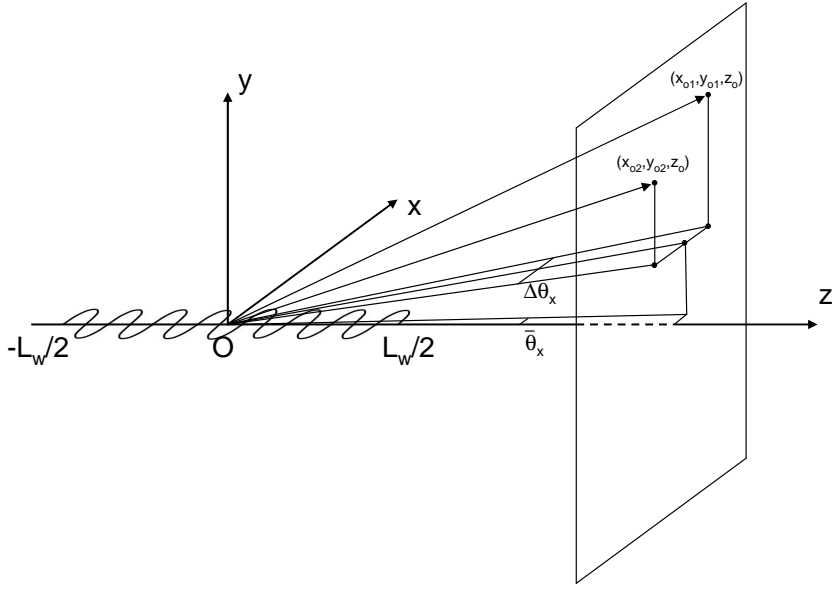}% Here is how to import EPS art
\caption{\label{geo} Illustration of the undulator geometry and of
the observation plane. }
\end{center}
\end{figure}

In our work \cite{OURS} we presented an expression for the reduced
field $\tilde{E}_\bot(\omega)$ $= \bar{E}_\bot(\omega)
\exp{(-i\omega z_o/c)}$ of a \textit{single particle} with offset
$\vec{l}$ and deflection $\vec{\eta}$ with respect to the optical
axis $z$ in an undulator. In order to derive our result, we used a
Green's function approach to solve the paraxial Maxwell equations
for the Fourier transform of the electric field and we took
advantage of a consistent use of the resonance approximation. The
field $\tilde{E}_\bot$ differs from $\bar{E}_\bot$ for a phase
factor which depends on the variable $z_o$ and on the frequency
$\omega$ only: therefore, the use of one expression instead of the
other in the equation for $G$ does not change the result. In
\cite{OURS}, we presented results in normalized units in the far
field zone for a particle with offset and deflection. Based on
that work we can calculate the field in normalized units for a
particle with offset and deflection at any distance from the exit
of the undulator, where the center of the undulator is taken at
$z=0$, as specified in Fig. \ref{geo}:

\begin{eqnarray}
\hat{E}_{s\bot}=  \hat{z}_o  \int_{-1/2}^{1/2} d\hat{z}'
\frac{1}{\hat{z}_o-\hat{z}'} \exp \left\{i
\left[\left(\hat{C}+\frac{\left.\vec{\hat{\eta}}\right.^2}{2}\right)\hat{z}'
+ \frac{\left(\vec{\hat{{r}}}_{\bot
o}-\vec{{\hat{l}}}-\vec{\hat{\eta}}\hat{z}' \right)^2 }{2
(\hat{z}_o-\hat{z}')}\right] \right\} . \label{undunormfin}
\end{eqnarray}
Eq. (\ref{undunormfin}) is valid for the system tuned at resonance
with the fundamental harmonic $\omega_o$. This means that we are
considering a large number of undulator periods $N_w \gg 1$ and
that we are looking at frequencies near the fundamental and at
angles within the main lobe of the directivity diagram. In this
situation one can neglect the vertical $y$-polarization of the
field with an accuracy $(4\pi N_w)^{-1}$. This constitutes a great
simplification of the problem since, at any position of the
observer, we may consider the electric field Fourier transform,
$\hat{E}_{s\bot}$, as a complex scalar quantity corresponding to
the surviving $x$-polarization component of the original vector
quantity. Normalized units were defined as

\begin{eqnarray} \hat{E}_{s\bot} = -{c^2 z_o \gamma
\tilde{E}_{s\bot} \over{K \omega e L_w A_{JJ}}} ~,&&\cr
\vec{\hat{\eta}} =\vec{{\eta}}\sqrt{\frac{\omega L_w}{c}}  ~,&&\cr
\hat{C} = L_w C = 2 \pi N_w \frac
{\omega-\omega_o}{\omega_o}~,&&\cr\vec{\hat{r}}_{\bot o}
={\vec{r}}_{\bot o} \sqrt{{\omega \over{L_w
c}}}~,&&\cr\vec{\hat{l}} ={\vec{l}}\sqrt{{\omega \over{L_w
c}}}~,&&\cr \hat{z}={z\over{L_w}}~.\label{Cnorm}
\end{eqnarray}
$K$ being the deflection parameter, $L_w$ being the undulator
length,

\begin{equation}
A_{JJ} = J_0\left( \frac{K^2}{4+2K^2}\right) - J_1\left(
\frac{K^2}{4+2K^2}\right) ~,\label{AJJdef}
\end{equation}
\begin{equation}
\omega_o = \frac{4\pi c \gamma^2}{\lambda_w \left(1+K^2/2\right)}
\label{omegazerodef}
\end{equation}
being the resonant frequency, $J_n$ the Bessel function of the
first kind of order $n$, $\lambda_w$ the undulator period, $(-e)$
the electron charge and $\gamma$ the relativistic Lorentz factor.
$\vec{\hat{l}}$ is the normalized offset in the center of the
undulator. Finally, the parameter $\hat{C}$ represents the
normalized detuning, which accounts for small deviation in
frequency from resonance.

\begin{figure}
\begin{center}
\includegraphics*[width=110mm]{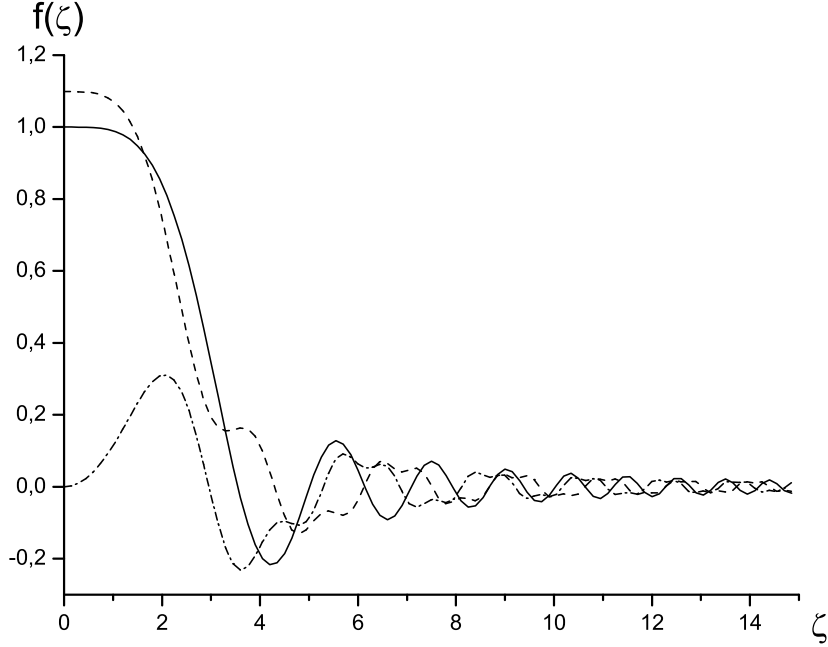}% Here is how to import EPS art
\caption{\label{splot} Comparison between
$f(\zeta)=\mathrm{sinc}(\zeta^2/4)$ (solid line), the real (dashed
line) and the imaginary (dash-dotted line) parts of
$f(\zeta)=S(0,\hat{z}_o,\zeta^2)$ at $\hat{z}_o = 1$.}
\end{center}
\end{figure}

As it is shown in Appendix B, Eq. (\ref{undunormfin}) can be
rewritten as

\begin{eqnarray}
\hat{E}_{s\bot} =    \int_{-1/2}^{1/2} \frac{\hat{z}_o d\hat{z}'
}{\hat{z}_o-\hat{z}'} \exp \left\{i \left[\Phi_U +\hat{C} \hat{z}'
+ \frac{\hat{z}_o \hat{z}'}{2 (\hat{z}_o-\hat{z}')}
\left(\vec{\hat{\theta}}- \frac{\vec{\hat{l}}}{\hat{z}_o}-
\vec{\hat{\eta}}\right)^2 \right] \right\} \label{undunormfinult}
\end{eqnarray}
where

\begin{eqnarray}
\vec{\hat{\theta}}= \frac{{\vec{\hat{r}}}_{\bot
o}}{\hat{z}_o}~\label{Cnorm2}
\end{eqnarray}
represents the observation angle and $\Phi_U$ is given by

\begin{equation}
\Phi_U =
\left(\vec{\hat{\theta}}-\frac{\vec{\hat{l}}}{\hat{z}_o}\right)^2
 \frac{\hat{z}_o}{2} ~.\label{phisnorm}
\end{equation}
Eq. (\ref{undunormfinult}) is of the form

\begin{equation}
\hat{E}_{s\bot}\left(\hat{C},\hat{z}_o,\vec{\hat{\theta}}-
\frac{\vec{\hat{l}}}{\hat{z}_o}- \vec{\hat{\eta}}\right) =
\exp{(i\Phi_U)} S\left[\hat{C},\hat{z}_o,\left(\vec{\hat{\theta}}-
\frac{\vec{\hat{l}}}{\hat{z}_o}-
\vec{\hat{\eta}}\right)^2\right]~. \label{Esum}
\end{equation}
Starting from the next Section we will restrict our attention to
the case $\hat{C}=0$ for simplicity. Therefore, it may be
interesting to note that in the particular case $\hat{C} =0$, the
function $S$ can be represented in terms of the exponential
integral function Ei as:

\begin{equation}
S\left(0,\hat{z}_o,\zeta^2\right) = \exp (-i \hat{z}_o \zeta^2/2)
\hat{z}_o \left[ \mathrm{Ei} \left(\frac{i \hat{z}_o^2 \zeta^2
}{-1+2\hat{z}_o}\right)- \mathrm{Ei} \left(\frac{i \hat{z}_o^2
\zeta^2 }{1+2\hat{z}_o}\right) \right] \label{SfuncEi}
\end{equation}
It is easy to show that the expression for the function $S(\cdot)$
reduces to a $\mathrm{sinc(\cdot)}$ function as $\hat{z}_o \gg
1$~. In fact, in this limiting case, the expression for the
electric field from a single particle, given in Eq.
(\ref{undunormfin}) is simplified to

\begin{eqnarray}
\hat{E}_{s\bot} =  \exp{(i\Phi_U)} \int_{-1/2}^{1/2} d\hat{z}'
\exp \left\{i \hat{z}' \left[\hat{C} +
\frac{1}{2}\left(\vec{\hat{{\theta}}}-\frac{\vec{\hat{l}_x}}{\hat{z}_o}-
\vec{\hat{\eta}}\right)^2 \right]\right\} ~,~ \label{undunormfin2}
\end{eqnarray}
Eq. (\ref{undunormfin2}) can be integrated analytically giving

\begin{equation}
\hat{E}_{s\bot}= \exp{(i \Phi_U)}~~
\mathrm{sinc}\left(\frac{\hat{C}}{2}+\frac{\zeta^2}{4}\right) ~,
\label{endangle}
\end{equation}
where

\begin{equation}
\zeta =
\vec{\hat{{\theta}}}-\frac{\vec{\hat{l}}}{\hat{z}_o}-\vec{\hat{\eta}}~.
\label{zeta}
\end{equation}
A comparison between $\mathrm{sinc}(\zeta^2/4)$ and the real and
imaginary parts of $S(0,\hat{z}_o,\zeta^2)$ for $\hat{z}_o = 1$ is
given in Fig. \ref{splot}.

Let us now go back to the general case for $\hat{z}_o \geqslant
1/2$ and use Eq. (\ref{undunormfinult}) to calculate the
cross-spectral density. The cross-spectral density $G_\omega$ is
given Eq. (\ref{coore}) in  dimensional units and as a function of
dimensional variables. Since the field in Eq.
(\ref{undunormfinult}) is given in normalized units and as a
function of normalized variables $\hat{z}_o$,
$\vec{\hat{\theta}}_{x,y}$ and $\hat{C}$, it is convenient to
introduce a version of $G_\omega$ defined by means of the field in
normalized units:

\begin{eqnarray}
\hat{G}(\hat{z}_o,\vec{\hat{\theta}}_1,\vec{\hat{\theta}}_2,\hat{C})
 = \Bigg \langle \hat{E}_{s\bot}\left(\hat{C},\hat{z}_o,\vec{\hat{\theta}}_1-
\frac{\vec{\hat{l}}}{\hat{z}_o}- \vec{\hat{\eta}}\right)
\hat{E}_{s\bot}^*\left(\hat{C},\hat{z}_o,\vec{\hat{\theta}}_2-
\frac{\vec{\hat{l}}}{\hat{z}_o}- \vec{\hat{\eta}}\right)
\Bigg\rangle_{\vec{\eta},\vec{l}} ~.&& \cr\label{Gnormdef1}
\end{eqnarray}
Transformation of  $G_\omega$ in Eq. (\ref{coore}) to $\hat{G}$
(and viceversa) can be easily performed shifting from dimensional
to normalized variables and multiplying $G_\omega$ by an
inessential factor:

\begin{equation}
\hat{G} =  \left(\frac{c^2 z_o \gamma}{K \omega e L_w
A_{JJ}}\right)^2 G_\omega ~.\label{inetransf}
\end{equation}
As a result, we can always use $\hat{G}$ in stance of  $G_\omega$.
Substituting Eq. (\ref{undunormfinult}) in Eq. (\ref{Gnormdef1})
we obtain:

\begin{eqnarray}
\hat{G}(\hat{z}_o,\vec{\hat{\theta}}_1,\vec{\hat{\theta}}_2,\hat{C})
 = \Bigg \langle S\left[\hat{C},\hat{z}_o,\left(\vec{\hat{\theta}}_1-
\frac{\vec{\hat{l}}}{\hat{z}_o}- \vec{\hat{\eta}}\right)^2\right]
S^*\left[\hat{C},\hat{z}_o,\left(\vec{\hat{\theta}}_2-
\frac{\vec{\hat{l}}}{\hat{z}_o}- \vec{\hat{\eta}}\right)^2\right]
&&\cr \times
\exp\left\{i\left[\left(\vec{\hat{\theta}}_{1}-\frac{\vec{\hat{l}}}{\hat{z}_o}\right)^2
-
\left(\vec{\hat{\theta}}_{2}-\frac{\vec{\hat{l}}}{\hat{z}_o}\right)^2
\right]
 \frac{\hat{z}_o}{2}  \right\}
\Bigg\rangle_{\vec{\eta},\vec{l}} \label{Gzlarge}
\end{eqnarray}
Expanding the exponent in the exponential factor in the right hand
side of Eq. (\ref{Gzlarge}), it is easy to see that terms in
$\hat{l}_{x,y}^2$ cancel out. Terms in $\hat{\theta}_{x,y}^2$
contribute for a common factor, and only linear terms in
$\hat{l}_{x,y}$ remain inside the ensemble average sign.
Substitution of the ensemble average with integration over the
beam distribution function leads to

\begin{eqnarray}
\hat{G}(\hat{z}_o,\vec{\hat{\theta}}_1,\vec{\hat{\theta}}_2,\hat{C})
=
\exp{\left[i\left(\vec{\hat{\theta}}_1^2-\vec{\hat{\theta}}_2^2\right)\frac{\hat{z}_o}
{2}\right]}
 \int d \vec{\hat{\eta}} d
\vec{\hat{l}}~
F_{\vec{\hat{\eta}},\vec{\hat{l}}}\left(\vec{\hat{\eta}},\vec{\hat{l}}\right)
\exp\left[i (\vec{\hat{\theta}}_{2}-\vec{\hat{\theta}_{1}})\cdot
\vec{ \hat{l}} \right] &&\cr \times
S\left[\hat{C},\hat{z}_o,\left(\vec{\hat{\theta}}_1-
\frac{\vec{\hat{l}}}{\hat{z}_o}-
\vec{\hat{\eta}}\right)^2\right]S^*\left[\hat{C},\hat{z}_o,\left(\vec{\hat{\theta}}_2-
\frac{\vec{\hat{l}}}{\hat{z}_o}- \vec{\hat{\eta}}\right)^2\right]
~.~ \label{Gzlarge2}
\end{eqnarray}
Here integrals $d \vec{\hat{\eta}}$ and in $d \vec{\hat{l}}$ are
to be intended as integrals over the entire plane spanned by the
$\vec{\hat{\eta}}$ and $\vec{\hat{l}}$ vectors. Eq.
(\ref{Gzlarge2}) is very general and can be used as a starting
point for computer simulations.

We already  assumed that the distribution in the horizontal and
vertical planes are not correlated, so that
$F_{\vec{\hat{\eta}},\vec{\hat{l}}} = F_{\hat{\eta}_x,\hat{l}_x}
F_{\hat{\eta}_y,\hat{l}_y}$. If the transverse phase space is
specified at position $\hat{z}_o = 0$ corresponding to the minimal
values of the $\beta$-functions, we can write
$F_{\hat{\eta}_x,\hat{l}_x}= F_{\hat{\eta}_x} F_{\hat{l}_x}$ and
$F_{\hat{\eta}_y,\hat{l}_y}= F_{\hat{\eta}_y} F_{\hat{l}_y}$ with

\begin{eqnarray}
F_{\hat{\eta}_x}(\hat{\eta}_x) = \frac{1}{\sqrt{2\pi \hat{D}_x}}
\exp{\left(-\frac{\hat{\eta}_x^2}{2 \hat{D}_x}\right)}~,&&\cr
F_{\hat{\eta}_y}(\hat{\eta}_y)  = \frac{1}{\sqrt{2\pi \hat{D}_y}}
\exp{\left(-\frac{\hat{\eta}_y^2}{2 \hat{D}_y}\right)}~,&&\cr
F_{\hat{l}_x}(\hat{l}_x) =\frac{1}{\sqrt{2\pi\hat{N}_x} }
\exp{\left(-\frac{\hat{l}_x^2}{2 \hat{N}_x}\right)}~, && \cr
F_{\hat{l}_y}(\hat{l}_y)=\frac{1}{\sqrt{2\pi\hat{N}_y} }
\exp{\left(-\frac{\hat{l}_y^2}{2 \hat{N}_y}\right)}~.\label{distr}
\end{eqnarray}
From Eq. (\ref{Cnorm}) and Eq. (\ref{Cnorm2}) it is easy to see
that

\begin{equation}
\hat{D}_{x,y} = \sigma_{x',y'}^2 {\frac{\omega L_w}{c}}\label{sig}
\end{equation}
\begin{equation}
\hat{N}_{x,y} = \sigma^2_{x,y} \frac{\omega}{c L_w}\label{enne}
\end{equation}
where $\sigma_{x,y}$ and $\sigma_{x',y'}$ are the rms transverse
bunch dimension and angular spread. Parameters $\hat{N}_{x,y}$
will be indicated as the beam diffraction parameters and are, in
fact, analogous to Fresnel numbers and correspond to the
normalized square of the electron beam sizes, whereas
$\hat{D}_{x,y}$ represent the normalized square of the electron
beam divergences.

Substitution of relations (\ref{distr}) in Eq. (\ref{Gzlarge2})
yields, at perfect resonance ($\hat{C} = 0$):

\begin{eqnarray}
\hat{G}(\hat{z}_o,\vec{\hat{\theta}}_1,\vec{\hat{\theta}}_2)
=\frac{\exp{\left[i\left(\vec{\hat{\theta}}_1^2-
\vec{\hat{\theta}}_2^2\right){\hat{z}_o}/{2}\right]}}{4\pi^2
\sqrt{\hat{D}_x\hat{D}_y\hat{N}_x\hat{N}_y}}
\int_{-\infty}^{\infty} d \hat{\eta}_{x }
\exp{\left(-\frac{\hat{\eta}_x^2}{2 \hat{D}_x}\right)} &&\cr
\times \int_{-\infty}^{\infty} d \hat{\eta}_{y }
\exp{\left(-\frac{\hat{\eta}_y^2}{2 \hat{D}_y}\right)}
\int_{-\infty}^{\infty} d \hat{l}_{x }
\exp{\left(-\frac{\hat{l}_x^2}{2 \hat{N}_x}\right)}
\int_{-\infty}^{\infty} d \hat{l}_{y
}\exp{\left(-\frac{\hat{l}_y^2}{2 \hat{N}_y}\right)}&& \cr
S\left[\hat{z}_o,\left(\vec{\hat{\theta}}_1-
\frac{\vec{\hat{l}}}{\hat{z}_o}-
\vec{\hat{\eta}}\right)^2\right]S^*\left[\hat{z}_o,\left(\vec{\hat{\theta}}_2-
\frac{\vec{\hat{l}}}{\hat{z}_o}- \vec{\hat{\eta}}\right)^2\right]
\exp\left[i (\vec{\hat{\theta}}_{2}-\vec{\hat{\theta}_{1}})\cdot
\vec{ \hat{l}} \right] ~.\label{Gzlarge3}
\end{eqnarray}
For notational simplicity, in Eq. (\ref{Gzlarge3}) we have
substituted the proper notation
$\hat{G}(\hat{z}_o,\vec{\hat{\theta}}_1,\vec{\hat{\theta}}_2,
\hat{C})$ with the simplified dependence
$\hat{G}(\hat{z}_o,\vec{\hat{\theta}}_1,\vec{\hat{\theta}}_2)$
because we will be treating the case $\hat{C}=0$ only.
Consistently, also $S {[\hat{z}_o,
(\vec{\hat{\theta}}-\vec{\hat{l}}/ {\hat{z}_o}-\vec{\hat{\eta}})^2
]}$ is to be understood as a shortcut notation for
$S[\hat{C},\hat{z}_o,(\vec{\hat{\theta}}-
\vec{\hat{l}}/\hat{z}_o-\vec{\hat{\eta}})^2]$ calculated at
$\hat{C}=0$.

\section{\label{sec:quasi} Undulator radiation as a quasi-homogeneous source}

When describing physical principles it is always important to find
a model which provides the possibility of an analytical
description without loss of essential information about the
feature of the random process.

In order to get a feeling for some realistic magnitude of
parameters we start noting that the geometrical emittances of the
electron beam are simply given by ${\epsilon}_{x,y} =
{\sigma}_{x,y} {\sigma}_{x',y'}$. Here they will be normalized as
$\hat{\epsilon}_{x,y} = {2 \pi} \epsilon_{x,y}/\lambda$.  Then
${\sigma}_{x,y}^2 = \beta^o_{x,y} \epsilon_{x,y}$, where
$\beta^o_{x,y}$ are the minimal values of the horizontal and
vertical betatron functions. In this paper we will assume that the
betatron functions will have their minimal value at the undulator
center. Therefore we have ${\epsilon}_{x,y} = {\sigma}_{x',y'}^2
\beta^o_{x,y}$ or, in normalized units, $\hat{\epsilon}_{x,y} =
\hat{D}_{x,y} \hat{\beta}_{x,y}$, where $\hat{\beta}_{x,y} =
\beta_{x,y}^o/L_w$. Equivalently we can write
$\hat{\epsilon}_{x,y} =  \sqrt{\hat{D}_{x,y}\hat{N}_{x,y}}$. It
follows that $\hat{N}_{x,y} = \hat{\epsilon}_{x,y}
\hat{\beta}_{x,y}$. Now taking $\lambda = 1~\AA$, $\epsilon_x = 1
\div 3$ nm, $\epsilon_y = 10^{-2} \epsilon_x$ and $\beta^o_x =
10^{-1} \div 10 L_w$ one obtains, in normalized units,
$\hat{\epsilon}_x = 10^2 \div 3 \cdot 10^2$, $\hat{\epsilon}_y = 1
\div 3$ and $\hat{\beta}_x = 10^{-1} \div 10$: therefore
$\hat{D}_x \gg 1$, $\hat{N}_{x} \gg 1$. This is always the case in
situations of practical interest, with $\hat{N}_x$ which may range
from values much smaller to much larger than $\hat{D}_x$.

Assuming $\hat{D}_x \gg 1$ and $\hat{N}_{x} \gg 1$, independently
on the values of $\hat{D}_y$ and $\hat{N}_{y}$, introduces
simplifications in the expression for the cross-correlation
function and allows further analytical investigations. As we will
see, in particular, a model of the electron beam based on these
assumptions contains (but it is not limited to) the class of
quasi-homogeneous sources discussed in Section \ref{sub:seco}.

In the next Sections \ref{sub:oned} and \ref{sub:disc} and later
in Section \ref{sec:twod}, we will see what are the conditions in
terms of the dimensionless parameters $\hat{N}_{x,y}$ and
$\hat{D}_{x,y}$ for some undulator radiation wavefront at position
$\hat{z}_o$, to be quasi-homogeneous in the usual and in the weak
sense (according to the definition in Section \ref{sub:seco}), we
will justify the introduction of the concept of weak
quasi-homogeneity itself and we will discuss the applicability
regions of the VCZ (and "anti" VCZ) theorem. Then, in Section
\ref{sec:nonh}, we will also discuss some case characterized by
non weakly quasi-homogenous fields.

\subsection{\label{sub:oned} A simple model}

To provide a first analysis of the problem we adopt some
simplifying assumptions that are only occasionally met in
practice.

As already assumed vertical emittance is much smaller than
horizontal emittance.
%Working under the assumptions
%$\hat{\epsilon}_y \hat{\beta}_y \ll 1$ and
%$\hat{\epsilon}_y/\hat{\beta}_y \ll 1$, attention can be
%concentred completely on transverse coherence effects in the
%horizontal direction.
For notational simplicity we will make the assumptions $\hat{N}_y
\ll 1$ and $\hat{D}_y \ll 1$. This means that we theoretically
assume  $\hat{\eta}_y \ll 1$ and $\hat{l}_y \ll 1$. As a result,
the terms in $\hat{\eta}_y$ and $\hat{l}_y$ can be neglected in
the $S(\cdot)$ term on the right hand side of Eq.
(\ref{Gzlarge3}). Although this model includes obvious
schematization it is still close to reality in many situations,
and it is only to be considered as a provisory model for physical
understanding to be followed, below, by more comprehensive
generalizations. In this Section we will restrict our attention at
the correlation function for $\hat{\theta}_{y1} =
\hat{\theta}_{y2}$ that is on any horizontal plane. Here again,
for notational simplicity, we will substitute the proper notation
$\hat{G}(\hat{z}_o,{\hat{\theta}}_{x1},0,{\hat{\theta}}_{x2},0)$
with $\hat{G}(\hat{z}_o,{\hat{\theta}}_{x1},{\hat{\theta}}_{x2})$.
Eq. (\ref{Gzlarge3}) can be greatly simplified leading to

\begin{eqnarray}
\hat{G}(\hat{z}_o,{\hat{\theta}}_{x1},{\hat{\theta}}_{x2})
=\frac{1}{2\pi \sqrt{\hat{D}_x\hat{N}_x}}
\exp{\left[i\left({\hat{\theta}}_{x1}^2-{\hat{\theta}}_{x2}^2\right)\hat{z}_o/2\right]}
&&\cr \times  \int_{-\infty}^{\infty} d \hat{\eta}_{x }
\exp{\left(-\frac{\hat{\eta}_x^2}{2 \hat{D}_x}\right)}
\int_{-\infty}^{\infty} d \hat{l}_{x }
\exp{\left(-\frac{\hat{l}_x^2}{2 \hat{N}_x}\right)}\exp\left[i
(\hat{\theta}_{x2}-\hat{\theta}_{x1})\hat{l}_x\right] &&\cr
\times\Bigg\{ S {\left[\hat{z}_o,
\left(\hat{{\theta}}_{x1}-{\hat{l}_x}/
{\hat{z}_o}-\hat{\eta}_x\right)^2 \right]} S^*
{\left[\hat{z}_o,\left(\hat{{\theta}}_{x2}-{\hat{l}_x}/
{\hat{z}_o}-\hat{\eta}_x\right)^2 \right]} \Bigg\}
~.\label{Gzlarge4}
\end{eqnarray}
Let us now introduce

\begin{equation}
\Delta \hat{\theta} =
\frac{\hat{\theta}_{x1}-\hat{\theta}_{x2}}{2} \label{deltadd}
\end{equation}
\begin{equation}
\bar{\theta} = \frac{\hat{\theta}_{x1}+\hat{\theta}_{x2}}{2}
\label{thetatt}
\end{equation}
With this variables redefinition we obtain
\begin{eqnarray}
\hat{G}(\hat{z}_o,\bar{\theta},\Delta \hat{\theta}) =\frac{1}{2\pi
\sqrt{\hat{D}_x\hat{N}_x}} \exp{\left(i 2 \bar{\theta}\Delta
\hat{\theta} \hat{z}_o\right)}  \int_{-\infty}^{\infty} d
\hat{\eta}_{x } \exp{\left(-\frac{\hat{\eta}_x^2}{2
\hat{D}_x}\right)} &&\cr \times \int_{-\infty}^{\infty} d
\hat{l}_{x } \exp{\left(-\frac{\hat{l}_x^2}{2
\hat{N}_x}\right)}\exp\left[- 2 i \Delta \hat{\theta}
\hat{l}_x\right] \left\{ S {\left[\hat{z}_o,
\left(\bar{\theta}+\Delta \hat{\theta}-{\hat{l}_x}/
{\hat{z}_o}-\hat{\eta}_x\right)^2 \right]} \right.&&\cr \left.
\times S^* {\left[\hat{z}_o,\left(\bar{\theta}-\Delta
\hat{\theta}-{\hat{l}_x}/ {\hat{z}_o}-\hat{\eta}_x\right)^2
\right]} \right\} ~.\label{Gzlarge5}
\end{eqnarray}
A double change of variables $\hat{\eta}_x \longrightarrow
\hat{\eta} + \bar{\theta}$ followed by $\hat{l}_x/\hat{z}_o
\longrightarrow \hat{\phi} - \hat{\eta}$ yields

\begin{eqnarray}
\hat{G}(\hat{z}_o,\bar{\theta},\Delta \hat{\theta})
=\frac{\exp{\left(i 2 \bar{\theta}\Delta \hat{\theta}
\hat{z}_o\right)}}{2\pi \sqrt{\hat{D}\hat{N}/\hat{z}_o^2}}
\int_{-\infty}^{\infty} d \hat{\eta}
\exp{\left(-\frac{(\hat{\eta}+\bar{\theta})^2}{2 \hat{D}}+2 i
\Delta \hat{\theta} \hat{z}_o \hat{\eta}\right)} &&\cr \times
\int_{-\infty}^{\infty} d \hat{\phi}
\exp{\left(-\frac{(\hat{\phi}-\hat{\eta})^2}{2
\hat{N}/\hat{z}_o^2}\right)}\exp\left(- 2 i \Delta \hat{\theta}
\hat{z}_o \hat{\phi}\right)&&\cr\times S^*
{\left[\hat{z}_o,(\hat{\phi}-\Delta \hat{\theta})^2\right]}
S{\left[\hat{z}_o,(\hat{\phi}+\Delta
\hat{\theta})^2\right]}~.\label{Gzlarge7}
\end{eqnarray}
where  we have posed $\hat{D} = \hat{D}_x$ and $\hat{N} =
\hat{N}_x$ for notational simplicity. Eq. (\ref{Gzlarge7}) can be
also written as

\begin{eqnarray}
\hat{G}(\hat{z}_o,\bar{\theta},\Delta \hat{\theta})
=\frac{\exp{\left(i 2 \bar{\theta}\Delta \hat{\theta}
\hat{z}_o\right)}}{2\pi \sqrt{\hat{D}\hat{N}/\hat{z}_o^2}}
\exp{\left(-\frac{\bar{\theta}^2}{2\hat{D}}\right)}
 \int_{-\infty}^{\infty} d \hat{\phi}\left[
\exp{\left(-\frac{\hat{\phi}^2}{2
\hat{N}/\hat{z}_o^2}\right)}\right.&& \cr\left. \times\exp\left(-
2 i \Delta \hat{\theta} \hat{z}_o \hat{\phi}\right) S^*
{\left[\hat{z}_o,(\hat{\phi}-\Delta \hat{\theta})^2\right]} S
{\left[\hat{z}_o,(\hat{\phi}+\Delta
\hat{\theta})^2\right]}\right.&& \cr\left. \times
\int_{-\infty}^{\infty} d \hat{\eta}
\exp{\left(-\frac{\hat{N}/\hat{z}_o^2 +\hat{D} }{2\hat{D}
\hat{N}/\hat{z}_o^2}\hat{\eta}^2+\frac{\hat{\phi}}{\hat{N}/\hat{z}_o^2}\hat{\eta}
-\frac{\bar{\theta}}{ \hat{D}}\hat{\eta}+2 i \Delta \hat{\theta}
\hat{z}_o \hat{\eta}\right)}\right] ~.\label{Gzlarge8}
\end{eqnarray}
The integral in $\hat{\eta}$ can be performed analytically thus
leading to

\begin{eqnarray}
\hat{G}(\hat{z}_o,\bar{\theta},\Delta \hat{\theta})
=\frac{\exp{\left(i 2 \bar{\theta}\Delta \hat{\theta}
\hat{z}_o\right)}}{\sqrt{2\pi(\hat{N}/\hat{z}_o^2 + \hat{D})}}
&&\cr\times\exp{\left[-\frac{\bar{\theta}^2+ 4 \hat{N} \Delta
\hat{\theta}^2 \hat{D} + 4 i (\hat{N}/\hat{z}_o)\bar{\theta}\Delta
\hat{\theta} }{2(\hat{N}/\hat{z}_o^2+\hat{D})}\right]}
 &&\cr
\times \int_{-\infty}^{\infty} d \hat{\phi}
\exp{\left[-\frac{\hat{\phi}^2+2\hat{\phi}\left(\bar{\theta}+2i
(\hat{N}/\hat{z}_o) \Delta \hat{\theta} \right)}{2
(\hat{N}/\hat{z}_o^2+\hat{D})}\right]} S^*
{\left[\hat{z}_o,(\hat{\phi}-\Delta \hat{\theta})^2\right]}
&&\cr\times S {\left[\hat{z}_o,(\hat{\phi}+\Delta
\hat{\theta})^2\right]}~.\label{Gzlarge9gen}
\end{eqnarray}
It is important to remember again that an asymptotic formula for
$\hat{z}_o \gg 1$ can be obtained from Eq. (\ref{Gzlarge9gen})
simply substituting  $S[\hat{z}_o,(\hat{\phi} \pm \Delta
\hat{\theta})^2]$ with $\mathrm{sinc}[(\hat{\phi} \pm \Delta
\hat{\theta})^2/4]$. Then, it is easy to understand that $S$ is
bound to go to zero for values of $(\hat{\phi} \pm \Delta
\hat{\theta})^2$ larger than unity, exactly as the asymptotic
terms in $\mathrm{sinc}(\cdot)$ would do. In fact, once $\hat{C}$
is set to zero, $S$ depends parametrically on the normalized
distance $\hat{z}_o$ alone, that is $S = S[\hat{z}_o, (\hat{\phi}
\pm  \Delta \hat{\theta})^2]$, and gives the previously found
asymptotic expression of $\mathrm{sinc}[(\hat{\phi} \pm \Delta
\hat{\theta})^2/4]$ in the limit for $\hat{z}_o \gg 1$. Since here
$\hat{z}_o$ is supposed to be at least of order unity ($\hat{z}_o
> 1/2$), we can conclude that $S$ must be different from zero only
for values of $(\hat{\phi} \pm \Delta \hat{\theta})^2$ of order
unity (to be more precise, for values $(\hat{\phi} \pm \Delta
\hat{\theta})^2\simeq 4$, $(\hat{\phi} \pm \Delta \hat{\theta})^2/
4$ being the arguments of the sinc function) as it can be seen,
for instance, from Fig. \ref{splot} for a particular case. Thus
Eq. (\ref{Gzlarge9gen}) and its asymptotic equivalent for
$\hat{z}_o \gg 1$ share the same mathematical structure.

Let us now introduce the non-restrictive assumptions:

\begin{eqnarray}
\hat{N} \gg1 ~,~~~ \hat{D} \gg 1~ \label{assunonr}
\end{eqnarray}
and define

\begin{equation}
\hat{A} = \frac{\hat{N}}{\hat{z}_o^2} ~.\label{barNdef}
\end{equation}
The physical interpretation of $\hat{A}$ follows from that of
$\hat{\sigma}/\hat{z}_o$: $\hat{A}$ is the dimensionless square of
the apparent angular size of the source at the observer point
position, calculated as if the source was positioned at
$\hat{z}_o=0$. If $\hat{N} \gg1$ and $\hat{D} \gg 1$ we have $( 2
\hat{A} \hat{z}_o^2 \hat{D})/(\hat{A}+\hat{D}) \gg 1$ for any
value of $\hat{z}_o$ and any choice of $\hat{N}$ and $\hat{D}$. As
a result, from the exponential factor $\exp{[-2 \hat{A}
\hat{z}_o^2 \Delta \hat{\theta}^2 \hat{D}/(\hat{A}+\hat{D})]}$
outside the integral sign in Eq. (\ref{Gzlarge9gen}) we have that
$G_{\omega}(\hat{z}_o,\bar{\theta},\Delta \hat{\theta})$ is
different from zero only for $\Delta \hat{\theta} \ll 1$. Then we
can neglect terms in $\Delta\hat{\theta}$ in the factors
$S(\cdot)$ within the integral sign thus getting

\begin{eqnarray}
\hat{G}(\hat{z}_o,\bar{\theta},\Delta \hat{\theta})
=\frac{\exp{\left(i 2 \bar{\theta}\hat{z}_o \Delta
\hat{\theta}\right)}}{\sqrt{2\pi(\hat{A} + \hat{D})}}
\exp{\left[-\frac{\bar{\theta}^2+ 4 \hat{A} \hat{z}_o^2\Delta
\hat{\theta}^2 \hat{D} + 4 i \hat{A} \bar{\theta} \hat{z}_o \Delta
\hat{\theta} }{2(\hat{A}+\hat{D})}\right]}
 &&\cr
\times \int_{-\infty}^{\infty} d \hat{\phi} ~
{\exp{\left[-\frac{\hat{\phi}^2+2\hat{\phi}\bar{\theta}}{2
(\hat{A}+\hat{D})}\right]}} \exp{\left[-i\frac{2 \hat{\phi}
\hat{A} \hat{z}_o\Delta \hat{\theta} }{\hat{A}+\hat{D}}\right]}
\left | S \left[\hat{z}_o,{{\hat{\phi}^2}}\right] \right |^2
~.\label{G11}
\end{eqnarray}
The maximal value of $\bar{\theta}$ is related with the width of
the exponential function $\exp {[- \bar{\theta}^2}$ ${/(2\hat{A}}$
${+2 \hat{D})]}$ outside the integral sign in Eq. (\ref{G11}). It
follows that in the limit for $\hat{D} \gg 1$ we can neglect the
exponential factor
$\exp{[-{(\hat{\phi}^2+2\hat{\phi}\bar{\theta})}/({2
\hat{A}+2\hat{D})}]}$ within the integral sign: in fact, its
argument assumes values of order unity for $\hat{\phi}\gg 1$, but
the factor $\mid S [\hat{z}_o,\hat{\phi}^2] \mid^2$ cuts off the
integrand for $\hat{\phi} \gtrsim 1$. Therefore Eq. (\ref{G11})
can be simplified as follows:

\begin{eqnarray}
\hat{G}(\hat{z}_o,\bar{\theta},\Delta \hat{\theta})
=\frac{\exp{\left(i 2 \bar{\theta}\hat{z}_o\Delta
\hat{\theta}\right)}}{\sqrt{2\pi(\hat{A} + \hat{D})}}
\exp{\left[-\frac{\bar{\theta}^2 }{2(\hat{A}+\hat{D})}\right]}
 \exp{\left[-\frac{2 i \hat{A} \bar{\theta}\hat{z}_o
\Delta \hat{\theta} }{\hat{A}+\hat{D}}\right]} &&\cr
\times\exp{\left[-\frac{2 \hat{A} \hat{D} \hat{z}_o^2\Delta
\hat{\theta}^2}{\hat{A}+\hat{D}}\right]} \int_{-\infty}^{\infty} d
\hat{\phi} ~ \exp{\left[i\left(-\frac{2 \hat{A}
}{\hat{A}+\hat{D}}\hat{z}_o\Delta
\hat{\theta}\right)\hat{\phi}\right]} \left | S \left[\hat{z}_o,
{{\hat{\phi}^2}}\right] \right|^2 ~.\label{G12}
\end{eqnarray}
The integral in Eq. (\ref{G12}) is simply the Fourier transform of
the function $f(\hat{\phi}) = \mid S [\hat{z}_o,\hat{\phi}^2]
\mid^2$ with respect to the variable $- 2\hat{A}\hat{z}_o\Delta
\hat{\theta} / (\hat{A}+\hat{D})$. Since the function
$f(\hat{\phi})$ has values sensibly different from zero only as
$\hat{\phi}$ is of order unity or smaller, its Fourier Transform
will also be suppressed for values of $2\hat{A}\hat{z}_o|\Delta
\hat{\theta}| / (\hat{A}+\hat{D})$ larger than unity, by virtue of
the Bandwidth Theorem. This means that the integral in Eq.
(\ref{G12}) gives non-negligible contributions only up to some
maximal value of $|\Delta \hat{\theta}|$:

\begin{equation}
|\Delta \hat{\theta}|_{\mathrm{max}} \sim
\frac{1}{2\hat{z}_o}\left(1+\frac{\hat{D} }{\hat{A}}\right)~.
\label{conG12}
\end{equation}
On the other hand, the exponential factor outside the integral in
Eq. (\ref{G12}) will cut off the function $\hat{G}$ around some
other value

\begin{equation}
|\Delta \hat{\theta}|_{\mathrm{max}2} \sim \frac{1}{2 \hat{z}_o
}\left(\frac{1}{\hat{D}}+\frac{1}{\hat{A}} \right)^{1/2}
\label{conG122}~.
\end{equation}
It is easy to see that, for any value of $\hat{z}_o$, $|\Delta
\hat{\theta}|_{\mathrm{max}} \gg |\Delta
\hat{\theta}|_{\mathrm{max}2}$. In fact we have

\begin{equation}
\frac{|\Delta \hat{\theta}|_{\mathrm{max}}}{|\Delta
\hat{\theta}|_{\mathrm{max}2}} \sim  \sqrt{\hat{D}}
\sqrt{1+(\hat{D}/\hat{A})} > \sqrt{\hat{D}} \gg 1~,\label{demo}
\end{equation}
in the limit for $\hat{D} \gg 1$. As a result the Fourier
transform in Eq. (\ref{G12}) is significant only for values of the
variable  $- 2\hat{A}\hat{z}_o\Delta \hat{\theta} /
(\hat{A}+\hat{D})$ near to zero and contributes to $\hat{G}$ only
by the inessential factor

\begin{equation}
\int_{-\infty}^{\infty} d \hat{\phi} ~ \left| S \left[\hat{z_o},
{{\phi}^2}\right]\right |^2 = \mathrm{constant}~. \label{contrin}
\end{equation}
In order to use the correlation function $\hat{G}$ for calculation
of coherence length and other statistical properties, one has to
use the spectral degree of coherence $g$, which can be presented
as a function of $\bar{\theta}$ and $\Delta \hat{\theta}$ instead
of $x_{\bot o 2}$ and $x_{\bot o 1}$:

\begin{equation}
g\left(\bar{\theta},\Delta \hat{\theta}\right) =
\frac{\hat{G}\left(\bar{\theta},\Delta
\hat{\theta}\right)}{\left\langle \left
|\hat{E}_{s\bot}\left(\bar{\theta}+\Delta
\hat{\theta}\right)\right|^2\right \rangle ^{1/2} \left \langle
\left|E\left(\bar{\theta}-\Delta
\hat{\theta}\right)\right|^2\right\rangle^{1/2}} ~.
\label{normfine}
\end{equation}
From Eq. (\ref{G12}) we obtain :

\begin{eqnarray}
g(\hat{z}_o,\bar{\theta},\Delta \hat{\theta})= \exp{\left(i 2
\bar{\theta}\hat{z}_o\Delta \hat{\theta}\right)}
\exp{\left[-\frac{2 i \hat{A}\bar{\theta}\hat{z}_o \Delta
\hat{\theta} }{\hat{A}+\hat{D}}\right]}\exp{\left[-\frac{ 2
\hat{A} \hat{D}\hat{z}_o^2\Delta
\hat{\theta}^2}{\hat{A}+\hat{D}}\right]} . \label{G13a}
\end{eqnarray}
In the asymptotic limit for a large value of $\hat{z}_o$, $\hat{A}
\ll 1$,  Eq. (\ref{G13a}) can be simplified to

\begin{eqnarray}
g(\hat{z}_o,\bar{\theta},\Delta \hat{\theta})= \exp{\left(i 2
\bar{\theta}\hat{z}_o\Delta \hat{\theta}\right)}
\exp{\left[-\frac{2 i \hat{A} \bar{\theta}\hat{z}_o \Delta
\hat{\theta} }{\hat{D}}\right]}\exp{\left[- 2 \hat{A}
\hat{z}_o^2\Delta \hat{\theta}^2\right]} ~. \label{G13asimplif}
\end{eqnarray}
From Eq. (\ref{G12}) it is easy to see that the region of interest
where the field intensity is not negligible is when $\bar{\theta}
\lesssim \sqrt{\hat{D}}$. Therefore, Eq. (\ref{G13asimplif}) can
be further approximated to

\begin{eqnarray}
g(\hat{z}_o,\bar{\theta},\Delta \hat{\theta})= \exp{\left(i 2
\bar{\theta}\hat{z}_o\Delta \hat{\theta}\right)} \exp{\left[- 2
\hat{A} \hat{z}_o^2\Delta \hat{\theta}^2\right]} ~.
\label{G13asimplif2}
\end{eqnarray}
It is interesting to calculate the transverse coherence length
$\hat{\xi}_c$ as a function of the observation distance
$\hat{z}_o$. For any experiment, complete information on the
coherence properties of light are given by the function $g$. When
calculating the coherence length one applies a certain algorithm
to $g$ thus extracting a single number. This number does not
include all information about the coherence properties of light,
and the algorithm applied to $g$ is simply a convenient
definition. Then, in order to calculate a coherence length one
has, first, to choose a definition among all the possible
convenient ones. In this paper we will simply follow the approach
by Mandel, originally developed for the time domain, but trivially
extensible to any domain of interest, in our case the angular
domain. The coherence length, naturally normalized to the
diffraction length $\sqrt{L_w c/\omega}$ is defined as

\begin{equation}
\hat{\xi}_c(\hat{z}_o) = 2 \int_{-\infty}^{\infty}
|g(\Delta\hat{\theta})|^2 d(\hat{z}_o\Delta\hat{\theta})~,
\label{cohlen}
\end{equation}
where the factor $2$ in front of the integral on the right hand
side is due to the fact that we chose Mandel's approach and that
our definition of $\Delta \hat{\theta}$ differs of a factor $1/2$
from his definition. Performing the integration in Eq.
(\ref{cohlen}) with the help of Eq. (\ref{G13a}) yields:

\begin{equation}
\hat{\xi}_c(\hat{z}_o) =  {\sqrt{\pi}}
\left(\frac{1}{\hat{A}}+\frac{1}{\hat{D}}\right)^{1/2}~.\label{cohlen2}
\end{equation}
\begin{figure}
\begin{center}
\includegraphics*[width=140mm]{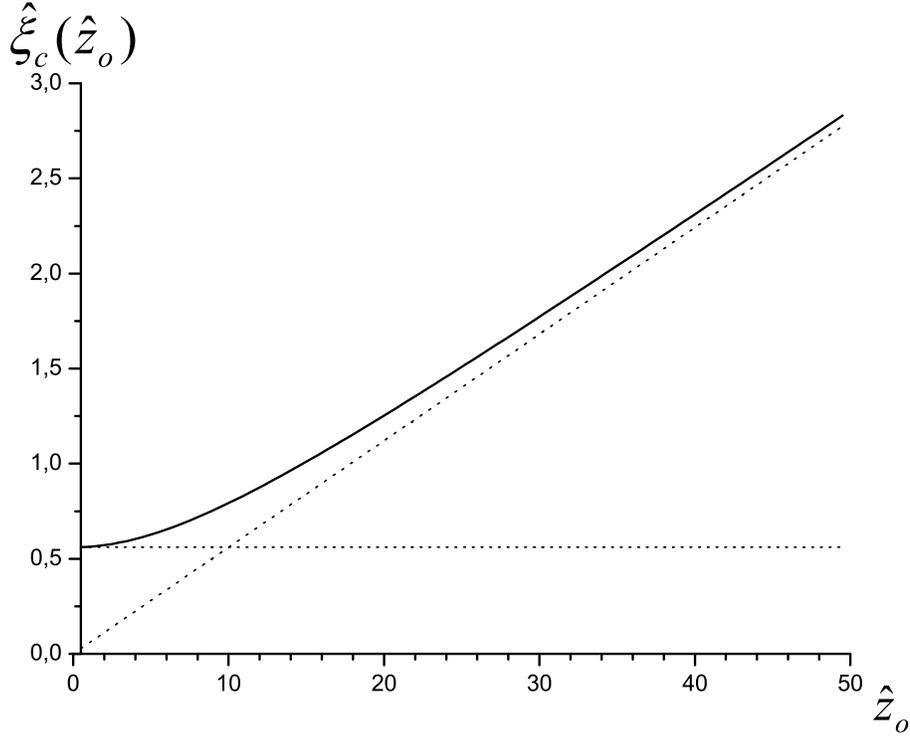}% Here is how to import EPS art
\caption{\label{uno} Coherence length $\hat{\xi}_c$ as a function
of $\hat{z}_o$ and asymptotic behaviors for $\hat{z}_o
\longrightarrow 1/2$ and $\hat{z}_o \gg 1$. Here $\hat{N} = 10^3$
and $\hat{D} = 10$.}
\end{center}
\end{figure}

\subsection{\label{sub:disc} Discussion}

\begin{figure}
\begin{center}
\includegraphics*[width=140mm]{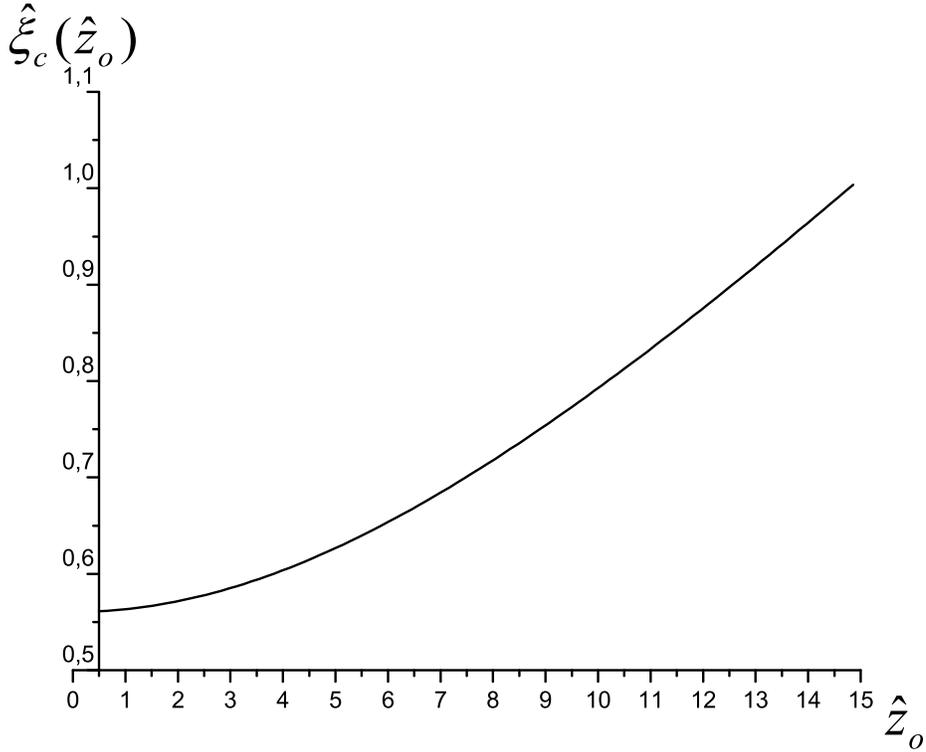}% Here is how to import EPS art
\caption{\label{due} Enlarged view of the initial part of Fig.
\ref{uno}.}
\end{center}
\end{figure}

The coherence length in Eq. (\ref{cohlen2}) exhibits linear
dependence on $\hat{z}_o$, that is $\hat{\xi}_c \longrightarrow
\sqrt{\pi/ \hat{N}}~{\hat{z}_o}$ while for $\hat{z}_o
\longrightarrow 1/2$ that is at the end of the undulator, it
converges to a constant $\hat{\xi}_c \longrightarrow
[\pi/(4\hat{N})+\pi/\hat{D}]^{1/2}$. Eq. (\ref{cohlen2}) and its
asymptotes are presented in Fig. \ref{uno} and Fig. \ref{due} for
the case $\hat{N} = 10^3$, $\hat{D} = 10$. It is evident that at
the exit of the undulator, $\hat{\xi}_c \sim 1/\sqrt{\hat{D}}$,
because $\hat{N} \gg \hat{D}$. On the other hand, horizontal
dimension of the light spot is simply proportional to
$\sqrt{\hat{N}}$ as it is evident from Eq. (\ref{G13a}). This
means that the horizontal dimension of the light spot is
determined by the electron beam size, as is intuitive, while the
beam angular distribution is printed in the fine structures of the
intensity function, that are of the dimension of the coherence
length.  In the limit for $\hat{z}_o \gg 1$ the situation is
reversed. The radiation field at the source can be presented as a
superposition of plane waves, all at the same frequency
$\omega_o$, but with different propagation angles with respect to
the $z$-direction. Since the radiation at the exit of the
undulator is partially coherent, a spiky angular spectrum is to be
expected. The nature of the spikes is easily described in terms of
Fourier transform theory, in perfect analogy with what has been
said about the frequency spectrum in Section \ref{sub:def}. From
Fourier transform theorem or, directly, from Eq. (\ref{G12}) or
from geometrical optics arguments we can expect an angular
spectrum envelope with Gaussian distribution and rms width of
$\sqrt{\hat{D}}$.

\begin{figure}
\begin{center}
\includegraphics*[width=140mm]{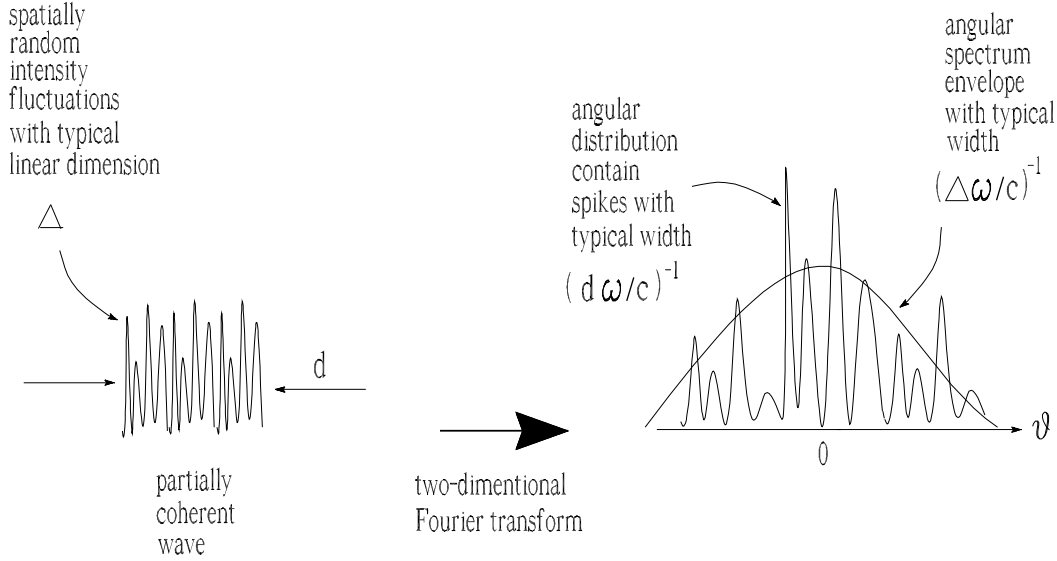}% Here is how to import EPS art
\caption{\label{spikesp} Physical interpretation of the
generalized van Cittert-Zernike theorem. If the radiation beyond
the source plane is partially coherent, a spiky angular spectrum
is expected. The nature of the spikes in the angular spectrum is
easily described in Fourier transform notations. We can expect
that typical width of the angular spectrum should be of order
$(\omega \Delta/c)^{-1}$, where $\Delta$ is the typical linear
dimension of spatially random intensity fluctuations. Also an
angular spectrum of the source having transverse size $d$ should
contain spikes with typical width of about $(\omega d/c)^{-1}$, a
consequence of the reciprocal width relations of Fourier transform
pairs.}
\end{center}
\end{figure}

Also, the angular spectrum should contain spikes with
characteristic width $1/\sqrt{\hat{N}}$, as a consequence of the
reciprocal width relations of Fourier transform pairs (see Fig.
\ref{spikesp}). This can be seen realized in mathematical form
from the expression for the cross-spectral density, Eq.
(\ref{G13a}) and from the equation for the coherence length, Eq.
(\ref{cohlen2}). Since $\hat{N} \gg 1$, the horizontal width of
the coherence spot is much smaller than the vertical one.

It is also important to remark that the asymptotic behavior for
$\hat{A} \ll 1$ of $g$ in Eq. (\ref{G13asimplif2}) and
$\hat{\xi}_c$

\begin{equation}
\hat{\xi}_c \longrightarrow \sqrt{\frac{\pi}{
\hat{N}}}{\hat{z}_o}\label{cohlen2vcz}
\end{equation}
are direct application of van Cittert-Zernike theorem.  In fact,
the last exponential factor on the right hand side of Eq.
(\ref{G13asimplif}) is simply linked with the Fourier transform of
$F_{\hat{l}_x}(\hat{l}_x)$. We derived Eq. (\ref{G13asimplif}) for
$\hat{N} \gg 1$ and $\hat{D} \gg 1$, with $ \hat{z}_o^2 \hat{D}\gg
\hat{N}$: in non-normalized units these conditions mean that the
VCZ theorem is applicable when the electron beam divergence is
much larger than the diffraction angle, i.e. $\sigma_{x'}^2 \gg
\lambda /(2\pi L_w)$, the electron beam dimensions are much larger
than the diffraction size, i.e. $\sigma_{x}^2 \gg \lambda L_w/2\pi
$, and $({\sigma}_{x'} {z}_o)^2 \gg {\sigma}_x^2$. On the
contrary, authors of \cite{TAKA} state that, in order for the van
Cittert-Zernike theorem to be applicable, "the electron-beam
divergence must be much smaller than the photon divergence", that
is our diffraction angle, i.e. $\sigma_{x'} \ll \sqrt{\lambda
/(2\pi L_w)}$ (reference \cite{TAKA}, page 571, Eq. (57)). Our
derivation shows that this conclusion is incorrect.

In \cite{GOOD} (paragraph 5.6.4) a rule of thumb is given for the
applicability region of the generalization of the VCZ theorem to
quasi-homogeneous sources. The rule of thumb requires $z_o > 2 d
\Delta/\lambda$ where $d$ is "the maximum linear dimension of the
source", that is the diameter of a source with uniform intensity
and $\Delta$ "represents the maximum linear dimension of a
coherence area of the source". In our case $d\simeq 2\sigma_x$,
since $\sigma_x$ is the rms source dimension, and from Eq.
(\ref{cohlen2}) we have $\Delta = \xi_c \simeq \lambda/(2
\sqrt{\pi} \sigma_{x'})$. The rule of thumb then requires $z_o >2
\sigma_x/( \sqrt{\pi} \sigma_{x'})$: in dimensionless this reads $
\hat{z}_o \gtrsim \sqrt{\hat{N}/\hat{D}}$. This is parametrically
in agreement with our limiting condition $\hat{z}_o^2 \hat{D} \gg
\hat{N}$, even though these two conditions are obviously different
when it come to actual estimations: our condition is, in fact,
only an asymptotic one. To see how well condition $ \hat{z}_o
\gtrsim \sqrt{\hat{N}/\hat{D}}$ works in reality we might consider
the plot in Fig. \ref{uno}. There $\hat{N} = 10^3$ and  $\hat{D} =
10$ so that, following \cite{GOOD} we may conclude that a good
condition for the applicability of the VCZ theorem should be
$\hat{z}_o \gtrsim 10$. However as it is seen from the figure, the
linear asymptotic behavior is not yet a good approximation at
$\hat{z}_o \simeq 10$. This may be ascribed to the fact that the
derivation in \cite{GOOD} is not generally valid, but has been
carried out for sources which drop to zero very rapidly outside
the maximum linear dimension $d$ and whose correlation function
also drops rapidly to zero very rapidly outside maximum linear
dimension $\Delta$.

However, at least parametrically, the applicability of the VCZ
theorem in the asymptotic limit $\hat{z}_o^2 \hat{D} \gg \hat{N}$
can be also expected from the condition $z_o > 2 d \Delta/\lambda$
in \cite{GOOD}. In other words, with the help of our approach we
were able to specify an asymptotic region where the VCZ theorem
holds. Such a region overlaps with predictions from Statistical
Optics. Statistical Optics can describe propagation of the
cross-spectral density only once it is known at some source plane
position. Our treatment allows us to specify the cross-spectral
density at the exit of the undulator, but it should be noted that
we do not need to use customary results of Statistical Optics and
propagate the cross-spectral density from the exit of the
undulator in order to obtain the cross-spectral density at some
distance along the beamline. In fact our approach, which consists
in taking advantage of the system Green's function in paraxial
approximation and, subsequently, of the resonance approximation,
allows us to calculate the cross-spectral density directly at any
distance from the exit of the undulator.

Let us now consider the structure of Eq. (\ref{G13a}) and discuss
the meaning of the phase terms in $\bar{\theta} \Delta
\hat{\theta}$. These are important in relation with the condition
for quasi-homogeneous source: their presence couples the two
variables $\bar{\theta}$ and $\Delta \hat{\theta}$ and prevents
the source to be quasi-homogeneous at any given value of
$\hat{z}_o$\footnote{Note, again, that the definition of "source
plane" is just conventional. One may define the source plane at
the exit of the undulator, that is at $\hat{z}_o = 1/2$, but there
is no fundamental reason for such a definition: one may pick any
value of $\hat{z}_o$ as the source position.}, unless they
compensate each other in some parameter region.

Let us discuss the limit, $\hat{A} \gg 1$. We may consider two
subcases. First, consider $\hat{A}\gg \hat{D} \gg 1$. In this
case, inspection of Eq. (\ref{G13a}) shows that the two phase
terms compensate and the source is quasi-homogeneous, because the
cross-spectral density is factorized in a function of
$\bar{\theta}$ and a function of $\Delta \hat{\theta}$. It should
be noted that if condition $\hat{A} \gg 1$ is not satisfied at the
exit of the undulator, where $\hat{z}_o \sim 1$, then it is never
satisfied. If $\hat{N} \gg \hat{D} \gg 1$ we have a
quasi-homogeneous source at the exit of the undulator.

\begin{figure}
\begin{center}
\includegraphics*[width=140mm]{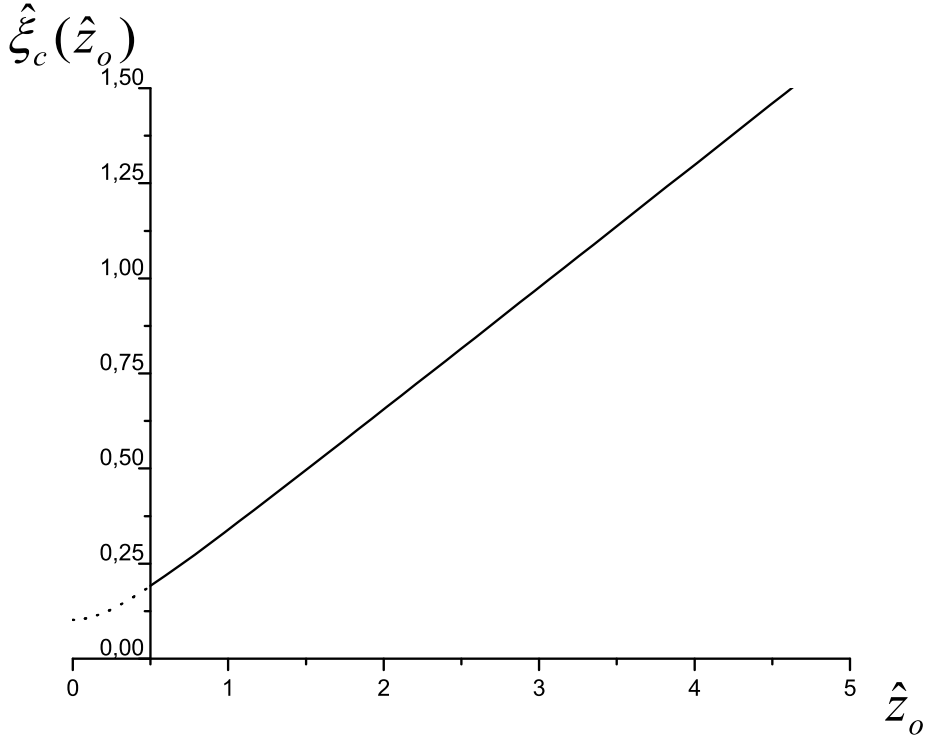}% Here is how to import EPS art
\caption{\label{plotDMNM1} Coherence length $\hat{\xi}_c$ as a
function of $\hat{z}_o$ in the case $\hat{D} \gg \hat{N} \gg 1$.
In particular, here, $\hat{N} = 30$ and $\hat{D}=300$. The linear
dependence on $\hat{z}_o$ starts already from the exit of the
undulator, because $\hat{D} \gg \hat{N}$.}
\end{center}
\end{figure}
Second, consider $\hat{D} \gg \hat{N} \gg 1$. This correspond to a
situation with a low value of the normalized betatron function in
the horizontal direction. Figure \ref{plotDMNM1} shows a numerical
example with $\hat{\epsilon}_x=100$ and $\hat{\beta}_x=0.3$ that
is $\hat{N} = 30$ and $\hat{D}=300$: the value for the horizontal
betatron function is similar to the low-$\beta$ case reported at
page 12, Table 2.2.2 in \cite{PETR}, where $\beta_x = 1.3$ m for a
$5$m-long insertion device. The value $\hat{\epsilon}_{x} = 100$
corresponds to a wavelength of about $0.6 \AA$ for the PETRA III
case. When $\hat{D} \gg \hat{N} \gg 1$ no compensation of the
phase terms in Eq. (\ref{G13a}) is possible, not even at the exit
of the undulator. In this case, whatever the value of $\hat{z}_o$
we can never have a quasi-homogeneous wavefront. This constitutes
no problem. Simply, the wavefront is not-quasi-homogeneous in this
case. However, we may interpret the situation by saying that a
"virtual" quasi-homogeneous source placed in the center of the
undulator would result in the non-homogeneous source described by
Eq. (\ref{G13a}) at the exit of the undulator. Although it
physically makes no sense to discuss about Eq. (\ref{G13a}) inside
the undulator, the "virtual" source analogy is suggested by the
fact that setting $\hat{z}_o=0$ in Eq. (\ref{G13a}), both phase
terms become zero.

Note that in general, whatever the values of $\hat{N} \gg 1$ and
$\hat{D} \gg 1$, one never has quasi-homogeneous sources in the
limit for $\hat{A} \ll 1$. In fact, in the asymptotic $\hat{A} \ll
1$   only the phase factor $\exp( i 2 \bar{\theta} \Delta
\hat{\theta} \hat{z}_o)$ contributes, which couples $\bar{\theta}$
and $\Delta \hat{\theta}$. Such a factor is connected with phase
of the field from a single electron in an undulator in the far
zone, $\omega (x_o^2+y_o^2) / (2c z_o)$, which represents, in
paraxial approximation, the phase difference between the point
$(x_o, y_o, z_o)$ and the point $(0,0,z_o)$: in the asymptotic for
large values of $\hat{z}_o$, the electric field generated by a
single electron with offset and deflection in an undulator has a
spherical wavefront (see \cite{OURS}). When one calculates the
field correlation function at two different points, he ends up
with a contribution equal to the difference (due to complex
conjugation) between $\omega (x_{o2}^2+y_{o2}^2) / (2c z_o)$ and
$\omega (x_{o1}^2+y_{o1}^2) / (2c z_o)$, that for the vertical
(and separately, the horizontal) direction gives exactly the shift
$2 \bar{\theta} \Delta \hat{\theta} \hat{z}_o$ in normalized
units. It should be noted that such a reasoning is not limited to
Synchrotron Radiation sources, but it is quite general since, as
already discussed, it relies on the fact that the wavefront of a
single radiator (in our case, an electron) produces a spherical
wavefront in the far field. This is, for instance, the case of
thermal sources as well. In other words, if the far field
radiation of a quasi-homogeneous source is taken as a new source,
that new source will never be quasi-homogeneous.

A common property of all situations with $\hat{N} \gg 1$ and
$\hat{D} \gg 1$ is that, for any value of $\hat{z}_o$, the modulus
of $g$, i.e. $|g|$, is always independent on $\bar{\theta}$.
Moreover, it is always possible to apply the VCZ theorem starting
either from a virtual quasi-homogeneous source placed at center of
the undulator when $\hat{D} \gg \hat{N} \gg 1$, or otherwise from
a real one placed at the exit of the undulator (or at any other
position $\hat{z}_o$ close enough to the exit of the undulator to
guarantee a quasi-homogeneous wavefront). These observations
suggest to extend the concept of quasi-homogeneity, and introduce
the new concept of "weak quasi-homogeneity" as discussed before.
With respect to the new coordinates $\bar{\theta}$ and $\Delta
\hat{\theta}$, a given wavefront at fixed position $\hat{z}_o$ is
said to be weakly quasi-homogeneous when $|g|$ is independent of
$\bar{\theta}$. With this new definition at hand we can restate
some of our conclusions in a slightly different language. We have
seen that in the far field, when the VCZ theorem holds, the
wavefronts are weakly quasi-homogeneous, but never
quasi-homogeneous in the usual sense. In the case $\hat{N} \gg
\hat{D} \gg 1$ we pass from quasi-homogeneous wavefronts (in the
usual sense) in the near field to weakly-quasi homogeneous
wavefronts (but not quasi-homogenous in the usual sense) in the
far field. Note that the wavefronts are always weakly
quasi-homogeneous, even during the transition from near to far
zone. In the case $\hat{D} \gg \hat{N} \gg 1$ instead, the VCZ
theorem is applicable already from the exit of the undulator, as
it can be seen from Fig. \ref{plotDMNM1}, and the wavefront is not
quasi-homogeneous in the usual sense, but still weakly
quasi-homogeneous from the very beginning.

The weak quasi-homogeneity of the wavefronts at any value of
$\hat{z}_o$, i.e. the fact that $|g|$ is independent of
$\bar{\theta}$ for any value of $\hat{z}_o$ guarantees that the
plot in Fig. \ref{plotDMNM1} is universal. It should be noted that
this fact depends on the choice of large parameters $\hat{N} \gg
1$ and $\hat{D} \gg 1$, but it is also strictly related with the
Gaussian nature of the electron distribution in angles and
offsets, that is a well-established fact for storage-ring beams.
If angles or offsets were obeying different distribution laws, in
general, one could not perform the integral in $\hat{\eta}$ in Eq.
(\ref{Gzlarge7}) and, in general, $|g|$ would have shown a
dependence on $\bar{\theta}$: our noticeable result is linked with
the properties of the exponential elementary function. However, it
should be clear that even in the case when angles or offsets were
obeying different distribution laws, i.e when the plot in Fig.
\ref{plotDMNM1} is not universal, we could have situations when
wavefronts are quasi-homogeneous in the usual sense near the exit
of the undulator and are weakly quasi-homogeneous in the far field
limit, but not along the transition between these two zones. A
more detailed discussion of this issue will be given in Section
\ref{sec:twod}, where we will be discussing conditions for the
source to be quasi-homogeneous.

Another remark to be made pertains the applicability of the VCZ
theorem. As we deal with a quasi-homogenous source (in the usual
sense) the knowledge of $I(\bar{\theta})$ and $g(\Delta
\hat{\theta})$ in the far zone allow, respectively, the
calculation at the source plane of $g(\Delta \hat{x})$ through the
"anti" VCZ theorem and of $I(\bar{x})$ through the VCZ theorem
(here we consider only one dimension, the horizontal one $x$).
Viceversa the knowledge of $I$ and $g$ at the source allow
calculation of $I$ and $g$ in the far field. In terms of
intensity, all information regarding wavefront evolution (assuming
a quasi-homogeneous source, in the usual sense) is included in
$I(\bar{\theta})$ and $I(\bar{x})$. For instance, the knowledge of
$I(\bar{\theta})$ allows calculation of $g(\Delta \hat{x})$ at the
source plane through the "anti" VCZ theorem. Then, the knowledge
of $g(\Delta \hat{x})$ and $I(\bar{x})$ allow the calculation of
the cross-spectral density, which can be propagated at any
distance, and allow to recover $g(\Delta \hat{\theta})$. So,
complete characterization of the undulator source is given when
$I(\bar{\theta})$ and $I(\bar{x})$, when the source is assumed
quasi-homogenous.

Yet we have seen that, when the electron distribution in angles
and offsets are Gaussian and  $\hat{N} \gg 1$ and $\hat{D} \gg 1$,
the VCZ theorem holds also in the case $\hat{D} \gg \hat{N} \gg
1$, when the source is non quasi-homogeneous in the usual sense.
We have seen that the situation can be equivalently described with
the help of a virtual quasi-homogeneous source in the middle of
the undulator. However, such an interpretation is only valid
\textit{a posteriori}. For the case $\hat{N} \gg 1$ and $\hat{D}
\gg 1$ there is a non quasi-homogeneous wavefront at the undulator
exit; before our approach was presented one would have concluded
that the VCZ theorem cannot be applied, since the spectral degree
of coherence does not form a Fourier pair with the intensity
distribution at the undulator exit. Our approach is based on the
simplification of mathematical results through the use of small
and large parameters and subsequent understanding and
interpretation of these simplified results:  in our analysis we
were never limited to the treatment of quasi-homogenous cases
alone.

As a closing remark about the coherence length we like to draw the
reader's attention on the fact that the dimensional form $\xi_c$
of the coherence length, given in normalized units by
$\hat{\xi}_c$ in Eq. (\ref{cohlen2}), does not include the
undulator length. This is to be expected since, in the limit
$\hat{N} \gg 1$ and $\hat{D} \gg 1$, the typical size and
divergence of the electron beam are much larger than the
diffraction size $\sqrt{c L_w/\omega}$ and angle $\sqrt{c/(\omega
L_w)}$, which are intrinsic properties of the undulator radiation.
As a result, in this limit, the evolution of the radiation beam is
a function of the electron beam parameters only, and does not
depend on the undulator length. In the following Section
\ref{sec:twod}, where we will extend our model to a
two-dimensional case, we will see that the quasi-homogeneous
approximation is valid in many practical situations, but we will
have to account for diffraction of undulator radiation in the
vertical direction. In this case the dimensional coherence $\xi_c$
length will be a function of the undulator length $L_w$ as well.

\section{\label{sec:twod} Effect of the vertical emittance on the cross-spectral density}

Up to now we were dealing with the field correlation function $g$
within the framework of a one-dimensional model.

In fact we considered the limit for $\hat{\epsilon}_y
\hat{\beta}_y \ll 1$ and $\hat{\epsilon}_y/ \hat{\beta}_y \ll 1$
and we calculated $g$ for $\hat{\theta}_{y1} =
\hat{\theta}_{y2}=0$ and $\hat{C}=0$, so that our attention was
focused on coherent effects in the horizontal direction. We will
now extend our considerations to a two-dimensional model always
for $\hat{C}=0$. This can be done by a straightforward
generalization of Eq. (\ref{Gzlarge9gen}) which can be obtained
from Eq. (\ref{Gzlarge3}) following the same steps which lead to
Eq. (\ref{Gzlarge9gen}), but this time without assumptions on
$\hat{N}_y$, $\hat{D}_y$, $\hat{\theta}_{y1}$ and
$\hat{\theta}_{y2}$.  Finally, at the end of calculations, our
final expression for $\hat{G}$ should be normalized to

\begin{equation}
\hat{W} = \left\langle \left
|\hat{E}_{s\bot}\left({\vec{\hat{\theta}}}_{1}\right)\right|^2\right
\rangle ^{1/2} \left \langle
\left|\hat{E}_{s\bot}\left({\vec{\hat{\theta}}}_{2}\right)\right|^2\right\rangle^{1/2}~.
\label{normfine2}
\end{equation}
As has already been seen in Section \ref{sec:quasi},  after
normalization to $\hat{W}$ we will obtain the spectral degree of
coherence $g$. With this in mind we will neglect, step after step,
unnecessary multiplicative factors that, in any case, would be
finally disposed after normalization of the final result.
Retaining indexes $x$ and $y$ in our notation we obtain

\begin{eqnarray}
\hat{G}(\hat{z}_o,\bar{\theta}_x,\bar{\theta}_y,\Delta
\hat{\theta}_x,\Delta \hat{\theta}_y) ={\exp{\left[i 2\left(
\bar{\theta}_x\hat{z}_o\Delta
\hat{\theta}_x+\bar{\theta}_y\hat{z}_o\Delta \hat{\theta}_y\right)
\right]}}&&\cr \times \exp{\left[-\frac {\bar{\theta}_x^2 + 4
\hat{A}_x \hat{z}_o^2 \Delta \hat{\theta}_x^2 \hat{D}_x + 4 i
\hat{A}_x \bar{\theta}_x \hat{z}_o \Delta \hat{\theta}_x
}{2(\hat{A}_x+\hat{D}_x)}\right]} &&\cr\times \exp{\left[ - \frac
{\bar{\theta}_y^2 + 4 \hat{A}_y \hat{z}_o^2 \Delta
\hat{\theta}_y^2 \hat{D}_y + 4 i \hat{A}_y \bar{\theta}_y
\hat{z}_o \Delta \hat{\theta}_y }{2(\hat{A}_y+\hat{D}_y)}\right] }
&&\cr \times \int_{-\infty}^{\infty} d \hat{\phi}_x
\int_{-\infty}^{\infty} d \hat{\phi}_y
\exp{\left[-\frac{\hat{\phi}_x^2+2\hat{\phi}_x\left(\bar{\theta}_x+2i
\hat{A}_x \hat{z}_o \Delta \hat{\theta}_x \right)}{2
(\hat{A}_x+\hat{D}_x)}\right]} &&\cr \times
\exp{\left[-\frac{\hat{\phi}_y^2+2\hat{\phi}_y\left(\bar{\theta}_y+2i
\hat{A}_y \hat{z}_o \Delta \hat{\theta}_y \right)}{2
(\hat{A}_y+\hat{D}_y)}\right]}&&\cr \times
S^*{\left[\hat{z}_o,(\hat{\phi}_x-\Delta
\hat{\theta}_x)^2+(\hat{\phi}_y-\Delta \hat{\theta}_y)^2\right]}
S{\left[\hat{z}_o,(\hat{\phi}_x+\Delta
\hat{\theta}_x)^2+(\hat{\phi}_y+\Delta
\hat{\theta}_y)^2\right]}~.\label{G2D}
\end{eqnarray}
We will still assume $\hat{N}_x\gg 1$ and $\hat{D}_x\gg 1$. This
allows to factorize the right hand side of Eq. (\ref{G2D}) in the
product of contribution depending on horizontal
($\bar{\theta}_x,\Delta \hat{\theta}_x$) coordinates only with a
second depending on vertical ($\bar{\theta}_y,\Delta
\hat{\theta}_y$) coordinates only. In fact, from the exponential
factor outside the integral sign in Eq. (\ref{G2D}) it is possible
to see that the maximum value of $\Delta \hat{\theta}_x^2$ is of
order $(\hat{A}_x + \hat{D}_x)/(\hat{A}_x \hat{D}_x \hat{z}_o^2)
\ll 1$. As a result, $\Delta \hat{\theta}_x$ can be neglected
inside the $S$ functions in Eq. (\ref{G2D}). Moreover, since
$\hat{D}_x \gg 1$ one can also neglect the exponential factor in
$\hat{\phi}_x^2+2\hat{\phi}_x \bar{\theta}_x$ inside the integral
sign. This leads to

\begin{eqnarray}
\hat{G}(\hat{z}_o,\bar{\theta}_x,\bar{\theta}_y,\Delta
\hat{\theta}_x,\Delta \hat{\theta}_y) ={\exp{\left[i 2\left(
\bar{\theta}_x\hat{z}_o\Delta
\hat{\theta}_x+\bar{\theta}_y\hat{z}_o\Delta \hat{\theta}_y\right)
\right]}}&&\cr \times \exp{\left[-\frac {\bar{\theta}_x^2 + 4
\hat{A}_x \hat{z}_o^2 \Delta \hat{\theta}_x^2 \hat{D}_x + 4 i
\hat{A}_x \bar{\theta}_x \hat{z}_o \Delta \hat{\theta}_x
}{2(\hat{A}_x+\hat{D}_x)}\right]}&&\cr \exp{\left[ - \frac
{\bar{\theta}_y^2 + 4 \hat{A}_y \hat{z}_o^2 \Delta
\hat{\theta}_y^2 \hat{D}_y + 4 i \hat{A}_y \bar{\theta}_y
\hat{z}_o \Delta \hat{\theta}_y }{2(\hat{A}_y+\hat{D}_y)}\right] }
&&\cr \times \int_{-\infty}^{\infty} d \hat{\phi}_x  \exp{\left[i
\hat{\phi}_x\frac{
 2  \hat{A}_x \hat{z}_o \Delta \hat{\theta}_x
}{\hat{A}_x+\hat{D}_x}\right]} \int_{-\infty}^{\infty} d
\hat{\phi}_y
\exp{\left[-\frac{\hat{\phi}_y^2+2\hat{\phi}_y\left(\bar{\theta}_y+2i
\hat{A}_y \hat{z}_o \Delta \hat{\theta}_y \right)}{2
(\hat{A}_y+\hat{D}_y)}\right]}&&\cr \times
S^*{\left[\hat{z}_o,\hat{\phi}_x^2+(\hat{\phi}_y-\Delta
\hat{\theta}_y)^2\right]}
S{\left[\hat{z}_o,\hat{\phi}_x^2+(\hat{\phi}_y+\Delta
\hat{\theta}_y)^2\right]}~.\label{G2Dnewsimplif}
\end{eqnarray}
Based on the same reasoning in Section \ref{sub:oned}, that we
repeat here for completeness, we can also neglect the phase factor
in $\hat{\phi}_x$ under the integral in $d\hat{\phi}_x$ in Eq.
(\ref{G2Dnewsimplif}). Such integral in $d\hat{\phi}_x$ is simply
the Fourier transform of the function

\begin{eqnarray}
f(\hat{\phi}_x) = \int_{-\infty}^{\infty} d \hat{\phi}_y
\exp{\left[-\frac{\hat{\phi}_y^2+2\hat{\phi}_y\left(\bar{\theta}_y+2i
\hat{A}_y \hat{z}_o \Delta \hat{\theta}_y \right)}{2
(\hat{A}_y+\hat{D}_y)}\right]}&&\cr \times
S^*{\left[\hat{z}_o,\hat{\phi}_x^2+(\hat{\phi}_y-\Delta
\hat{\theta}_y)^2\right]}
S{\left[\hat{z}_o,\hat{\phi}_x^2+(\hat{\phi}_y+\Delta
\hat{\theta}_y)^2\right]}~. \label{difficultfunc}
\end{eqnarray}
with respect to the variable $- 2\hat{A}_x\hat{z}_o\Delta
\hat{\theta}_x / (\hat{A}_x+\hat{D}_x)$. In the argument of the
$S$ functions on the right hand side of Eq. (\ref{difficultfunc}),
$\hat{\phi}_x^2$ is always summed to positively defined
quantities. This remark allows one to conclude that
$f(\hat{\phi}_x)$ has values sensibly different from zero only as
$\hat{\phi}_x$ is of order unity or smaller. Therefore, its
Fourier Transform will also be suppressed for values of $
2\hat{A}_x\hat{z}_o |\Delta \hat{\theta}_x| /
(\hat{A}_x+\hat{D}_x)$ larger than unity, by virtue of the
Bandwidth Theorem. This means that the integral in $d\hat{\phi}_x$
in Eq. (\ref{G2Dnewsimplif}) gives non-negligible contributions
only up to some maximal value of $|\Delta \hat{\theta}_x|$:

\begin{equation}
|\Delta \hat{\theta}|_{x \mathrm{max}} \sim
\frac{1}{2\hat{z}_o}\left( 1+\frac{\hat{D}_x }{\hat{A}_x}\right)~.
\label{conG12tris}
\end{equation}
On the other hand, the exponential factor outside the integral in
Eq. (\ref{G2Dnewsimplif}) will cut off the function $\hat{G}$
around some other value

\begin{equation}
|\Delta \hat{\theta}|_{x \mathrm{max}2} \sim
\frac{1}{2\hat{z}_o}\left(\frac{1}{ \hat{D}_x}+\frac{1}{\hat{A}_x}
\right)^{1/2} \label{conG122bissss}~.
\end{equation}
It is easy to see that, for any value of $\hat{z}_o$, $|\Delta
\hat{\theta}|_{x \mathrm{max}} \gg |\Delta \hat{\theta}|_{x
\mathrm{max}2}$. In fact we have

\begin{equation}
\frac{|\Delta \hat{\theta}|_{x \mathrm{max}}}{|\Delta
\hat{\theta}|_{x \mathrm{max}2}} \sim  \sqrt{\hat{D}_x} \sqrt{1+(
\hat{D}_x/\hat{A}_x)} > \sqrt{\hat{D}_x} \gg 1~,\label{demobissss}
\end{equation}
in the limit for $\hat{D}_x \gg 1$. As a result the Fourier
transform in Eq. (\ref{G2Dnewsimplif}) is significant only for
values of the variable  $- 2\hat{A}_x \hat{z}_o \Delta
\hat{\theta}_x / (\hat{A}_x+\hat{D}_x)$ near to zero. As a result
we obtain the following equation for $\hat{G}$:

\begin{eqnarray}
\hat{G}(\hat{z}_o,\bar{\theta}_x,\bar{\theta}_y,\Delta
\hat{\theta}_x,\Delta \hat{\theta}_y) ={\exp{\left[i 2\left(
\bar{\theta}_x\hat{z}_o\Delta
\hat{\theta}_x+\bar{\theta}_y\hat{z}_o\Delta \hat{\theta}_y\right)
\right]}}&&\cr \times \exp{\left[-\frac {\bar{\theta}_x^2 + 4
\hat{A}_x \hat{z}_o^2 \Delta \hat{\theta}_x^2 \hat{D}_x + 4 i
\hat{A}_x \bar{\theta}_x \hat{z}_o \Delta \hat{\theta}_x
}{2(\hat{A}_x+\hat{D}_x)}\right]}&&\cr \exp{\left[ - \frac
{\bar{\theta}_y^2 + 4 \hat{A}_y \hat{z}_o^2 \Delta
\hat{\theta}_y^2 \hat{D}_y + 4 i \hat{A}_y \bar{\theta}_y
\hat{z}_o \Delta \hat{\theta}_y }{2(\hat{A}_y+\hat{D}_y)}\right] }
&&\cr \times \int_{-\infty}^{\infty} d \hat{\phi}_y
\exp{\left[-\frac{\hat{\phi}_y^2+2\hat{\phi}_y\left(\bar{\theta}_y+2i
\hat{A}_y \hat{z}_o \Delta \hat{\theta}_y \right)}{2
(\hat{A}_y+\hat{D}_y)}\right]}&&\cr \times \int_{-\infty}^{\infty}
d \hat{\phi}_x
S^*{\left[\hat{z}_o,\hat{\phi}_x^2+(\hat{\phi}_y-\Delta
\hat{\theta}_y)^2\right]}
S{\left[\hat{z}_o,\hat{\phi}_x^2+(\hat{\phi}_y+\Delta
\hat{\theta}_y)^2\right]}~,\label{G2Dnewsimplif2}
\end{eqnarray}
where horizontal and vertical  coordinates are obviously
factorized.

Eq. (\ref{G2Dnewsimplif2}) has been derived assuming $\hat{N}_x
\gg 1$ and $\hat{D}_x \gg 1$. Note that assuming setting $\Delta
\hat{\theta}_y = \bar{\theta}_y =0$ one can obtain Eq.
(\ref{G13a}) from Eq. (\ref{G2Dnewsimplif2}). This proves that Eq.
(\ref{G13a}) has a wider range of validity than that for
$\hat{N}_y \ll 1$ and $\hat{D}_y \ll 1$ (as the reader will
remember, these assumptions were made just for notational
simplicity). In fact, as $\hat{N}_x \gg 1$ and $\hat{D}_x \gg 1$
horizontal and vertical direction factorize and the horizontal
factor is always that in Eq. (\ref{G13a}), independently on
$\hat{N}_y$ and $\hat{D}_y$.

Under one of the two extra assumptions $\hat{N}_y \gg 1$ or
$\hat{D}_y \gg 1$, Eq. (\ref{G2Dnewsimplif2}) can be further
simplified and often describes a weakly quasi-homogeneous
wavefront according to the definition given at the end of Section
\ref{sub:oned}. At the end of the Section we will show that as
$\hat{D}_y \gg 1$ we always have a weakly quasi-homogeneous
wavefront, for any value of $\hat{N}_y$. It will also be seen that
the same applies when $\hat{N}_y \gg 1$ and $\hat{A}_y \ll 1$ (far
field) or $\hat{A}_y \gg 1$ (near field) for any value of
$\hat{D}_y$. However, as $\hat{N}_y \gg 1$, $\hat{D}_y \lesssim 1$
and $\hat{A}_y \sim 1$ we have an intermediate region between the
near and far region were, in general, wavefronts are not
quasi-homogeneous, not even in the weak case.

For simplicity of discussion we will set $\bar{\theta}_x =
\bar{\theta}_y = 0$ thus obtaining from Eq. (\ref{G2Dnewsimplif2})

\begin{eqnarray}
\hat{G}(\hat{z}_o,\Delta \hat{\theta}_x,\Delta \hat{\theta}_y) =
\exp{\left[-\frac{2 \hat{A}_x \hat{z}_o^2 \Delta \hat{\theta}_x^2
\hat{D}_x }{(\hat{A}_x+\hat{D}_x)} \right]} \exp{\left[-\frac{2
\hat{A}_y \hat{z}_o^2\Delta \hat{\theta}_y^2 \hat{D}_y
}{(\hat{A}_y+ \hat{D}_y)}\right]} &&\cr \times
\int_{-\infty}^{\infty} d \hat{\phi}_y
\exp{\left[-\frac{\hat{\phi}_y^2+2\hat{\phi}_y\left(2i \hat{A}_y
\hat{z}_o \Delta \hat{\theta}_y \right)}{2
(\hat{A}_y+\hat{D}_y)}\right]} &&\cr \times
\int_{-\infty}^{\infty} d \hat{\phi}_x
S^*{\left[\hat{z}_o,\hat{\phi}_x^2+(\hat{\phi}_y-\Delta
\hat{\theta}_y)^2\right]}
S{\left[\hat{z}_o,\hat{\phi}_x^2+(\hat{\phi}_y+\Delta
\hat{\theta}_y)^2\right]}~,~~~~~~\label{G2D2}
\end{eqnarray}
which is easier to manipulate. It should be reminded that, if one
is interested in ascertaining the weak quasi-homogeneity of a
wavefront, one has to deal with the full Eq.
(\ref{G2Dnewsimplif2}). Moreover, it should be noted that on the
one hand,  within the weak quasi-homogeneous case, Eq.
(\ref{G2D2}) is quite general and we can extract from it all
important information on the transverse coherence independently on
the values of $\bar{\theta}_x$ and $\bar{\theta}_y$ . On the other
hand though, in the case the weakly quasi-homogeneous assumption
fails, Eq. (\ref{G2D2}), e.g. when $\hat{N}_y \gg 1$, $\hat{D}_y
\lesssim 1$ and $\hat{A}_y \sim 1$ as we will see, $|g|$ depends
on $\bar{\theta}_{y}$ and the study of Eq. (\ref{G2D2}) has a more
restricted range of validity, namely for the particular value of
$\bar{\theta}_y=0$.

In Section \ref{sub:N} we will assume $\hat{N}_y \gg 1$ and
arbitrary $\hat{D}_y$, while in Section \ref{sub:D} we will study
the case with arbitrary $\hat{N}_y$ and $\hat{D}_y \gg 1$.

In general, the coherence length in the $\hat{y}$ direction
(calculated at $\Delta \hat{\theta}_x = 0$, but trivially
extendible to the case $\Delta \hat{\theta}_x \neq 0$),
$\hat{\xi}_{c y}$ is a function of $\hat{D}_y$, $\hat{N}_y$ and
$\hat{z}_o$, as it can be concluded by inspection of the general
expression for $\hat{G}$ in Eq. (\ref{G2D}).

As $\hat{N}_y \gg 1$ and $\hat{D}_y$ is arbitrary we will
demonstrate that $\hat{\xi}_{c y}$ can be approximated as

\begin{equation}
\hat{\xi}_{c y} = \Phi\left[\hat{D}_y, \hat{A}_y \right]
\label{xicdarb} ~,
\end{equation}
where $\Phi$ is a universal function of the dimensionless
parameters $\hat{D}_y$ and $\hat{A}_y$. We will first study the
asymptotic cases $\hat{D}_y \gg 1$ and $\hat{D}_y \ll 1$, which
are useful for physical understanding. Then we will generalize our
results accounting for the influence of a finite divergence of the
electron beam on the cross-spectral density. An analytical
approximation for the function ${\Phi}$ will be proposed. This
will be chosen to match in a very simple way both asymptotic
expressions for $\hat{D}_y \gg 1$ and $\hat{D}_y \ll 1$: we will
demonstrate that for any value of $\hat{D}_y$, discrepancies
between the approximated and the actual (numerically calculated)
expression are less than $10 \%$, though there is no theoretical
reason to assume, \textit{a priori}, this relatively good
accuracy. Also, at $\hat{A}_y \ll 1$ (and $\hat{N}_y \gg 1$) we
will see that the VCZ theorem always hold.

As $\hat{N}_y$ is arbitrary and $\hat{D}_y \gg 1$ we will show,
instead, that the coherence length (calculated again at $\Delta
\hat{\theta}_x = 0$, but trivially extendible to the case $\Delta
\hat{\theta}_x \neq 0$) can be approximated as

\begin{equation}
\hat{\xi}_{c y} = \Psi\left[\hat{D}_y,
\hat{z}_o^2/\hat{N}_\mathrm{eff}(\hat{N}_y)\right]
~,\label{xicnarb}
\end{equation}
$\hat{N}_\mathrm{eff}(\hat{N}_y)$ being a universal function of
the Fresnel number $\hat{N}_y$. $\Psi$ is a universal function of
the dimensionless parameters $\hat{D}_y$, $\hat{N}_y$ and
$\hat{z}_o$. As usual in this paper, we will first study the
asymptotic cases $\hat{N}_y \gg 1$ and $\hat{N}_y \ll 1$. Then we
will generalize our results accounting for any value of
$\hat{N}_y$. As before, an analytical approximation for $\Psi$
will be proposed. This will be chosen to match in a very simple
way both asymptotic expressions for $\hat{N}_y \gg 1$ and
$\hat{N}_y \ll 1$: again, we will demonstrate that for any value
of $\hat{N}_y$, discrepancies between the approximated and the
actual (numerically calculated) expression are less than $13 \%$
though there is no theoretical reason to assume, \textit{a
priori}, this relatively good accuracy.

The case for a large Fresnel number with a finite electron beam
divergence, or viceversa of a large beam divergence and a finite
Fresnel number, is of practical importance. To give a numerical
example, let us put $\lambda = 1 ~\AA$, and consider a typical
vertical emittance (for third generation sources in operation)
$\epsilon_y \simeq 3 \cdot 10^{-11}$ m, that is $\hat{\epsilon}_y
\simeq 2$. On the one hand, if $\hat{\beta}_y = 3$, we have
$\hat{D}_y \simeq 0.6$ and $\hat{N}_y \simeq 6$. On the other
hand, if $\hat{\beta}_y = 0.3$ we have, viceversa, $\hat{D}_y
\simeq 6$ and $\hat{N}_y \simeq 0.6$.

\subsection{\label{sub:N} Very large Fresnel number
$\hat{N}_y\gg 1$, arbitrary divergence parameter $\hat{D}_y$}

As a matter of fact, the only important assumptions used to derive
results Section \ref{sub:oned} were that $\hat{N}_x \gg 1$ and
$\hat{D}_x \gg 1$ that are valid here as well. When these
assumptions are granted, results can be factorized as a product of
factors dependent separately  on the $x$ and on the $y$
directions. Then the spectral degree of coherence in the
horizontal direction will always be the same as in Section
\ref{sub:oned}. Differences will arise here, of course, due to
$\hat{\theta}_{y1} \ne \hat{\theta}_{y2}$.

\subsubsection{\label{paragr:1} Case with divergence parameter
$\hat{D}_y \gg 1$.}

This case is the easiest to analyze, because one can follow step
by step the derivation for the one dimensional model given in the
previous Section \ref{sub:oned}. Calculations in the vertical
direction $y$ simply follow the derivation for the horizontal
direction $x$. As a result, one can start from Eq. (\ref{G2D}) and
perform, separately for the $x$ and the $y$ directions, the same
simplifications which hold for the one-dimensional model. By
comparison with Eq. (\ref{G13a}), one obtains directly the result

\begin{eqnarray}
g(\hat{z}_o,\Delta \hat{\theta}_x, \Delta \hat{\theta}_y)=
\exp{\left[-\frac{ 2 \hat{A}_x
\hat{D}_x}{\hat{A}_x+\hat{D}_x}\hat{z}_o^2\Delta
\hat{\theta}_x^2\right]} \exp{\left[-\frac{ 2 \hat{A}_y
\hat{D}_y}{\hat{A}_y+\hat{D}_y}\hat{z}_o^2\Delta
\hat{\theta}_y^2\right]} \label{G142D} ~.\end{eqnarray}
Normalization of Eq. (\ref{G142D}) according to
Eq.(\ref{normfine2}) has been obtained simply setting
$\hat{G}(\hat{z}_o,0,0) = 1$.  In this case as well,
two-dimensional Fourier Transform of $|S
[\hat{z}_o,{\hat{\phi}_x^2+\hat{\phi}_y^2}] |^2$ calculated with
respect the variables $- 2\hat{A}_x \hat{z}_o \Delta
\hat{\theta}_x$ $/ (\hat{A}_x+\hat{D}_x)$ and $- 2\hat{A}_y
\hat{z}_o\Delta \hat{\theta}_y$ $/ (\hat{A}_y+\hat{D}_y)$ gives,
in analogy with Eq. (\ref{contrin}), an unessential factor

\begin{equation}
\int_{-\infty}^{\infty} d \hat{\phi}_x \int_{-\infty}^{\infty} d
\hat{\phi}_y ~ \left |
S\left[\hat{z}_o,{\hat{\phi}_x^2+\hat{\phi}_y^2}\right] \right
|^2= \mathrm{constant}~, \label{contrin3}
\end{equation}
which has been included in the normalization. Again, similarly as
before one can calculate the coherence area
$\hat{\Omega}_c(\hat{z}_o)$ defined in analogy with
$\hat{\xi}_c(\hat{z}_o)$ as

\begin{equation}
\hat{\Omega}_c(\hat{z}_o) = \hat{\xi}_{cx}(\hat{z}_o)
\hat{\xi}_{cy}(\hat{z}_o)\label{cohlen2D}
\end{equation}
Performing the integration yields:

\begin{equation}
\hat{\Omega}_c =  {\pi}
\left(\frac{1}{\hat{A}_x}+\frac{1}{\hat{D}_x}\right)^{1/2}
\left(\frac{1}{\hat{A}_y}+\frac{1}{\hat{D}_y}\right)^{1/2}
\label{cohlen22D}
\end{equation}
In the limit for a large value of $\hat{z}_o$ the coherence area
exhibits quadratic dependence on $\hat{z}_o$, that is
$\hat{\Omega}_c \longrightarrow {\pi \hat{z}_o^2}/
(\hat{N}_x\hat{N}_y)^{1/2}$ while for $\hat{z}_o \longrightarrow
1/2$, that is at the end of the undulator, it converges to the
constant value $\hat{\Omega}_c \longrightarrow $ $\pi
\left[{1}/(4{\hat{N}_x})+ {1}/{\hat{D}_x}\right]^{1/2}
\left[{1}/(4{\hat{N}_y})+ {1}/{\hat{D}_y}\right]^{1/2}$.

It should be noted as before that the asymptotic behavior for
$\hat{z}_o \gg 1$ of $g$

\begin{eqnarray}
g(\hat{z}_o,\Delta \hat{\theta}_x,\Delta \hat{\theta}_y)=
\exp{\left[- 2 \hat{A}_x \hat{z}_o^2 \Delta
\hat{\theta}_x^2\right]} \exp{\left[- 2 \hat{A}_y \hat{z}_o^2
\Delta \hat{\theta}_y^2\right]}\label{G14vcz2D} ~\end{eqnarray}
and $\hat{\Omega}_c$

\begin{equation}
\hat{\Omega}_c =\frac{\pi \hat{z}_o^2}{\sqrt{\hat{N}_x \hat{N}_y}}
\label{cohlen2vcz2D}
\end{equation}
are direct application of van Cittert-Zernike theorem, as it must
be since $\hat{A}_{x,y} \ll 1$. In fact, Eq. (\ref{G14vcz2D}) is
simply linked with the two-dimensional Fourier transform of
$F_{\hat{l}_x}(\hat{l}_x) F_{\hat{l}_y}(\hat{l}_y)$.

\subsubsection{\label{paragr:2} Case with divergence parameter
$\hat{D}_y \ll 1$.}

With respect to the situation treated in Paragraph \ref{paragr:1},
this case requires a more careful analysis of the relations
between different parameters. In fact, on the one hand $\hat{D}_y
\ll 1$ implies that the electron beam divergence drops out of the
problem parameters, but on the other hand in this case the
divergence of the radiation is described by the intrinsic
divergence of undulator radiation, that is described by a more
complicate mathematical function, compared with a Gaussian. In
relation with this, it should be noted that simplifications in
Section \ref{sub:oned} pr \ref{paragr:1} were based on the very
specific properties of the Gaussian function representing the
electron distributions in offset and deflection. Luckily, this is
a realistic description in storage ring beam physics.

Let us consider Eq.(\ref{G2D2}). In order to derive it we only
used the assumptions $\hat{N}_x \gg 1$ and $\hat{D}_x \gg 1$, for
$\bar{\theta}_x = \bar{\theta}_y=0$. The result of operations on
the right hand side of Eq. (\ref{G2D2}) depend on how $\hat{N}_y$
scales with respect to $\hat{z}_o^2$. The cases for $\hat{A}_y >1$
cannot be dealt with fully analytically.  In the following we will
analyze the asymptotic situation $\hat{A}_y \ll 1$ and then we
will treat semi-analytically the generic situation for all values
of $\hat{A}_y$. As we will see, as soon as $\hat{A}_y < 1$ we
start to be in the applicability region of the VCZ theorem.

It should be noted that the dependence in $\Delta \hat{\theta}_x$
and $\Delta \hat{\theta}_y$ in Eq. (\ref{G2D2}) are already
separated.  Therefore, what has been said in Section
\ref{sub:oned} regarding the behavior of coherence properties in
the horizontal direction hold independently of the behavior of
coherence properties in the vertical direction.

\textit{$~~~~~$(A) Far zone case: $\hat{A}_y \ll 1$.$~$---} Since
$\hat{N}_y \gg 1$ it must be $\hat{z}_o^2 \gg \hat{N}_y \gg 1 $ in
order to allow for $\hat{A}_y \ll 1$. Since we are working in
quasi-homogeneous source condition ($\hat{N}_{x,y} \gg 1$) in the
limiting situation $\hat{A}_y \ll 1$ we should recover VCZ
theorem: this is the far field case. Even in the presence of the
extra parameter $\hat{D}_y$ we can treat the generic case
$\hat{A}_y \ll 1$ independently on how $\hat{A}_y$ compares with
respect to $\hat{D}_y$.

Eq. (\ref{G2D2}) can be simplified on the assumptions $\hat{A}_y
\ll 1$ and $\hat{D}_y \ll 1$. In fact, the Gaussian exponential
factor inside the integral in $d \hat{\phi}_y$ in Eq. (\ref{G2D2})
imposes a maximal value $\hat{\phi}_y^2 \sim \hat{A}_y + \hat{D}_y
\ll 1$.

Simultaneously, from the oscillating factor, always inside the
integral in $d \hat{\phi}_y$, we have a condition for the maximal
value of $\Delta \hat{\theta}_y^2 \sim 1/\hat{z}_o^2 \ll 1$: in
fact, if this condition is not fulfilled the integrand will be
highly oscillatory. Alternatively, we can obtain a similar
condition from the Gaussian exponential factor in $\Delta
\hat{\theta}_y$ outside the integral, since $\Delta
\hat{\theta}_y^2 \sim 1/(\hat{A}_y \hat{z}_o^2) \ll 1$.

As a result, the dependence of $S$ on $(\hat{\phi} + \Delta
\hat{\theta}_y)^2$ can be dropped giving

\begin{eqnarray}
\hat{G}(\hat{z}_o,\Delta \hat{\theta}_x,\Delta \hat{\theta}_y) =
\exp{\left[-\frac{2 \hat{A}_x \hat{z}_o^2 \Delta \hat{\theta}_x^2
\hat{D}_x }{(\hat{A}_x+\hat{D}_x)} \right]} \exp{\left[-\frac{2
\hat{A}_y \hat{z}_o^2 \Delta \hat{\theta}_y^2 \hat{D}_y
}{(\hat{A}_y+ \hat{D}_y)}\right]} &&\cr \times
\int_{-\infty}^{\infty} d \hat{\phi}_x
 \left | S\left[{\hat{z}_o,\hat{\phi}_x^2}\right] \right |^2&&\cr \times
\int_{-\infty}^{\infty} d \hat{\phi}_y
\exp{\left[-\frac{\hat{\phi}_y^2+2\hat{\phi}_y\left(2i \hat{A}_y
\hat{z}_o \Delta \hat{\theta}_y \right)}{2
(\hat{A}_y+\hat{D}_y)}\right]} ~.\label{G2D50}
\end{eqnarray}
The integral in $d \hat{\phi}_y$ can be performed giving

\begin{eqnarray}
\hat{G}(\hat{z}_o,\Delta \hat{\theta}_x,\Delta \hat{\theta}_y) =
 \exp{\left[-\frac{2 \hat{A}_x \hat{z}_o^2 \Delta \hat{\theta}_x^2
\hat{D}_x }{\hat{A}_x+\hat{D}_x} \right]} &&\cr \times
\exp{\left[-\frac{2 \hat{A}_y \hat{z}_o^2 \Delta \hat{\theta}_y^2
\hat{D}_y }{\hat{A}_y+ \hat{D}_y}\right]} \exp{\left[-\frac{2
\hat{A}_y^2 \hat{z}_o^2 \Delta \hat{\theta}_y^2}
{\hat{A}_y+\hat{D}_y} \right]}&&\cr \times \int_{-\infty}^{\infty}
d \hat{\phi}_x  \left | S\left[{\hat{z}_o,\hat{\phi}_x^2}\right]
\right |^2 ~.\label{G2D51}
\end{eqnarray}
Normalizing $\hat{G}$ in such a way that $\hat{G}(\hat{z}_o,0,0) =
1$ we obtain the following expression for the spectral degree of
coherence $g$:

\begin{eqnarray}
g(\hat{z}_o,\Delta \hat{\theta}_x,\Delta \hat{\theta}_y) =
\exp{\left[-\frac{2 \hat{A}_x \hat{z}_o^2 \Delta \hat{\theta}_x^2
\hat{D}_x }{\hat{A}_x+\hat{D}_x} \right]} &&\cr \times
\exp{\left[-\frac{2 \hat{A}_y \hat{z}_o^2 \Delta \hat{\theta}_y^2
\hat{D}_y }{\hat{A}_y+ \hat{D}_y}\right]} \exp{\left[-\frac{2
\hat{A}_y^2 \hat{z}_o^2 \Delta \hat{\theta}_y^2}
{\hat{A}_y+\hat{D}_y} \right]} ~.\label{G2D51last}
\end{eqnarray}
Finally, combination of the second and the third exponential
functions yields the result

\begin{eqnarray}
g(\hat{z}_o,\Delta \hat{\theta}_x,\Delta \hat{\theta}_y) =
\exp{\left[-\frac{2 \hat{A}_x \hat{z}_o^2\Delta \hat{\theta}_x^2
\hat{D}_x }{(\hat{A}_x+\hat{D}_x)} \right]} \exp{\left[-2
\hat{A}_y \hat{z}_o^2 \Delta \hat{\theta}_y^2 \right]}
~,\label{G2D51lastreal}
\end{eqnarray}
that is, again, a direct application of the van Cittert-Zernike
theorem. For $\Delta \hat{\theta}_x = 0$ we have

\begin{equation}
\xi_{c y} = \left(\frac{\pi}{\hat{A}_y}\right)^{1/2}~.
\label{limitebaseta}
\end{equation}

\textit{$~~~~~$(B) Case $\hat{A}_y \gg 1$.$~$---} This case
encompasses situations with $\hat{z}_o \sim 1$ as well as
situations with $\hat{z}_o \gg 1$.

Let us first consider $\hat{z}_o \gg 1$. Looking at the integral
in $\hat{\phi}_y$ in Eq. (\ref{G2D2}) it is easy to recognize that
its integrand is highly oscillatory in $\hat{\phi}_y$ when $2
\hat{\phi}_y \hat{z}_o \Delta \hat{\theta}_y \gg 1$, since
$\hat{A}_y/(\hat{A}_y+\hat{D}_y) < 1$. Therefore, the integrand
will contribute to the integral significatively only up to values
$2 \hat{\phi}_y \hat{z}_o \Delta \hat{\theta}_y \lesssim 1$. On
the other hand, the terms in $S$ give non negligible contributions
only for values of $\hat{\phi}_y$ up to order unity. As a result,
it must be $2 \hat{z}_o \Delta \hat{\theta}_y \lesssim 1$. As
$\hat{z}_o \sim 1$ the width of $g$ in $\Delta \hat{\theta}_y$ is
then of order unity. When $\hat{z}_o$ becomes larger than unity,
the angular width $\Delta \hat{\theta}_y$ will decrease and
asymptotically, as $\hat{z}_o \gg 1$, one will have a rapidly
oscillating integrand for $\Delta \hat{\theta}_y \sim 1/\hat{z}_o
\ll 1$. However note that $\Delta \hat{\theta}_y$ must be
multiplied by $\hat{z}_o$ in order to obtain the correlation
length which means that this remains constant and comparable with
the diffraction length $\sqrt{c L_w/\omega}$ as $\hat{z}_o$
increases. As a result, under the assumption $\hat{z}_o \gg 1$,
terms in $\Delta \hat{\theta}_y$ can be dropped in the functions
$S(\cdot)$ of Eq. (\ref{G2D2}), and the functions $S$ can be
substituted with their limiting form $\mathrm{sinc}$. Moreover, in
the limit for $\hat{A}_y \gg 1$ and $\hat{D}_y \ll 1$ we have

\begin{eqnarray}
\hat{G}(\hat{z}_o,\Delta \hat{\theta}_x,\Delta \hat{\theta}_y) =
\exp{\left[-\frac{2 \hat{A}_x \hat{D}_x \hat{z}_o^2 \Delta
\hat{\theta}_x^2 }{\hat{A}_x+\hat{D}_x} \right]} \exp{\left[-{2
\hat{D}_y  \hat{z}_o^2 \Delta \hat{\theta}_y^2 } \right]}&&\cr
\times \int_{-\infty}^{\infty}d \hat{\phi}_y \exp{\left[-i\left(2
 \hat{z}_o \Delta \hat{\theta}_y\right) \hat{\phi}_y\right]}
\int_{-\infty}^{\infty} d \hat{\phi}_x ~\mathrm{sinc}^2
\left[\left({\hat{\phi}_x^2+\hat{\phi}_y^2}\right)/4\right]
~,\label{G2D3bis1}
\end{eqnarray}
where the simplification in the phase under the integral in
$d\hat{\phi}_y$ is possible for $\hat{z}_o \Delta \hat{\theta}_y
\hat{\phi}_y  \hat{D}_y/\hat{A}_y \ll 1$ and the exponential
function $\exp{[-{\hat{\phi}_y^2 }/2 (\hat{A}_y+\hat{D}_y)]}$
under the integral in $d\hat{\phi}_y$ can be neglected because the
$\mathrm{sinc} (\cdot)$ function has characteristic length in
$\hat{\phi}_y$ of order unity. Eq. (\ref{G2D3bis1}) is therefore
valid in the limit for $\hat{A}_y \gg 1$, $\hat{D}_y \ll 1$ and
$\hat{z}_o \gg 1$. The integral in $d \hat{\phi}_y$ in Eq.
(\ref{G2D3bis1}) is simply the Fourier transform of the universal
function

\begin{equation}
{I_S}(\hat{\phi}_y) = \int_{-\infty}^{\infty} d \hat{\phi}_x ~
\mathrm{sinc}^2
\left[\left({\hat{\phi}_x^2+\hat{\phi}_y^2}\right)/4\right]
~.\label{bingo2}
\end{equation}
done with respect to the variable $2 \hat{z}_o \Delta
\hat{\theta}_y$, conjugate to $\hat{\phi}_y$, that is

\begin{eqnarray}
\hat{G}(\hat{z}_o,\Delta \hat{\theta}_x,\Delta \hat{\theta}_y)
=\exp{\left[-\frac{2 \hat{A}_x \hat{D}_x \hat{z}_o^2\Delta
\hat{\theta}_x^2 }{\hat{A}_x+\hat{D}_x} \right]} \exp{\left[-{2
\hat{D}_y  \hat{z}_o^2 \Delta \hat{\theta}_y^2 } \right]}&&\cr
\times \int_{-\infty}^{\infty} d \hat{\phi}_y
 \exp{\left[-i\left(2 \hat{z}_o \Delta \hat{\theta}_y
\right)\hat{\phi}_y\right]}
{I_S}(\hat{\phi}_y)~.\label{G2D3lastlast0}
\end{eqnarray}
It is not difficult to see that the width of this Fourier
transform in $\Delta \hat{\theta}_y \hat{z}_o$ is much smaller
than the characteristic width imposed by the Gaussian exponentials
outside the integration sign, their ratio being of order
$\hat{D}_y \ll 1$. This means that the Gaussian in $\Delta
\hat{\theta}_y^2$ outside the integral sign is almost constant
with respect to the behavior of the Fourier transform, and can be
neglected, to obtain

\begin{eqnarray}
\hat{G}(\hat{z}_o,\Delta \hat{\theta}_x,\Delta \hat{\theta}_y)
=\exp{\left[-\frac{2 \hat{A}_x \hat{D}_x \hat{z}_o^2\Delta
\hat{\theta}_x^2 }{\hat{A}_x+\hat{D}_x} \right]}
\int_{-\infty}^{\infty} d \hat{\phi}_y
 \exp{\left[-i\left(2 \hat{z}_o \Delta \hat{\theta}_y
\right)\hat{\phi}_y\right]}
{I_S}(\hat{\phi}_y)~.&&\cr\label{G2D3lastlast}
\end{eqnarray}
The assumption $\hat{z}_o \gg 1$ was vital for the derivation of
Eq. (\ref{G2D3lastlast}). However, for all values of $\hat{z}_o$
such that $\hat{A}_y \gg 1$ (and, therefore, up to the exit of the
undulator at $\hat{z}_o = 1/2$), it is easy to see from Eq.
(\ref{G2D}) that the source is quasi-homogeneous, so that the
"anti" VCZ theorem applies and the cross-spectral density of the
field at the source plane forms a Fourier couple with the angular
distribution of the radiant intensity, which is simply given by
$I_S$ as defined in Eq. (\ref{bingo2}).

\begin{figure}
\begin{center}
\includegraphics*[width=140mm]{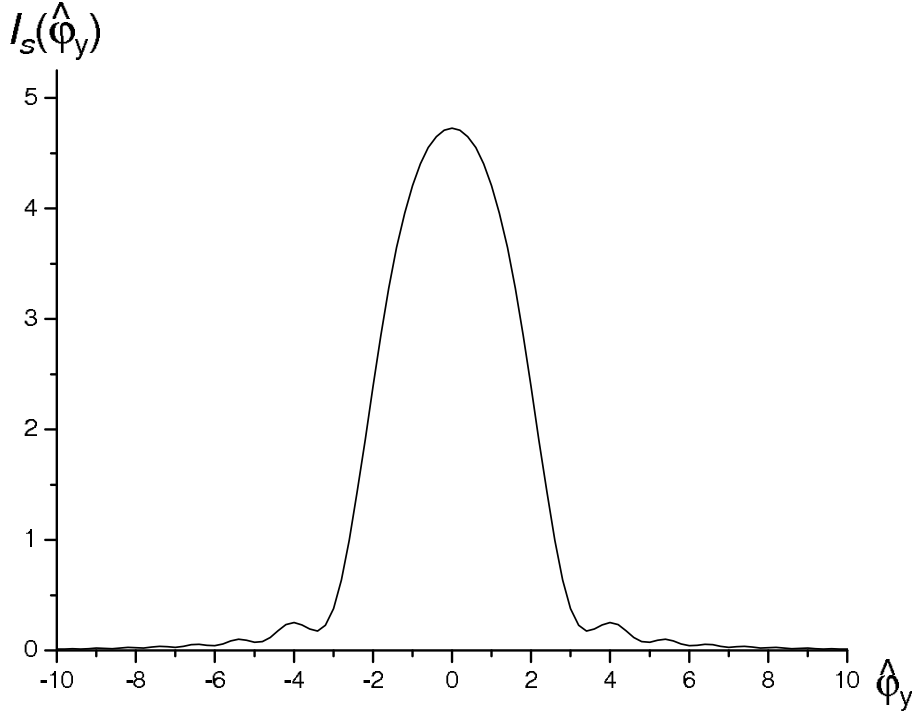}% Here is how to import EPS art
\caption{\label{ISnorm} The normalized radiant intensity $I_S$,
calculated from Eq. (\ref{bingo2}),  as a function of the
normalized vertical angle $\hat{\phi}_y$.}
\end{center}
\end{figure}
It is interesting to justify the fact that $I_S$ is in fact the
angular distribution of the radiant intensity. To this purpose, it
is sufficient to note that $I_S$ is simply, normalization factors
aside, Eq. (176) of \cite{OURS}, which represents the intensity
from a beam with $\hat{\epsilon}_x \rightarrow \infty$ and
$\hat{\epsilon}_y \rightarrow 0$. A representation of
${I_S}(\hat{\phi}_y)$ is given in Fig. \ref{ISnorm}.

As a result, one may therefore conclude that Eq.
(\ref{G2D3lastlast}) is valid in general, for any value of
$\hat{z}_o$ such that $\hat{A}_y \gg 1$, in the asymptotic limit
$\hat{D}_y \ll 1$.

\begin{figure}
\begin{center}
\includegraphics*[width=140mm]{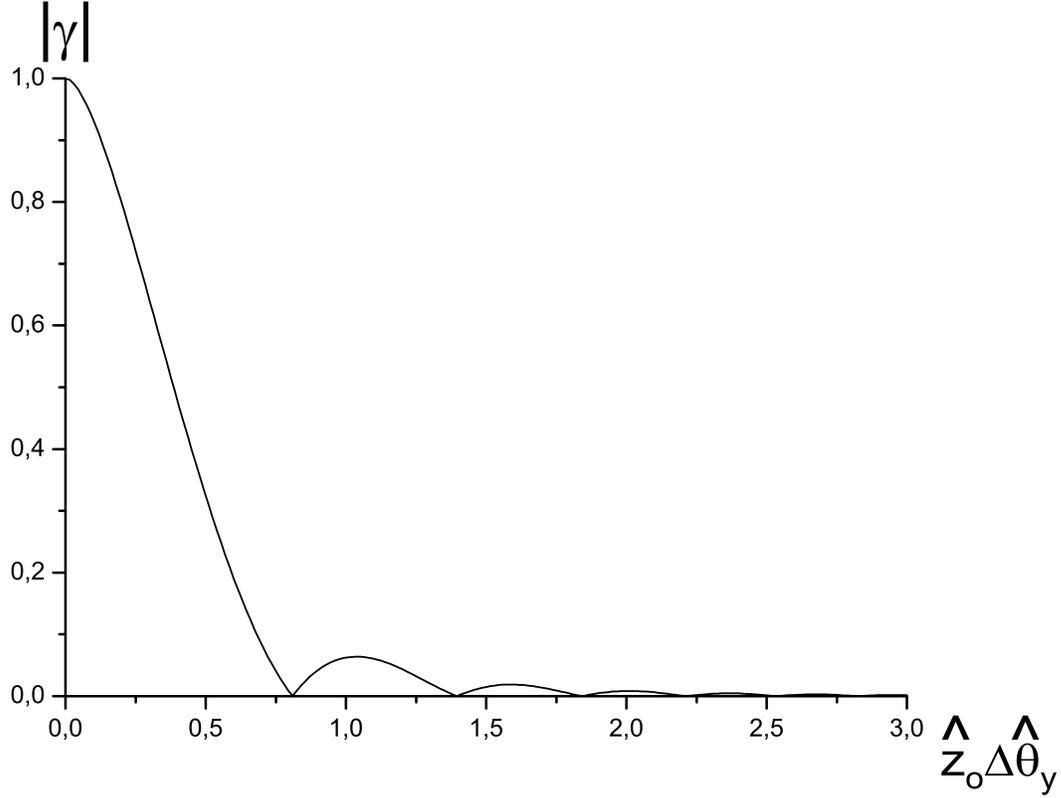}% Here is how to import EPS art
\caption{\label{GAM} Behavior of $|\gamma|$ as a function of
$\hat{z}_o \Delta \hat{\theta}_y$. This universal plot illustrates
the absolute value of the spectral degree of coherence $|g|$
calculated for $\Delta \hat{\theta}_x=0$ in the limit for
$\hat{A}_y \gg 1$ and $\hat{D}_y \ll 1$ .}
\end{center}
\end{figure}

It is important to note that Eq. (\ref{G2D3lastlast}) can be
written as

\begin{eqnarray}
g(\hat{z}_o,\Delta \hat{\theta}_x,\Delta \hat{\theta}_y) =
\exp{\left[-\frac{2 \hat{A}_x \hat{D}_x \hat{z}_o^2\Delta
\hat{\theta}_x^2 }{(\hat{A}_x+\hat{D}_x)} \right]}
\gamma(\hat{z}_o\Delta \hat{\theta}_y) ~,\label{G2D3lastlastno}
\end{eqnarray}
where $\gamma(\hat{z}_o\Delta \hat{\theta}_y)$, given by

\begin{eqnarray}
\gamma(\hat{z}_o\Delta \hat{\theta}_y) = \frac{1}{2 \pi^2}
\int_{-\infty}^{\infty} d \hat{\phi}_y
 \exp{\left[-i\left(2 \hat{z}_o \Delta \hat{\theta}_y
\right)\hat{\phi}_y\right]}
{I_S}(\hat{\phi}_y)~,\label{G2D3lastlastno2}
\end{eqnarray}
is a universal function normalized to unity. It is possible to
calculate Eq. (\ref{G2D3lastlastno2}) analytically. To this
purpose, it is sufficient to note that the Fourier Transform

\begin{eqnarray}
{\gamma_1}(\xi,\eta) = \int_{-\infty}^{\infty} d \hat{\phi}_x
\int_{-\infty}^{\infty} d \hat{\phi}_y
 \exp{\left[i(\xi \hat{\phi}_x+\eta \hat{\phi}_y)\right]}
\mathrm{sinc}^2\left(\frac{\hat{\phi}_x^2+\hat{\phi}_y^2}{4}\right)
~\label{G2D3lastlastno3}
\end{eqnarray}
can be evaluated with the help of the Bessel-Fourier formula as

\begin{eqnarray}
{\gamma_1}(\lambda) = 2\pi \int_{0}^{\infty} d {\phi} ~{\phi}
J_0\left({\phi} \lambda\right)
\mathrm{sinc}^2\left(\frac{{\phi}^2}{4}\right)~\label{G2D3lastlastno4}
&& \cr = 2\pi \left[\pi+ \lambda^2
\mathrm{Ci}\left(\frac{\lambda^2}{2}\right)- 2
\sin\left(\frac{\lambda^2}{2}\right)- 2
\mathrm{Si}\left(\frac{\lambda^2}{2}\right) \right]\end{eqnarray}
where $\lambda^2 = \xi^2+\eta^2$, $\phi^2 = \phi_x^2+\phi_y^2$,
$\mathrm{Si}(\cdot)$ is the sine integral function and
$\mathrm{Ci}(\cdot)$ is the cosine integral function. As a result
one has

\begin{eqnarray}
\gamma(\hat{z}_o\Delta \hat{\theta}_y) =\frac{2}{\pi}
\left[\frac{\pi}{2}+ 2\hat{z}_o^2\Delta \hat{\theta}_y^2
\mathrm{Ci}\left(2\hat{z}_o^2\Delta \hat{\theta}_y^2\right)-
\sin\left(2\hat{z}_o^2\Delta \hat{\theta}_y^2\right)-
\mathrm{Si}\left(2\hat{z}_o^2\Delta \hat{\theta}_y^2\right)
\right]~.&&\cr\label{G2D3lastlastno5}
\end{eqnarray}
The function $\gamma(\hat{z}_o\Delta \hat{\theta}_y)$ is
illustrated in Fig. \ref{GAM}.

It should be noted that $\gamma(\hat{z}_o\Delta \hat{\theta}_y)$
is the spectral degree of coherence $g$ calculated for $\Delta
\hat{\theta}_x=0$ in the limit for $\hat{A}_y \gg 1$ and
$\hat{D}_y \ll 1$, in agreement with Eq. (\ref{G2D3lastlastno}).

%
%Its generalization for $\hat{A}_y \gg 1$ and finite $\hat{D}_y$
%can be found comparing with Eq. (\ref{G2D3bis}) in the limit for
%$\hat{z}_o \gg 1$ and for arbitrary values of $\hat{D}_y$.by
%
%\begin{eqnarray}
%{g}(\hat{z}_o,\Delta \hat{\theta}_x,\Delta \hat{\theta}_y) =
%\exp{\left[-\frac{2 \hat{A}_x \hat{D}_x \hat{z}_o^2\Delta
%\hat{\theta}_x^2 }{(\hat{A}_x+\hat{D}_x)} \right]}
%\gamma(\hat{z}_o\Delta \hat{\theta}_y)
%~,\label{G2D3lastlastnonono}
%\end{eqnarray}
%%

\begin{figure}
\begin{center}
\includegraphics*[width=140mm]{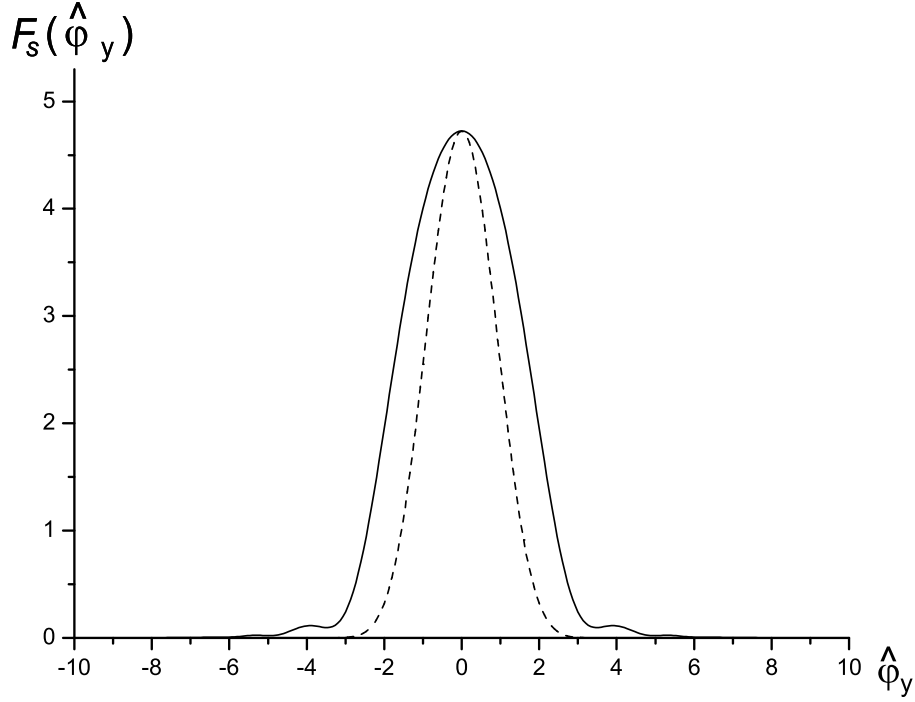}% Here is how to import EPS art
\caption{\label{IS05} Values of $\exp[-\hat{\phi}_y/(2 \hat{A}_y)]
I_S(\hat{\phi}_y)$ as a function of $\hat{\phi}_y$. Solid line:
$\hat{A}_y = 10$, the plot is still similar to Fig. \ref{ISnorm}.
Dashed line: $\hat{A}_y = 1.0$.}
\end{center}
\end{figure}
\begin{figure}
\begin{center}
\includegraphics*[width=140mm]{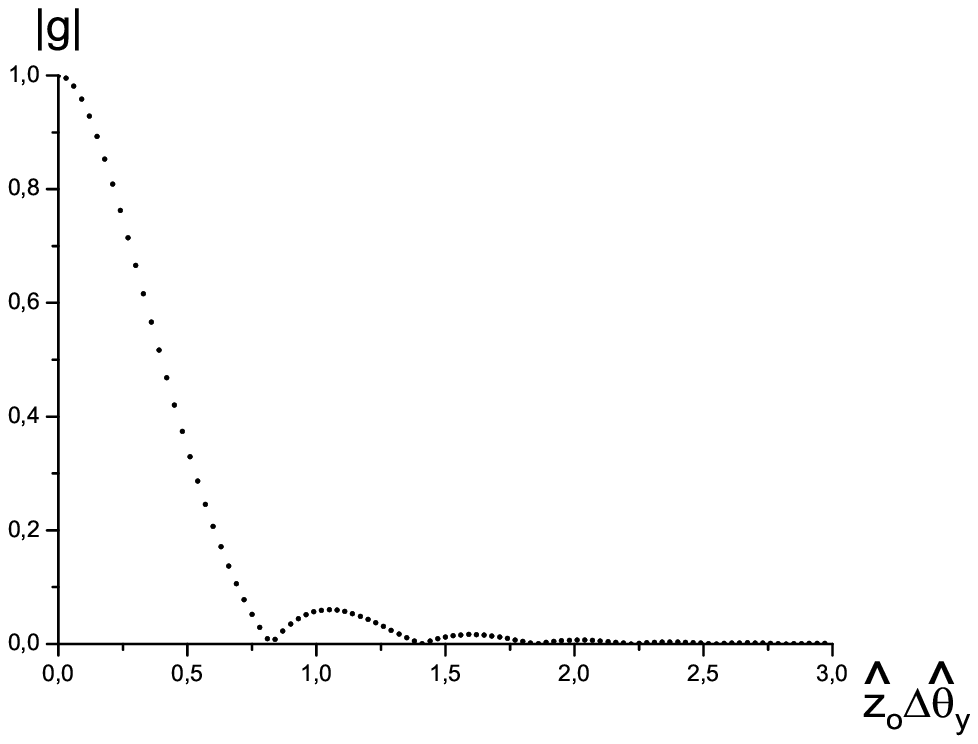}% Here is how to import EPS art
\caption{\label{G05} The behavior of the absolute value of the
spectral degree of coherence $g$ (at perfect resonance) as a
function of $\hat{z}_o \Delta \hat{\theta}_y$. Here $\hat{z}_o =
1/2$, $\hat{N}_y = 10$ and $\hat{D}_y$ is negligible. The black
circles represent actual numerical data.}
\end{center}
\end{figure}
%
%\begin{figure}
%\begin{center}
%\includegraphics*[width=140mm]{G3.EPS}% Here is how to import EPS art
%\caption{\label{G3} The behavior of the absolute value of the
%spectral degree of coherence $g$ (at perfect resonance) as a
%function of $\hat{z}_o \Delta \hat{\theta}_y$. Here $\hat{z}_o =
%3$, $\hat{N}_y = 10$ and $\hat{D}_y$ is negligible. The black
%circles represent actual numerical data.}
%\end{center}
%\end{figure}
%%
\textit{$~~~~~$(C) Case $\hat{A}_y \sim 1.~$---} In case
\textit{(B)} we have shown that terms in $\Delta \hat{\theta}_y$
can be dropped in the functions $S(\cdot)$ of Eq. (\ref{G2D2}),
for every value of $\hat{z}_o$, and the functions $S$ can be
substituted with their limiting form $\mathrm{sinc}$. In the case
of $\hat{A}_y \sim 1$ and $\hat{D}_y \ll 1$ we have

\begin{eqnarray}
\hat{G}(\hat{z}_o,\Delta \hat{\theta}_x,\Delta \hat{\theta}_y) =
\exp{\left[-\frac{2 \hat{A}_x \hat{D}_x \hat{z}_o^2 \Delta
\hat{\theta}_x^2 }{\hat{A}_x+\hat{D}_x} \right]} \exp{\left[-{2
 \hat{D}_y \hat{z}_o^2 \Delta \hat{\theta}_y^2
} \right]}&&\cr \times \int_{-\infty}^{\infty}d \hat{\phi}_y
\exp{\left[-i{2 \hat{z}_o \Delta \hat{\theta}_y}
\hat{\phi}_y\right]} \exp{\left[-\frac{\hat{\phi}_y^2 }{2
\hat{A}_y}\right]} && \cr \times \int_{-\infty}^{\infty} d
\hat{\phi}_x  ~\mathrm{sinc}^2
\left[\left({\hat{\phi}_x^2+\hat{\phi}_y^2}\right)/4\right]
~.\label{G2D3bis}
\end{eqnarray}
The integral in $d \hat{\phi}_y$ in Eq. (\ref{G2D3bis}) is  the
Fourier transform of

\begin{equation}
{F_S}(\hat{\phi}_y) = I_S(\hat{\phi}_y)
\exp{\left[-\frac{\hat{\phi}_y^2 }{2 \hat{A}_y}\right]}
\label{bingo}
\end{equation}
done with respect to the variable $2 \hat{z}_o \Delta
\hat{\theta}_y$, conjugate to $\hat{\phi}_y$. Similarly as before,
it is not difficult to see that the width of this Fourier
transform in $\Delta \hat{\theta}_y \hat{z}_o$ is much smaller
than the characteristic width imposed by the Gaussian exponentials
outside the integration sign, their ratio being of order
$\hat{D}_y \ll 1$. This means that the Gaussian in $\Delta
\hat{\theta}_y^2$ outside the integral sign is almost constant
with respect to the behavior of the Fourier transform, and can be
neglected. Using the definition of $F_S$ in Eq. (\ref{bingo}) we
have

\begin{eqnarray}
\hat{G}(\hat{z}_o,\Delta \hat{\theta}_x,\Delta \hat{\theta}_y)
=\exp{\left[-\frac{2 \hat{A}_x \hat{D}_x \hat{z}_o^2\Delta
\hat{\theta}_x^2 }{\hat{A}_x+\hat{D}_x} \right]}
\int_{-\infty}^{\infty} d \hat{\phi}_y \exp{\left[-i{2 \hat{z}_o
\Delta \hat{\theta}_y}\hat{\phi}_y\right]}
{F_S}(\hat{\phi}_y)~.&&\cr\label{G2D3last}
\end{eqnarray}
At this point, no simplification is possible and numerical
analysis of the problem should be undertaken in order to calculate
$\hat{G}$, followed by normalization according to

\begin{eqnarray}
\hat{G}(\hat{z}_o,0,0) =\int_{-\infty}^{\infty} d \hat{\phi}_y
{F_S}(\hat{\phi}_y)~,\label{G2D3lastnormfac}
\end{eqnarray}
as already said, in order to find an expression for the complex
degree of coherence $g$. This discussion underlines again how the
correlation angle $\Delta \hat{\theta}_y$ in the $y$ direction
behaves. It starts from a constant value equal to the diffraction
angle $\sqrt{c/(\omega L_w)}$ when $\hat{z}_o \sim 1$, which
corresponds to the maximal possible value of $\hat{A}_y$ once
$\hat{N}_y$ is fixed. Then it decreases as $\hat{z}_o$ grows.
Asymptotically in limit for $\hat{z}_o \gg 1$ (but still such that
$\hat{z}_o^2 \lesssim \hat{N}_y$), it behaves as $\sim
1/\hat{z}_o$, as it is clear from the fact that the function
${F_S}$, which is Fourier Transformed in Eq. (\ref{G2D3last}),
does not depend on any parameter. However, it should be noted that
$\Delta \hat{\theta}_y$ must be multiplied by $\hat{z}_o$ in order
to obtain the correlation length, which means that this remains
constant and comparable with the diffraction length $\sqrt{c
L_w/\omega}$ as $\hat{z}_o$ increases. Correlation length in the
$x$ direction is governed instead by the Gaussian exponential
function in ${\Delta} \hat{\theta}_x$ outside the integral sign in
Eq. (\ref{G2D3last}), exactly as in the simplified model treated
in Section \ref{sub:oned}.

We performed some numerical calculation with the aim of giving the
reader an exemplification. We set $\Delta \hat{\theta}_x=0$ and
$\hat{N}_y = 10$. Assuming that we are in the asymptotic limit
$\hat{D} \ll 1$ we can rely on what has been said in this
paragraph to  calculate $g$ and on Eq. (\ref{cohlen}) to calculate
the correlation length $\xi_{c y}$ in the vertical direction for
any value of $\hat{z}_o$. These numerical results must agree with
the Van Cittert-Zernike limit for $\hat{A}_y \ll 1$ treated in
paragraph \ref{paragr:1} \textit{(A)}: in fact, Eq.
(\ref{G2D51lastreal}) and Eq. (\ref{cohlen}) yield immediately a
linear dependence of $\xi_{c y}$ on $\hat{z}_o$ as $\hat{A}_y \ll
1$. Then, using paragraph \ref{paragr:1} \textit{(B)} we can
extend the function $\xi_{c y}(\hat{z}_o)$ to all values of
$\hat{z}_o$. In this way we obtain $\xi_{c y}(\hat{z}_o)$ for
every value of $\hat{z}_o$. In Fig. \ref{IS05} we plot Eq.
(\ref{bingo}) for two particular values of  $\hat{A}_y$:
$\hat{A}_y = 10$ (solid line) and $\hat{A}_y = 1$ (dashed line).
Note that the solid line Fig. \ref{IS05} is still similar to Fig.
\ref{ISnorm}, although the tails are changing and the width is
already smaller.

\begin{figure}
\begin{center}
\includegraphics*[width=140mm]{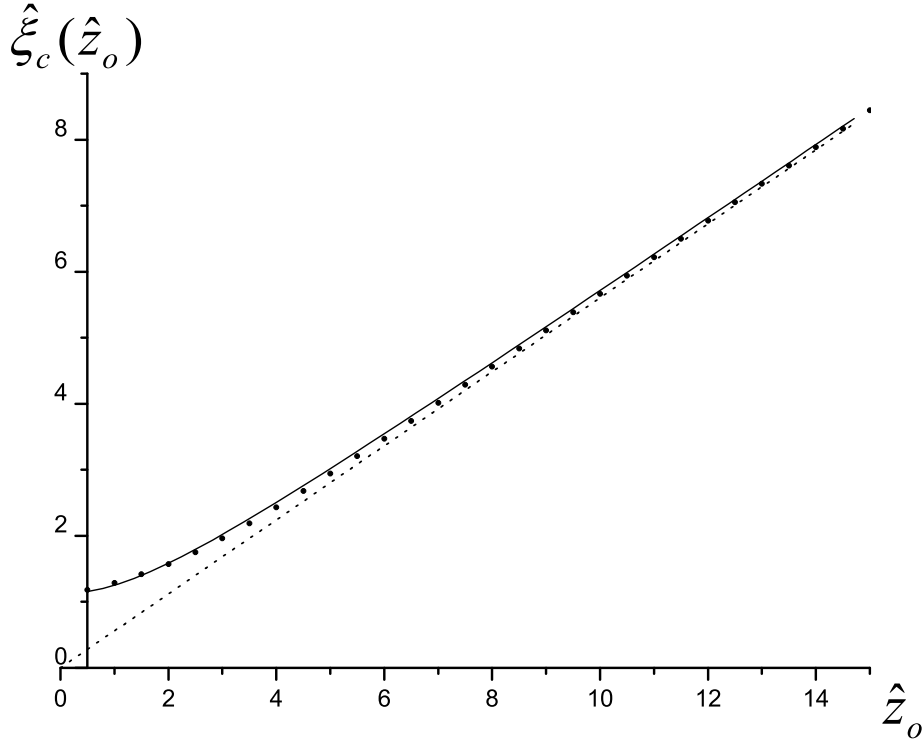}% Here is how to import EPS art
\caption{\label{CORLEN} Normalized coherence length $\hat{\xi}_{c
y}$ as a function of the normalized distance $\hat{z}_o$ when the
electron beam divergence is negligible. Here $\hat{N}_y = 10$. The
black circles represent actual numerical calculations. The
asymptotic limit for $\hat{A}_y \ll 1$ (VCZ theorem) is shown with
a dotted line. Finally, the solid line is calculated with the
approximated formula (\ref{FITXI2}). }
\end{center}
\end{figure}
The following step is to calculate the Fourier Transform of $F_S$
according to Eq. (\ref{G2D3last}) after normalization procedure
according to Eq. (\ref{G2D3lastnormfac}). This gives $g$
calculated for $\Delta \hat{\theta}_x=0$. Fig. \ref{G05}
illustrates $\mid g \mid$ for $\hat{z}_o = 1/2$ as a function of
$\hat{z}_o \Delta \hat{\theta}_y$ at $\Delta \hat{\theta}_x=0$.
Note that for this parameter choice we are still in the limit for
$\hat{A}_y \gg 1$, which explains why Fig. \ref{G05} is
practically identical to the universal plot Fig. \ref{GAM}.

One can calculate the coherence length $\hat{\xi}_{c y}
(\hat{z}_o)$ straightforwardly by means of Eq. (\ref{cohlen}). The
curve obtained can be then compared, in the limit for $\hat{A}_y
\ll 1$, with the van Cittert-Zernike behavior illustrated in
Paragraph \ref{paragr:2} \textit{(A)}. Fig. \ref{CORLEN} shows our
results. The black circles represent actual numerical
calculations. The asymptotic limit for $\hat{A}_y \ll 1$ (VCZ
theorem) is shown with a dotted line. When $\hat{N}_y \gg 1$ and
$\hat{D}_y \ll 1$, the coherence length $\hat{\xi}_{c y}
(\hat{z}_o)$ can be calculated with the approximated formula

\begin{equation}
\hat{\xi}_{c y}  \simeq \left[ a^2 + \frac{\pi}{\hat{A}_y}
\right]^{1/2}~. \label{FITXI2}
\end{equation}
Under the approximation of negligibly small electron beam
divergence in the vertical direction, the normalized coherence
length is thus a universal function of one dimensionless
parameter, $\hat{A}_y$. On the one hand, Eq. (\ref{FITXI2})
accounts for the asymptotic behavior as $\hat{A}_y \ll 1$ (VCZ
theorem). On the other hand, the value of $\hat{\xi}_{c y}$ in Eq.
(\ref{FITXI2}) approaches the constant value $\hat{\xi}_{c y}
\longrightarrow a$ for asymptotically large values of $\hat{A}_y$.
Beside accurately reproducing the asymptotes for small and large
values of $\hat{A}_y$, Eq. (\ref{FITXI2}) provides an accuracy of
several per cent with respect to the result of numerical
calculations (when $\hat{A}_y$ is within the limits $(0,\infty)$),
in the whole range of the parameter $\hat{A}_y$. The solid line in
Fig. \ref{CORLEN} is calculated with the  approximated formula
(\ref{FITXI2}), where calculated $a$ numerically using Eq.
(\ref{G2D3lastlast}) to calculate $\hat{\xi}_{c y}$ and obtaining
$a \simeq 1.12$. Therefore, according to Eq. (\ref{FITXI2}), when
$\hat{N}_y \gg 1$ and $\hat{D}_y \ll 1$, we have $\hat{\xi}_{c y}
= const = 1.12 $ at the undulator exit (with accuracy $\hat{D}_y
\ll 1$ and $1/\hat{N}_y \ll 1$) and, in dimensional units, $\xi_{c
y} = 1.12 \sqrt{L_w c/\omega}$. In this case the coherence length
is a function of undulator length and wavelength due to the
intrinsic divergence of the undulator radiation and $a = 1.12$ is
a universal constant.

\subsubsection{\label{paragr:3} Case with finite divergence
parameter $\hat{D}_y$.}

We will first discuss, in Paragraph \ref{paragr:3} \textit{(A)},
the limit for $\hat{A}_y \ll 1$: in this case, whatever the value
of $\hat{D}_y$, we will recover the VCZ theorem. Note that,
although we are discussing the limit $\hat{N}_y \gg 1$ there
always be values of $\hat{z}_o$ large enough so that $\hat{A}_y
\ll 1$. Further on, in Paragraph \ref{paragr:3} \textit{(B)}, we
will discuss the case $\hat{A}_y \gg 1$: note that, since we are
discussing the limit for $\hat{N}_y \gg 1$, this will always be
the case near the exit of the undulator at $\hat{z}_o = 1/2$. We
will be particularly interested to this situation; in fact, the
study of the case $\hat{A}_y \gg 1$ near the exit of the undulator
will allow us to give an explicit representation of Eq.
(\ref{xicdarb}).

\textit{$~~~~~$(A) Far zone case $\hat{A}_y \ll 1$.$~$---} Eq.
(\ref{G2D2}) is still valid and can be calculated numerically, in
principle, for any value of $\hat{z}_o$. In the case $\hat{A}_y
\ll 1$ it can be further simplified. In fact, in this situation,
the second exponential function of the right hand side of Eq.
(\ref{G2D2}) limits the possible values of $\Delta \hat{\theta}_y$
to $\Delta \hat{\theta}_y \ll 1$. Moreover $\hat{z}_o \gg 1$ so
that Eq. (\ref{G2D3bis}) is valid in this case. In particular,
remembering the definition of $F_S(\hat{\phi}_y)$ in Eq.
(\ref{bingo}) we have

\begin{eqnarray}
\hat{G}(\hat{z}_o,\Delta \hat{\theta}_x,\Delta \hat{\theta}_y) =
\exp{\left[-\frac{2 \hat{A}_x \hat{z}_o^2 \Delta \hat{\theta}_x^2
\hat{D}_x }{(\hat{A}_x+\hat{D}_x)} \right]} \exp{\left[-\frac{2
\hat{A}_y \hat{z}_o^2 \Delta \hat{\theta}_y^2 \hat{D}_y
}{(\hat{A}_y+\hat{D}_y)} \right]}&&\cr \times
\int_{-\infty}^{\infty} d \hat{\phi}_y \exp{\left[-i\frac{2
\hat{A}_y \hat{z}_o \Delta
\hat{\theta}_y}{\hat{A}_y+\hat{D}_y}\hat{\phi}_y\right]}
F_S(\hat{\phi}_y)~.~~~~~~\label{G2D2easy2}
\end{eqnarray}
It is not difficult to see that the ratio between the
characteristic width in $\Delta \hat{\theta}_y$ of the exponential
function outside the integral sign and the exponential function in
inside the integral sign in Eq. (\ref{G2D2easy2}) is of order
$\sqrt{\hat{A}_y/\hat{D}_y}$ and it is always much smaller than
unity unless $\hat{D}_y \ll 1$: such a case has already been
treated before in Paragraph \ref{paragr:1} \textit{(A)}, and has
been shown to obey the VCZ theorem \footnote{Alternatively one may
note directly that $\hat{\phi}_y$ can only range over values much
smaller than unity. As a result, the dependence of $I_S$ on
$\hat{\phi}_y$ can be dropped, giving an extra normalization
constant to be disposed of. Then, the integral in $d\hat{\phi}$
performed giving, as in Paragraph \ref{paragr:1} \textit{(A)},
$\exp{[-2 \hat{A}_y^2 \hat{z}_o^2 \Delta
\hat{\theta}_y^2/(\hat{A}_y+\hat{D}_y)]}$ to be combined with the
exponential function in $\Delta \hat{\theta}_y$ outside the
integral sign, giving $\exp{[-2 \hat{A}_y \hat{z}_o^2 \Delta
\hat{\theta}_y^2 ]}$.}. For all other values of $\hat{D}_y$ we
have, automatically, $\hat{D}_y \gg \hat{A}_y$, so that we can
neglect the integral in $d \hat{\phi}_y$ and we get back once more
the Van Cittert-Zernike regime. To sum up, we obtain the following
expression for $g$, which is valid for $\hat{A}_y \ll 1$ with no
restrictions on $\hat{D}_y$:

\begin{eqnarray}
g(\hat{z}_o,\Delta \hat{\theta}_x,\Delta \hat{\theta}_y) =
\exp{\left[-\frac{2 \hat{A}_x \hat{z}_o^2 \Delta \hat{\theta}_x^2
\hat{D}_x }{(\hat{A}_x+\hat{D}_x)} \right]} \exp{\left[-{2
\hat{A}_y \hat{z}_o^2 \Delta \hat{\theta}_y^2} \right]}
~.~~~~~~\label{G2D2easy3}
\end{eqnarray}
Calculation of the coherence length from Eq. (\ref{G2D2easy3}) at
$\Delta \hat{\theta}_x=0$ gives once more the behavior

\begin{equation}
\xi_{c y} = \left(\frac{\pi}{\hat{A}_y}\right)^{1/2}~,
\label{xioncemore}
\end{equation}
which is consistent with the partial result in Paragraph
\ref{paragr:1} (A).

\textit{$~~~~~$(B) Near zone, $\hat{A}_y \gg 1$.$~$---} Equation
(\ref{G2D3bis}) is still valid in this case and following the same
line of reasoning as paragraph \ref{paragr:2} (B) one gets

\begin{eqnarray}
g(\hat{z}_o,\Delta \hat{\theta}_x,\Delta \hat{\theta}_y)
=\exp{\left[-\frac{2 \hat{A}_x \hat{D}_x \hat{z}_o^2\Delta
\hat{\theta}_x^2 }{(\hat{A}_x+\hat{D}_x)} \right]}
\exp{\left[-\frac{2 \hat{A}_y \hat{D}_y \hat{z}_o^2\Delta
\hat{\theta}_y^2 }{(\hat{A}_y+\hat{D}_y)} \right]}
\gamma(\hat{z}_o\Delta \hat{\theta}_y)~,&&\cr\label{Gahia}
\end{eqnarray}
valid for $\hat{A}_y \gg 1$ and arbitrary $\hat{D}_y$. Note that
since we are working in the limit for $\hat{N}_y \gg 1$, for
$\hat{z}_o = 1/2$ we have $\hat{A}_y \gg 1$. We can see from Eq.
(\ref{Gahia}) that, for finite values of $\hat{D}_y$, the
cross-spectral density $g$, evaluated at the exit of the undulator
for $\Delta \hat{\theta}_x = 0$, is given by the product of the
exponential function $\exp{[-2 \hat{D}_y \hat{z}_o^2\Delta
\hat{\theta}_y^2 ]}$ with the function illustrated in Fig.
\ref{G05}. This remark is intuitively sound. Since $\hat{N}_y \gg
1$ in fact we have weakly quasi-homogeneous wavefronts near the
exit of the undulator, and we can use the "anti" VCZ theorem to
conclude that $g$ must for a Fourier couple with the intensity
distribution in the far zone. This will simply be, for any
arbitrary $\hat{D}_y$, a convolution between a Gaussian
distribution with rms width equal to $\sqrt{\hat{D}_y}$ and $I_S$,
which is the angular distribution of radiant intensity for a beam
with $\hat{\epsilon}_x \longrightarrow \infty$ and
$\hat{\epsilon}_y \longrightarrow 0$. Finally, the Fourier
transform of such a convolution between two function is simply
given by the product of the Fourier transforms of the two
functions.

\textit{$~~~~~$(C) Approximate formula.$~$---} With in mind Eq.
(\ref{cohlen22D}), Eq. (\ref{FITXI2}) and Eq. (\ref{xioncemore})
we make the working hypothesis that Eq. (\ref{xicdarb}) has the
form

\begin{equation}
\hat{\xi}_{c y} =
\left(\frac{\pi}{\hat{D}_{\mathrm{eff}}(\hat{D}_y)}+\frac{\pi}{\hat{A}_y}\right)^{1/2}~.
\label{xicdarbvera}
\end{equation}
Within the assumption $\hat{N}_y \gg 1$, we have seen that if
$\hat{D}_y \gg 1$ Eq. (\ref{cohlen22D}) is valid with relative
accuracy $1/\hat{D}_y$. This means that, in this limit,
$\hat{D}_{\mathrm{eff}}(\hat{D}_y) = \hat{D}_y$. Then we have seen
that if $\hat{D}_y \ll 1$ Eq. (\ref{FITXI2}) holds with accuracy
$1/\hat{N}_y$. This means that $\hat{D}_{\mathrm{eff}}(0) = 2.50$.
Moreover we have seen that in the far zone case, in the limit for
$\hat{A}_y \ll 1$, the VCZ theorem holds, in agreement with Eq.
(\ref{xioncemore}).

\begin{figure}
\begin{center}
\includegraphics*[width=140mm]{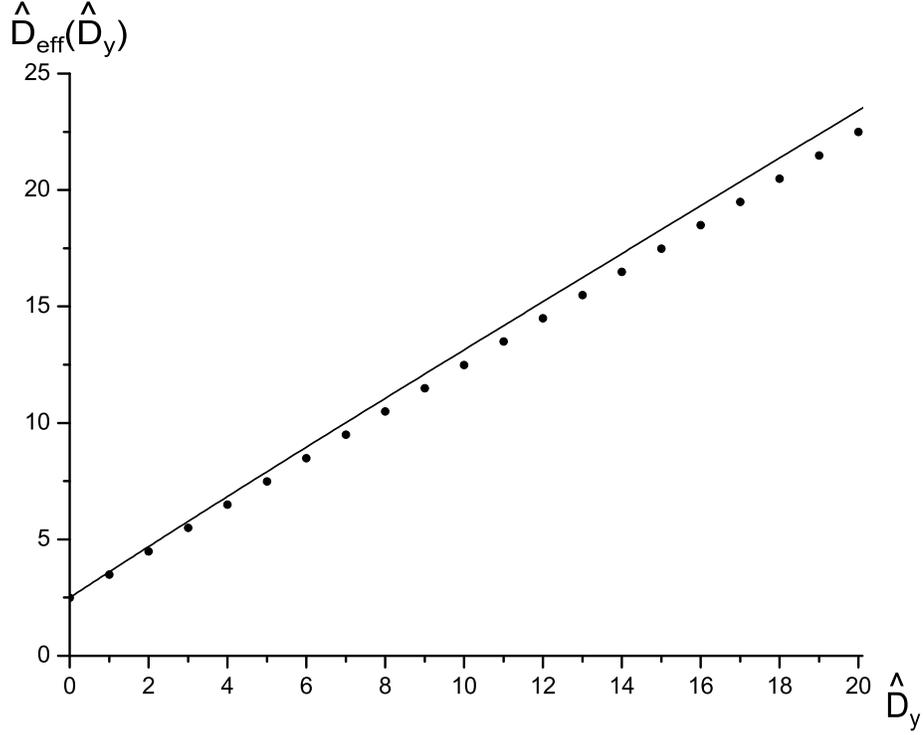}% Here is how to import EPS art
\caption{\label{BDY} Comparison between exact and interpolated
$\hat{D}_{\mathrm{eff}}$ functions. Solid line: plot of the exact
result. Circles: approximation according to Eq. (\ref{apprBDY}). }
\end{center}
\end{figure}

We are now in position to calculate
$\hat{D}_{\mathrm{eff}}=f(\hat{D}_y)$ for any value of
$\hat{D}_y$. Evaluation of Eq. (\ref{Gahia}) at $\hat{z}_o = 1/2$,
followed by normalization according to $\hat{G}(\hat{z}_o,0,0)=1$
gives the function $g$. Further integration of $|g|^2$ to
calculate the correlation function allows to recover
$\hat{D}_{\mathrm{eff}}=f(\hat{D}_y)$ as plotted in Fig.
\ref{BDY}.

\begin{figure}
\begin{center}
\includegraphics*[width=140mm]{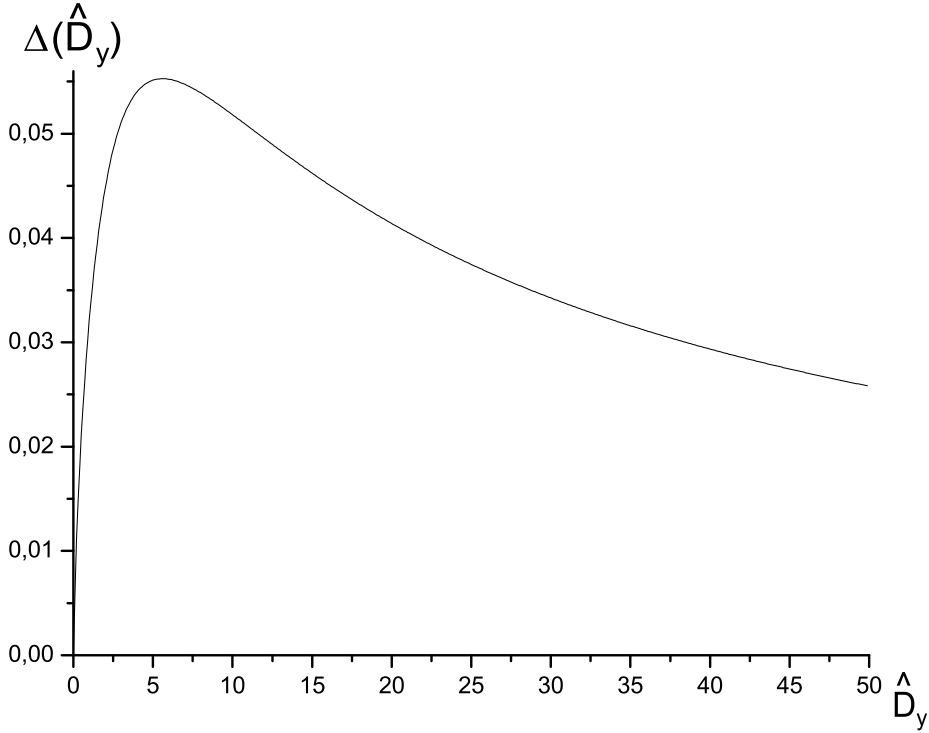}% Here is how to import EPS art
\caption{\label{BDYER} Relative accuracy $\Delta(\hat{D}_y)$ of
the match between true and interpolated $\hat{D}_{\mathrm{eff}}$
functions. }
\end{center}
\end{figure}
One may choose to calculate $\hat{D}_{\mathrm{eff}}=f(\hat{D}_y)$
numerically, but it is also possible to use, with reasonable
accuracy, the following analytical interpolation of
$\hat{D}_{\mathrm{eff}}$:

\begin{equation}
\hat{D}_{\mathrm{eff}} \simeq \hat{D}_{\mathrm{eff}}(0) +
\hat{D}_y = 2.50 + \hat{D}_y~. \label{apprBDY}
\end{equation}
There is of some interest to compare the exact and interpolated
$\hat{D}_\mathrm{eff}(\hat{D}_y)$ functions. Fig. \ref{BDYER}
shows the function

\begin{equation}
\Delta(\hat{D}_y)=\left|1-\frac{\hat{D}_{\mathrm{eff}}(\hat{D}_y)}{\hat{D}_{\mathrm{eff}}(0)
+ \hat{D}_y}\right| ~.\label{erBDY}
\end{equation}
There is seen to be good agreement between the interpolated and
exact $\hat{D}_{\mathrm{eff}}$ functions for small and large value
of $\hat{D}_y$. Noticeable discrepancies for $\hat{D}_y$ close to
unity are, anyway, less than $10 \%$.

Our conclusive result is, therefore, the following: when
$\hat{N}_y \gg 1$ and $\hat{D}_y$ assumes arbitrary values we
have:

\begin{equation}
\hat{\xi}_{c y} \simeq
\left(\frac{\pi}{2.50+\hat{D}_y}+\frac{\pi}{\hat{A}_y}\right)^{1/2}~.
\label{NM1DA}
\end{equation}
Also, it is important to remember that under conditions $\hat{N}_y
\gg 1$ and arbitrary $\hat{D}_y$ the spectral degree of coherence
is given by Eq. (\ref{Gahia}) in the near zone, and by Eq.
(\ref{G2D2easy3}) in the far zone.

The final step is to check that our main work hypothesis, i.e.
that the coherence length has the form in Eq. (\ref{xicdarbvera})
is correct. This can be done comparing Eq. (\ref{NM1DA}) with
numerical calculations for any given value of $\hat{N}_y \gg 1$
and a finite $\hat{D}_y$, which give a good agreement.

%For instance, in Fig. ....???...., we show such a comparison for
%the particular case $\hat{N}_y = 10 $ and $\hat{D}_y = 1$.

\subsection{\label{sub:D} Very large divergence $\hat{D}_y \gg 1$, arbitrary Fresnel number $\hat{N}_y$}

We now move on to treat the case with arbitrary beam transverse
size compared with the diffraction size (i.e. arbitrary Fresnel
number $\hat{N}_y$) and large divergence compared with the
diffraction angle (i.e. $\hat{D}_y \gg 1$). The particular case
for $\hat{N}_y \gg 1$ and $\hat{D}_y \gg 1$ overlaps with the
previous Section \ref{sub:N} and has already been treated in
Section \ref{paragr:1}. The conclusion was that $\xi_{c y} =
(\pi/\hat{D}_y + \pi/\hat{A}_y)^{1/2}$. In all the other remaining
cases Eq. (\ref{G2D2}) can be simplified as follows:

\begin{eqnarray}
\hat{G}(\hat{z}_o,\Delta \hat{\theta}_x,\Delta \hat{\theta}_y) =
\exp{\left[-\frac{2 \hat{A}_x \hat{z}_o^2 \Delta \hat{\theta}_x^2
\hat{D}_x }{\hat{A}_x+\hat{D}_x} \right]} \exp{\left[-\frac{2
\hat{A}_y \hat{z}_o^2 \Delta \hat{\theta}_y^2 \hat{D}_y
}{\hat{A}_y+\hat{D}_y} \right]} &&\cr \times
\int_{-\infty}^{\infty} d \hat{\phi}_y \int_{-\infty}^{\infty} d
\hat{\phi}_x
S^*{\left[\hat{z}_o,\hat{\phi}_x^2+(\hat{\phi}_y-\Delta
\hat{\theta}_y)^2\right]}
S{\left[\hat{z}_o,\hat{\phi}_x^2+(\hat{\phi}_y+\Delta
\hat{\theta}_y)^2\right]}~,~~~~~~\label{G2D2perDcase}
\end{eqnarray}
that is

\begin{eqnarray}
g(\hat{z}_o,\Delta \hat{\theta}_x,\Delta \hat{\theta}_y) =
\exp{\left[-\frac{2 \hat{A}_x \hat{z}_o^2 \Delta \hat{\theta}_x^2
\hat{D}_x }{\hat{A}_x+\hat{D}_x} \right]} \exp{\left[-\frac{2
\hat{A}_y \hat{z}_o^2 \Delta \hat{\theta}_y^2 \hat{D}_y
}{\hat{A}_y+\hat{D}_y} \right]}
\tilde{f}(\hat{z}_o,\Delta\hat{\theta}_y)
~,&&\cr\label{G2D2perDsimpl}
\end{eqnarray}
where

\begin{eqnarray}
\tilde{f}(\hat{z}_o,\Delta\hat{\theta}_y)=
\frac{1}{2\pi^2}\int_{-\infty}^{\infty} d \hat{\phi}_y
\int_{-\infty}^{\infty} d \hat{\phi}_x
S^*{\left[\hat{z}_o,\hat{\phi}_x^2+(\hat{\phi}_y-\Delta
\hat{\theta}_y)^2\right]}
S{\left[\hat{z}_o,\hat{\phi}_x^2+(\hat{\phi}_y+\Delta
\hat{\theta}_y)^2\right]}~,~~~~~~\label{G2D2perDF}
\end{eqnarray}
As is shown in Appendix C, having defined

\begin{eqnarray}
\beta(\Delta\hat{\theta}_y)=\frac{1}{2\pi^2}
\int_{-\infty}^{\infty} d \hat{\phi}_y \int_{-\infty}^{\infty} d
\hat{\phi}_x
\mathrm{sinc}{\left[\frac{\hat{\phi}_x^2+(\hat{\phi}_y-\Delta
\hat{\theta}_y)^2}{4}\right]}
\mathrm{sinc}{\left[\frac{\hat{\phi}_x^2+(\hat{\phi}_y+\Delta
\hat{\theta}_y)^2}{4}\right]}~,~~~~~~\label{G2D2perDG}
\end{eqnarray}
we have the important result

\begin{equation}
\tilde{f}(\hat{z}_o,\Delta\hat{\theta}_y) =
\beta(\Delta\hat{\theta}_y)\label{impores}
\end{equation}
for every choice of $\hat{z}_o$. $\beta$ is defined in such a way
to be normalized to unity. If we account for Eq. (\ref{impores})
we obtain the following expression for the spectral degree of
coherence $g$:

\begin{eqnarray}
g(\hat{z}_o,\Delta \hat{\theta}_x,\Delta \hat{\theta}_y) =
\exp{\left[-\frac{2 \hat{A}_x \hat{z}_o^2 \Delta \hat{\theta}_x^2
\hat{D}_x }{\hat{A}_x+\hat{D}_x} \right]}  \exp{\left[-\frac{2
\hat{A}_y \hat{z}_o^2 \Delta \hat{\theta}_y^2 \hat{D}_y
}{\hat{A}_y+\hat{D}_y} \right]} \beta(\Delta\hat{\theta}_y)
~.&&\cr\label{G2D2perDsimpl2}
\end{eqnarray}
It should be noted that, as $\hat{N}_y \gg 1$, the width of the
gaussian function in $\Delta \hat{\theta}_y$ in Eq.
(\ref{G2D2perDsimpl2}) becomes much smaller than unity and the
function $\beta$ can be considered constant and drops out of the
normalized expression for $g$. So, even if we did not analyze here
the limit for $\hat{N}_y \gg 1$ (we did it in Paragraph
\ref{paragr:1}), we see that the limit of  Eq.
(\ref{G2D2perDsimpl2}) for $\hat{N}_y \gg 1$ restitutes the
results found in Paragraph \ref{paragr:1}. Therefore we conclude
that Eq. (\ref{G2D2perDsimpl2}) is valid for any value of
$\hat{N}_y$.

\begin{figure}
\begin{center}
\includegraphics*[width=140mm]{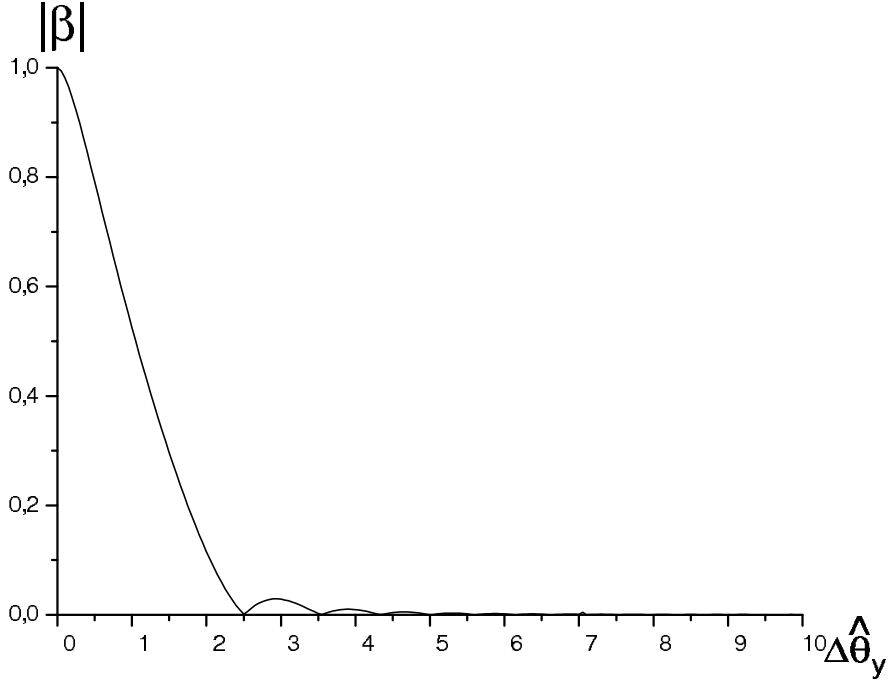}% Here is how to import EPS art
\caption{\label{gtilde} Plot of $|\beta|$ as a function of $\Delta
\hat{\theta}_y$. This universal plot illustrates the absolute
value of the spectral degree of coherence $|g|$ calculated for
$\Delta \hat{\theta}_x =0$ in the limit for $\hat{N}_y \ll 1$ and
$\hat{D}_y \gg 1$.}
\end{center}
\end{figure}

As is shown in Appendix C, the function $\beta(\Delta
\hat{\theta}_y)$ can also be calculated as

\begin{equation}
{\beta}(\Delta \hat{\theta}_y) = \frac{1}{\pi} \int_{0}^{\infty} d
\alpha ~ \alpha J_0\left(\alpha \frac{\Delta
\hat{\theta}_y}{2}\right) \left[\pi -
2\mathrm{Si}(\alpha^2)\right]^2~. \label{gtildedue}
\end{equation}
where $\mathrm{Si}$ indicates the sine integral function. The
representation of $\beta$ in Eq. (\ref{gtildedue}) is easier to
deal with numerically, because it involves a one-dimensional
integration only. Performing the integral, one can tabulate
$|\beta|$ to obtain the plot in Fig. \ref{gtilde}. This is an
universal plot. It should be noted that $\beta(\Delta
\hat{\theta}_y)$ is the spectral degree of coherence $g$
calculated for $\Delta \hat{\theta}_x =0$ in the limit for
$\hat{N}_y \ll 1$ and $\hat{D}_y \gg 1$. Its generalization for
arbitrary $\hat{N}_y$ is given by Eq. (\ref{G2D2perDsimpl2}).
Using Eq. (\ref{G2D2perDsimpl2}) and the tabulated values for the
universal function $\beta$ we can therefore calculate $g$
numerically for any choice of $\hat{N}_y$ and subsequently, we can
calculate the coherence length $\xi_{c y}(\hat{z}_o)$. For
instance, in the particular case $\hat{N}_y = 10$ and $\Delta
\hat{\theta}_x=0$ we obtain the simple linear behavior

\begin{equation}
\xi_{c y}(\hat{z}_o) = 0.54 \hat{z}_o \label{xicN10}
\end{equation}
For a generic value of $\hat{N}_y$, one can introduce an effective
function $\hat{N}_\mathrm{eff}(\hat{N}_y)$ so that

\begin{figure}
\begin{center}
\includegraphics*[width=140mm]{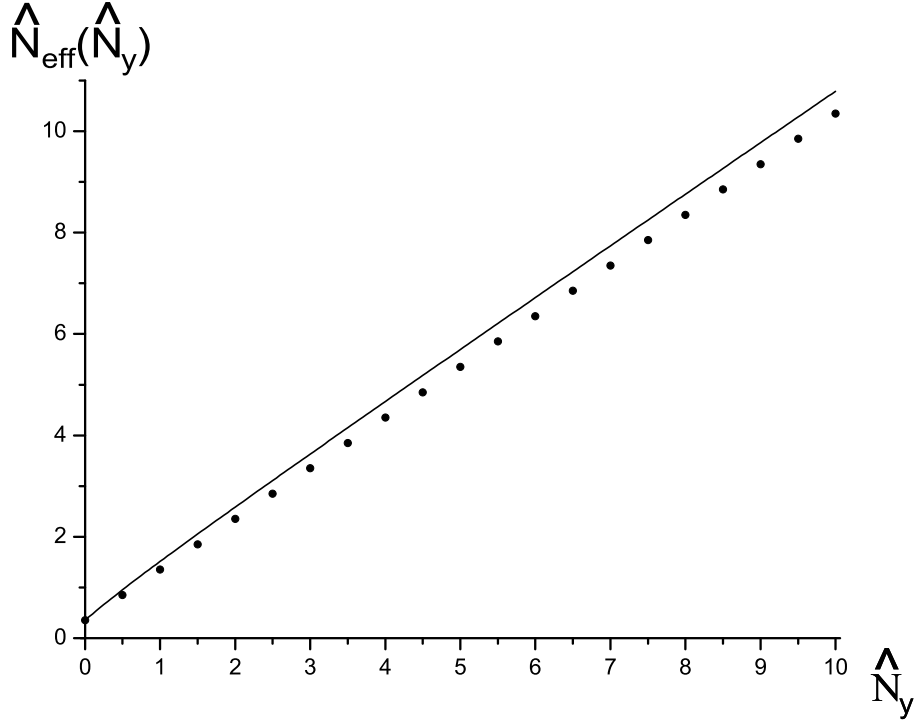}% Here is how to import EPS art
\caption{\label{NEFF} Comparison between exact and approximated
$\hat{N}_\mathrm{eff}(\hat{N}_y)$ functions. Solid line: plot of
the exact results. Circles: interpolation according to Eq.
(\ref{AssuNN}).}
\end{center}
\end{figure}
\begin{equation}
\xi_{c y}(\hat{z}_o) =
\sqrt{\frac{\pi}{\hat{N}_\mathrm{eff}(\hat{N}_y)}} \hat{z}_o
\label{xicN11}
\end{equation}
On the one hand, the function $\hat{N}_\mathrm{eff}(\hat{N}_y)$
can be computed numerically, similarly as we did for the
particular case $\hat{N}_y=10$. $\hat{N}_\mathrm{eff}(\hat{N}_y)$
is represented by the solid line in Fig. \ref{NEFF}.

On the other hand, one may also use an interpolation for
$\hat{N}_\mathrm{eff}(\hat{N}_y)$. First, numerical calculations
tell us that, in the particular case $\hat{N}_y \longrightarrow
0$, we have

\begin{equation}
\xi_{c y}(\hat{z}_o) = \sqrt{\frac{\pi}{0.35}} \hat{z}_o
~.\label{xicN12}
\end{equation}
Second, as  $\hat{D}_y \gg 1$ and $\hat{N}_y \gg \hat{D}_y$ we
have

\begin{equation}
\xi_{c y}(\hat{z}_o) \longrightarrow \sqrt{\frac{\pi}{\hat{N}_y}}
\hat{z}_o ~.\label{xicN12bissss}
\end{equation}
The simpler interpolated formula which satisfies both asymptotes
is therefore:

\begin{equation}
\hat{N}_\mathrm{eff}(\hat{N}_y) \simeq \hat{N}_y + 0.35~,
\label{AssuNN}
\end{equation}
The interpolation of $\hat{N}_\mathrm{eff}(\hat{N}_y)$ is
represented by black circles line in Fig. \ref{NEFF}.

\begin{figure}
\begin{center}
\includegraphics*[width=140mm]{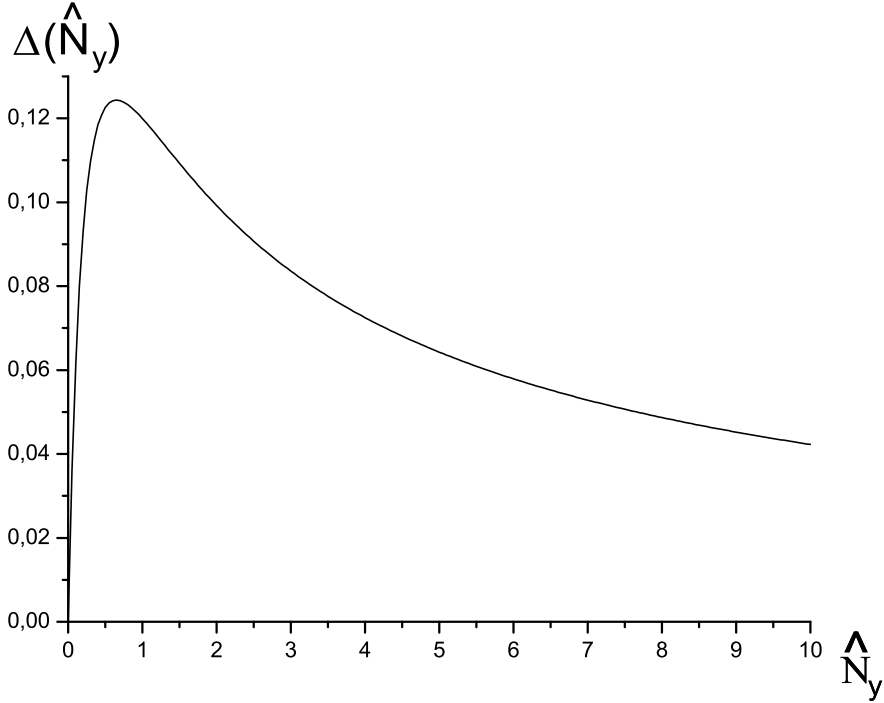}% Here is how to import EPS art
\caption{\label{BNYER} Relative accuracy $\Delta(\hat{N}_y)$ of
the match between exact and interpolated $\hat{N}_{\mathrm{eff}}$
functions. }
\end{center}
\end{figure}
There is of some interest to compare the exact and interpolated
$\hat{N}_\mathrm{eff}(\hat{N}_y)$ functions. Fig. \ref{BNYER}
shows the function

\begin{equation}
\Delta(\hat{N}_y)=\left|1-\frac{\hat{N}_{\mathrm{eff}}(\hat{N}_y)}{\hat{N}_{\mathrm{eff}}(0)
+ \hat{N}_y}\right| ~.\label{erBNY}
\end{equation}
There is seen to be good agreement between the interpolated and
exact $\hat{N}_{\mathrm{eff}}$ functions for small and large value
of $\hat{N}_y$. Noticeable discrepancies for $\hat{N}_y$ close to
unity are, anyway, less than $13 \%$.

Since in the case $\hat{D}_y \gg 1$ and $\hat{N}_y\gg1$ we
concluded that $\xi_{c y} = (\pi/\hat{D}_y +
\pi/\hat{A}_y)^{1/2}$, we can formulate the hypothesis that, for
$\hat{D}_y \gg 1$ and generic value of $\hat{N}_y$ one has

\begin{equation}
\xi_{c y}(\hat{z}_o) = \left( \frac{\pi}{\hat{D}_y}+
{\frac{\pi}{0.35+\hat{N}_y}} \hat{z}_o^2 \right)^{1/2}
~.\label{xicultimN}
\end{equation}
The final step is to check that such hypothesis is correct. This
can be done comparing Eq. (\ref{xicultimN}) with numerical
calculations for any given value of $\hat{D}_y \gg 1$ and finite
$\hat{N}_y$, which give a good agreement.

\subsection{\label{QHND} Conditions for the source to be quasi-homogeneous}

Up to this moment we discussed, for simplicity, the case
$\bar{\theta}_x = \bar{\theta}_y = 0$. We will now treat the
generic case with arbitrary $\bar{\theta}_x$ and $\bar{\theta}_y$.
This discussion will reduce to the relation between the weakly
quasi-homogeneous condition and our assumptions $\hat{N}_x \gg 1$,
$\hat{D}_x \gg 1$ and either $\hat{N}_y \gg 1$ or $\hat{D}_y \gg
1$. When we deal with weakly quasi-homogeneous wavefront, the
results found for $\bar{\theta}_x=\bar{\theta}_y =0$ have extended
validity for generic values of $\bar{\theta}_x$ and
$\bar{\theta}_y$. As already said before though, we will find that
the wavefronts are not always weakly quasi-homogeneous in the
vertical $y$ direction. In this case, previously found results are
only valid in the particular case $\bar{\theta}_y = 0$. Let us
consider the situation in more detail. We will start with Eq.
(\ref{G2Dnewsimplif2}), that may also be written as

\begin{eqnarray}
\hat{G}(\hat{z}_o,\bar{\theta}_x,\bar{\theta}_y,\Delta
\hat{\theta}_x,\Delta \hat{\theta}_y) ={\exp{\left[i 2\left(
\bar{\theta}_x\hat{z}_o\Delta
\hat{\theta}_x+\bar{\theta}_y\hat{z}_o\Delta \hat{\theta}_y\right)
\right]}}&&\cr \times \exp{\left[-\frac {\bar{\theta}_x^2 + 4
\hat{A}_x \hat{z}_o^2 \Delta \hat{\theta}_x^2 \hat{D}_x + 4 i
\hat{A}_x \bar{\theta}_x \hat{z}_o \Delta \hat{\theta}_x
}{2(\hat{A}_x+\hat{D}_x)}\right]}&&\cr \exp{\left[ - \frac { 2
\hat{A}_y \hat{z}_o^2 \Delta \hat{\theta}_y^2 \hat{D}_y + 2 i
\hat{A}_y \bar{\theta}_y \hat{z}_o \Delta \hat{\theta}_y
}{\hat{A}_y+\hat{D}_y}\right] } &&\cr \times
\int_{-\infty}^{\infty} d \hat{\phi}_y
\exp{\left[-\frac{(\bar{\theta}_y+\hat{\phi}_y)^2+4 i \hat{\phi}_y
\hat{A}_y \hat{z}_o \Delta \hat{\theta}_y}{2
(\hat{A}_y+\hat{D}_y)}\right]}&&\cr \times \int_{-\infty}^{\infty}
d \hat{\phi}_x
S^*{\left[\hat{z}_o,\hat{\phi}_x^2+(\hat{\phi}_y-\Delta
\hat{\theta}_y)^2\right]}
S{\left[\hat{z}_o,\hat{\phi}_x^2+(\hat{\phi}_y+\Delta
\hat{\theta}_y)^2\right]}~.\label{G2Dnewsimplif3}
\end{eqnarray}

\subsubsection{\label{casoD} Very large divergence parameter $\hat{D}_y\gg1$}

Because of the properties of the $S$ function $\hat{\phi}_y$ can
only change of a quantity $\Delta \hat{\phi}_y \sim 1$, otherwise
the $S$ functions will drop to zero. Then, since $\hat{D}_y \gg 1$
we have from Eq. (\ref{G2Dnewsimplif3})

\begin{eqnarray}
\hat{G}(\hat{z}_o,\bar{\theta}_x,\bar{\theta}_y,\Delta
\hat{\theta}_x,\Delta \hat{\theta}_y) ={\exp{\left[i 2\left(
\bar{\theta}_x\hat{z}_o\Delta
\hat{\theta}_x+\bar{\theta}_y\hat{z}_o\Delta \hat{\theta}_y\right)
\right]}}&&\cr \times \exp{\left[-\frac {\bar{\theta}_x^2 + 4
\hat{A}_x \hat{z}_o^2 \Delta \hat{\theta}_x^2 \hat{D}_x + 4 i
\hat{A}_x \bar{\theta}_x \hat{z}_o \Delta \hat{\theta}_x
}{2(\hat{A}_x+\hat{D}_x)}\right]}&&\cr \exp{\left[ - \frac { 2
\hat{A}_y \hat{z}_o^2 \Delta \hat{\theta}_y^2 \hat{D}_y + 2 i
\hat{A}_y \bar{\theta}_y \hat{z}_o \Delta \hat{\theta}_y
}{\hat{A}_y+\hat{D}_y}\right] } &&\cr \times
\exp{\left[-\frac{\bar{\theta}_y^2}{2
(\hat{A}_y+\hat{D}_y)}\right]}\int_{-\infty}^{\infty} d
\hat{\phi}_y \exp{\left[-\frac{2 i \hat{\phi}_y \hat{A}_y
\hat{z}_o \Delta \hat{\theta}_y}{
(\hat{A}_y+\hat{D}_y)}\right]}&&\cr \times \int_{-\infty}^{\infty}
d \hat{\phi}_x
S^*{\left[\hat{z}_o,\hat{\phi}_x^2+(\hat{\phi}_y-\Delta
\hat{\theta}_y)^2\right]}
S{\left[\hat{z}_o,\hat{\phi}_x^2+(\hat{\phi}_y+\Delta
\hat{\theta}_y)^2\right]}~.\label{G2Dnewsimplif4}
\end{eqnarray}
that is obviously weakly quasi-homogeneous.

\subsubsection{\label{casoN} Very large Fresnel number
$\hat{N}_y\gg1$.}

\textit{$~~~~~$(A) Case $\hat{A}_y \gg 1$.---$~$} In the case
$\hat{A}_y \gg 1$ we can follow the same line of reasoning in
Section \ref{casoD} simply replacing the roles of $\hat{D}_y$ with
$\hat{A}_y$, obtaining

\begin{eqnarray}
\hat{G}(\hat{z}_o,\bar{\theta}_x,\bar{\theta}_y,\Delta
\hat{\theta}_x,\Delta \hat{\theta}_y) ={\exp{\left[i 2\left(
\bar{\theta}_x\hat{z}_o\Delta
\hat{\theta}_x+\bar{\theta}_y\hat{z}_o\Delta \hat{\theta}_y\right)
\right]}}&&\cr \times \exp{\left[-\frac {\bar{\theta}_x^2 + 4
\hat{A}_x \hat{z}_o^2 \Delta \hat{\theta}_x^2 \hat{D}_x + 4 i
\hat{A}_x \bar{\theta}_x \hat{z}_o \Delta \hat{\theta}_x
}{2(\hat{A}_x+\hat{D}_x)}\right]}&&\cr \exp{\left[ - \frac { 2
\hat{A}_y \hat{z}_o^2 \Delta \hat{\theta}_y^2 \hat{D}_y + 2 i
\hat{A}_y \bar{\theta}_y \hat{z}_o \Delta \hat{\theta}_y
}{\hat{A}_y+\hat{D}_y}\right] } &&\cr \times
\exp{\left[-\frac{\bar{\theta}_y^2}{2
(\hat{A}_y+\hat{D}_y)}\right]}\int_{-\infty}^{\infty} d
\hat{\phi}_y \exp{\left[-\frac{2 i \hat{\phi}_y \hat{A}_y
\hat{z}_o \Delta \hat{\theta}_y}{
(\hat{A}_y+\hat{D}_y)}\right]}&&\cr \times \int_{-\infty}^{\infty}
d \hat{\phi}_x
S^*{\left[\hat{z}_o,\hat{\phi}_x^2+(\hat{\phi}_y-\Delta
\hat{\theta}_y)^2\right]}
S{\left[\hat{z}_o,\hat{\phi}_x^2+(\hat{\phi}_y+\Delta
\hat{\theta}_y)^2\right]}~.\label{G2Dnewsimplif4bissss}
\end{eqnarray}
that is obviously weakly quasi-homogeneous.

\textit{$~~~~~$(B) Case $\hat{A}_y \ll 1.$---$~$} In this
situation it must be $\hat{z}_o \gg 1$. Therefore we can
substitute all $S$ functions with $\mathrm{sinc}$ functions. The
case $\hat{D}_y \gg 1$ has been already treated. Let us,
therefore, first assume $\hat{D}_y \sim 1$. Using the fact that
$\Delta \hat{\theta}_y^2 \ll 1/\hat{z}_o^2 \ll 1$, Eq.
(\ref{G2Dnewsimplif3}) gives directly

\begin{eqnarray}
\hat{G}(\hat{z}_o,\bar{\theta}_x,\bar{\theta}_y,\Delta
\hat{\theta}_x,\Delta \hat{\theta}_y) ={\exp{\left[i 2\left(
\bar{\theta}_x\hat{z}_o\Delta
\hat{\theta}_x+\bar{\theta}_y\hat{z}_o\Delta \hat{\theta}_y\right)
\right]}}&&\cr \times \exp{\left[-\frac {\bar{\theta}_x^2 + 4
\hat{A}_x \hat{z}_o^2 \Delta \hat{\theta}_x^2 \hat{D}_x + 4 i
\hat{A}_x \bar{\theta}_x \hat{z}_o \Delta \hat{\theta}_x
}{2(\hat{A}_x+\hat{D}_x)}\right]} \exp{\left[ - {2 \hat{A}_y
\hat{z}_o^2 \Delta \hat{\theta}_y^2 }\right] } &&\cr \times
\int_{-\infty}^{\infty} d \hat{\phi}_y
\exp{\left[-\frac{(\hat{\phi}_y+\bar{\theta}_y)^2}{2\hat{D}_y}\right]}I_S(\hat{\phi}_y)~,\label{G2Dmeno1}
\end{eqnarray}
that is obviously weakly quasi-homogeneous.

Now let us consider the case $\hat{D}_y \ll 1$. Eq.
(\ref{G2Dnewsimplif3}) can be simplified on the assumptions
$\hat{A}_y \ll 1$ and $\hat{D}_y \ll 1$. In fact, the Gaussian
exponential factor inside the integral in $d \hat{\phi}_y$ in Eq.
(\ref{G2Dnewsimplif3}) imposes a maximal value
$(\bar{\theta}_y+\hat{\phi}_y)^2 \sim \hat{A}_y + \hat{D}_y \ll
1$.

Simultaneously,  from the Gaussian exponential factor in $\Delta
\hat{\theta}_y$ outside the integral, we have a condition for the
maximal value of $\Delta \hat{\theta}_y^2 \sim 1/\hat{z}_o^2 \ll
1$ since $\Delta \hat{\theta}_y^2 \sim 1/(\hat{A}_y \hat{z}_o^2)
\ll 1$.

As a result, the dependence of $S$ on $(\hat{\phi} + \Delta
\hat{\theta}_y)^2$ can be substituted by a dependence on
$\bar{\theta}_y$ and, as has already been said, the $S$ functions
can be substituted with $\mathrm{sinc}$ functions, giving

\begin{eqnarray}
\hat{G}(\hat{z}_o,\bar{\theta}_x,\bar{\theta}_y,\Delta
\hat{\theta}_x,\Delta \hat{\theta}_y) = {\exp{\left[i 2\left(
\bar{\theta}_x\hat{z}_o\Delta
\hat{\theta}_x+\bar{\theta}_y\hat{z}_o\Delta \hat{\theta}_y\right)
\right]}}&&\cr \times \exp{\left[-\frac {\bar{\theta}_x^2 + 4
\hat{A}_x \hat{z}_o^2 \Delta \hat{\theta}_x^2 \hat{D}_x + 4 i
\hat{A}_x \bar{\theta}_x \hat{z}_o \Delta \hat{\theta}_x
}{2(\hat{A}_x+\hat{D}_x)}\right]}&&\cr \times \exp{\left[ - \frac
{ 2 \hat{A}_y \hat{z}_o^2 \Delta \hat{\theta}_y^2 \hat{D}_y + 2 i
\hat{A}_y \bar{\theta}_y \hat{z}_o \Delta \hat{\theta}_y
}{\hat{A}_y+\hat{D}_y}\right] }  I_S(\bar{\theta}_y) &&\cr \times
\int_{-\infty}^{\infty} d \hat{\phi}_y
\exp{\left[-\frac{(\bar{\theta}_y+\hat{\phi}_y)^2+2\hat{\phi}_y\left(2i
\hat{A}_y \hat{z}_o \Delta \hat{\theta}_y \right)}{2
(\hat{A}_y+\hat{D}_y)}\right]} ~.\label{G2D50b}
\end{eqnarray}
The integral in $d \hat{\phi}_y$ can be performed giving

\begin{eqnarray}
\hat{G}(\hat{z}_o,\bar{\theta}_x,\bar{\theta}_y,\Delta
\hat{\theta}_x,\Delta \hat{\theta}_y) = {\exp{\left[i 2\left(
\bar{\theta}_x\hat{z}_o\Delta
\hat{\theta}_x+\bar{\theta}_y\hat{z}_o\Delta \hat{\theta}_y\right)
\right]}}&&\cr \times \exp{\left[-\frac {\bar{\theta}_x^2 + 4
\hat{A}_x \hat{z}_o^2 \Delta \hat{\theta}_x^2 \hat{D}_x + 4 i
\hat{A}_x \bar{\theta}_x \hat{z}_o \Delta \hat{\theta}_x
}{2(\hat{A}_x+\hat{D}_x)}\right]}&&\cr \times \exp{\left[ - \frac
{ 2 \hat{A}_y \hat{z}_o^2 \Delta \hat{\theta}_y^2 \hat{D}_y + 2 i
\hat{A}_y \bar{\theta}_y \hat{z}_o \Delta \hat{\theta}_y
}{\hat{A}_y+\hat{D}_y}\right] }  &&\cr \times \exp{\left[-\frac{2
\hat{A}_y^2 \hat{z}_o^2 \Delta \hat{\theta}_y^2}
{\hat{A}_y+\hat{D}_y} \right]} \exp{\left[\frac{2 i\hat{A}_y
\bar{\theta}_y \hat{z}_o \Delta
\hat{\theta}_y}{\hat{A}_y+\hat{D}_y}\right]} I_S(\bar{\theta}_y)
~.\label{G2D51b}
\end{eqnarray}
Normalizing $\hat{G}$ in such a way that
$\hat{G}(\hat{z}_o,\bar{\theta}_x,\bar{\theta}_y,0,0) = 1$ we
obtain the following expression for the spectral degree of
coherence $g$:

\begin{eqnarray}
g(\hat{z}_o,\Delta \hat{\theta}_x,\Delta \hat{\theta}_y) =
{\exp{\left[i 2\left( \bar{\theta}_x\hat{z}_o\Delta
\hat{\theta}_x+\bar{\theta}_y\hat{z}_o\Delta \hat{\theta}_y\right)
\right]}}&&\cr \times \exp{\left[-\frac {\bar{\theta}_x^2 + 4
\hat{A}_x \hat{z}_o^2 \Delta \hat{\theta}_x^2 \hat{D}_x + 4 i
\hat{A}_x \bar{\theta}_x \hat{z}_o \Delta \hat{\theta}_x
}{2(\hat{A}_x+\hat{D}_x)}\right]} \exp{\left[ - { 2 \hat{A}_y
\hat{z}_o^2 \Delta \hat{\theta}_y^2 }\right] }
 ~,\label{G2D51lastb}
\end{eqnarray}
which generalizes Eq. (\ref{G2D51last}) for any value of
$\bar{\theta}_x$ and $\bar{\theta}_y$, and shows weak
quasi-homogeneity of the wavefronts.

\textit{$~~~~~$(C) Case $\hat{A}_y \sim 1.$---$~$}  Since
$\hat{N}_y \gg 1$ and $\hat{z}_o \gg 1$, the exponential function
in $\Delta \hat{\theta}_y$ outside the integral sign of Eq.
(\ref{G2Dnewsimplif2}) impose $\Delta \hat{\theta}_y \ll 1$ so
that

\begin{eqnarray}
\hat{G}(\hat{z}_o,\bar{\theta}_x,\bar{\theta}_y,\Delta
\hat{\theta}_x,\Delta \hat{\theta}_y) ={\exp{\left[i 2\left(
\bar{\theta}_x\hat{z}_o\Delta
\hat{\theta}_x+\bar{\theta}_y\hat{z}_o\Delta \hat{\theta}_y\right)
\right]}}&&\cr \times \exp{\left[-\frac {\bar{\theta}_x^2 + 4
\hat{A}_x \hat{z}_o^2 \Delta \hat{\theta}_x^2 \hat{D}_x + 4 i
\hat{A}_x \bar{\theta}_x \hat{z}_o \Delta \hat{\theta}_x
}{2(\hat{A}_x+\hat{D}_x)}\right]}&&\cr \exp{\left[ - \frac {  2
\hat{A}_y \hat{z}_o^2 \Delta \hat{\theta}_y^2 \hat{D}_y + 2 i
\hat{A}_y \bar{\theta}_y \hat{z}_o \Delta \hat{\theta}_y
}{\hat{A}_y+\hat{D}_y}\right] } &&\cr \times
\int_{-\infty}^{\infty} d \hat{\phi}_y
\exp{\left[-\frac{(\bar{\theta}_y+\hat{\phi}_y)^2+4i\hat{\phi}_y
\hat{A}_y \hat{z}_o \Delta \hat{\theta}_y }{2
(\hat{A}_y+\hat{D}_y)}\right]}I_S(\hat{\phi}_y)~.\label{G2Dmeno2bis}
\end{eqnarray}
The intensity distribution can be found from Eq.
(\ref{G2Dmeno2bis}) setting $\Delta\hat{\theta}_x
=\Delta\hat{\theta}_y =0$ thus obtaining

\begin{eqnarray}
I(\hat{z}_o,\bar{\theta}_x,\bar{\theta}_y) =\exp{\left[-\frac
{\bar{\theta}_x^2 }{2(\hat{A}_x+\hat{D}_x)}\right]}
\int_{-\infty}^{\infty} d \hat{\phi}_y
\exp{\left[-\frac{(\bar{\theta}_y+\hat{\phi}_y)^2 }{2
(\hat{A}_y+\hat{D}_y)}\right]}
I_S(\hat{\phi}_y)~.&&\cr\label{G2Dmeno3}
\end{eqnarray}
It is evident by inspection that we cannot factorize Eq.
(\ref{G2Dmeno2bis}) to obtain $|\hat{G}| =
I(\hat{z}_o,\bar{\theta}_x,\bar{\theta}_y) w(\hat{z}_o,\Delta
\hat{\theta}_x,\Delta \hat{\theta}_y)$ (where have put
$|g|=w(\hat{z}_o,\Delta \hat{\theta}_x,\Delta \hat{\theta}_y)$).
As a result we conclude that, in this case, the wavefront is not
quasi-homogeneous, not even in the weak sense.

\subsubsection{\label{Analisidisc} Discussion.}

Let us discuss the results obtained in this analysis of
quasi-homogeneity.

In Section \ref{sub:oned} we have seen that weakly
quasi-homogeneous wavefronts which are not quasi-homogeneous in
the usual sense are present in the far field, when the VCZ theorem
holds. From our analysis in the one-dimensional framework, the
notion of far zone arises in the $x$ direction, when the apparent
angular dimension $\hat{A}_{x}$ of the source is much smaller than
the divergence of the radiation beam that, in this case, can be
identified with the electron beam divergence $\hat{D}_{x} \gg 1$.
Both $\hat{N}_{x}\gg 1$ and $\hat{D}_{x} \gg 1$. If $\hat{A}_{x}
\ll \hat{D}_{x}$ the wavefront is quasi-homogenous but only in the
weak sense, one is in the far zone and the VCZ theorem applies. If
$\hat{A}_{x} \gg \hat{D}_{x}$ one is in the near field zone and
the wavefront is quasi-homogeneous in the usual sense. Transition
from the near to the far zone always involves, in this case,
weakly quasi-homogeneous wavefront.

In the two-dimensional framework studied in the present Section,
as for the one-dimensional model, both $\hat{N}_{x}$ and
$\hat{D}_{x} \gg 1$. The $x$ and the $y$ coordinates appear
factorized and the far zone applies now separately to both $x$ and
$y$ directions, meaning that $\hat{A}_{x,y}$ of the source is much
smaller than the divergence of the radiation beam. In other words,
we are in the far zone as soon as the (square of the) beam size at
a given $\hat{z}_o$ begins to be much larger than the (square of
the) initial size of radiation (i.e. much larger than the Fresnel
numbers $\hat{N}_{x,y}$).

In the case $\hat{D}_y \gg 1$, independently on the value of
$\hat{N}_y$, we will always have weakly quasi-homogeneous (but not
always quasi-homogeneous in the usual sense!) wavefronts at any
distance $\hat{z}_o$. This is exactly the situation discussed for
the $x$ direction, where transitions from the near ($\hat{A}_y \gg
\hat{D}_y$) to the far field ($\hat{A}_y \ll \hat{D}_y$) involve
only weakly quasi-homogeneous wavefronts. Moreover, in the far
field zone, the VCZ theorem is valid.

In the case with $\hat{D}_y \lesssim 1$, $\hat{N}_y \gg 1$ we will
have quasi-homogeneous wavefronts in the usual sense at $\hat{z}_o
\sim 1$ (near zone) and weak quasi-homogeneous wavefronts in the
far zone, at $\hat{z}_o \gg 1$.

Finally, if $\hat{N}_y \gg 1$, $\hat{D}_y \lesssim 1$ and
$\hat{A}_y \sim 1$, i.e. for distances $\hat{z}_o \sim
\sqrt{\hat{N}_y}$, we have seen that the wavefront is non
quasi-homogenous, not even in the weak sense.

These results can be seen in terms of convolution between the
Gaussian distribution of intensity associated to the electron beam
emittance and the distribution of intensity due to intrinsic
properties of undulator radiation. The only case, among those
treated up to now, when such a convolution does not simplify into
the product of separate factors is when  $\hat{N}_y \gg 1$,
$\hat{D}_y$ not much larger than unity and $\hat{A}_y \sim 1$: in
this case we do not have quasi-homogeneous wavefronts, not even in
the weak sense.

\section{\label{sec:nonh} Radiation from some non-homogeneous undulator sources}

In Section \ref{sub:oned} we treated a simplified situation with
$\hat{N}_x \gg 1$, $\hat{D}_x \gg 1$ and
$\hat{\theta}_{y1}=\hat{\theta}_{y2}$. Moreover, for simplicity of
calculations we assumed $\hat{N}_y \ll 1$ and $\hat{D}_y \ll 1$.
In Section \ref{sec:twod}, instead, we treated the case of
electron beams with $\hat{N}_x \gg 1$, $\hat{D}_x \gg 1$ and
either $\hat{N}_y \gg 1$ or $\hat{D}_y \gg 1$ (or both). It is
important to note that, in the $x$ direction, the results obtained
in Section \ref{sub:oned} are the same as the one in  Section
\ref{sec:twod}. In fact as  $\hat{N}_x \gg 1$ and $\hat{D}_x \gg
1$, the cross-spectral density factorizes in the product of two
contributions depending separately on the $x$ and $y$ coordinates,
and under $\hat{N}_x \gg 1$ and $\hat{D}_x \gg 1$ the derivation
of the $x$-dependent factor is always the same.

We have seen that in some of the cases discussed in Section
\ref{QHND}, the assumptions $\hat{N}_x \gg 1$, $\hat{D}_x \gg 1$
and either $\hat{N}_y \gg 1$ or $\hat{D}_y \gg 1$  (or both) were
enough to guarantee that the wavefront is weakly quasi-homogeneous
in the sense specified by Eq. (\ref{introh2}). In this Section we
will extend our analytical investigations to some cases outside
the range of parameters treated before, where the weakly
quasi-homogeneous assumption is not fulfilled in the far zone. In
particular, we will  demonstrate that, under conditions $\hat{N}_x
\gg 1$, $\hat{D}_x \gg 1$ and both $\hat{N}_y \ll 1$ and
$\hat{D}_y \ll 1$, wavefronts are not weakly quasi-homogenous in
the vertical $y$ direction (although they are in the horizontal
$x$ direction).

First, in Section \ref{sub:caso1} we will analyze the case
$\hat{N}_x \sim 1$, $\hat{D}_x \ll 1$. Assuming a vertical
emittance of the electron beam much smaller than the horizontal
emittance $\hat{\epsilon}_y \ll \hat{\epsilon}_x$, we have,
automatically $\hat{N}_y \ll 1$ and $\hat{D}_y \ll 1$. This
corresponds to a practically important situation. For instance,
consider a VUV beamline at a third generation light source with
$\lambda = 30$ nm, $\epsilon_x = 3\cdot 10^{-9}$ m, and
$\epsilon_y = 0.03 \cdot 10^{-9}$ m, i.e. $\hat{\epsilon}_x = 0.6$
and $\hat{\epsilon}_y = 6\cdot 10^{-3}$. If $\hat{\beta}_x = 3$ we
would have $\hat{N}_x = 2$ and $\hat{D}_x = 0.2$. Second, in
Section \ref{sub:caso2} we will study the situation $\hat{N}_y
\lesssim 1$ and $\hat{D}_y \ll 1$ with  $\hat{N}_x \gg 1$ and
$\hat{D}_x \gg 1$, that will give us back also the limiting case
for $\hat{N}_y \ll 1$ and $\hat{D}_y \ll 1$ with  $\hat{N}_x \gg
1$ and $\hat{D}_x \gg 1$ (already discussed in Section
\ref{sub:oned}). The situation with finite vertical Fresnel number
and negligible vertical divergence (compared with the diffraction
angle) is a very practical one: for instance, given a third
generation light source with $\lambda = 1 \AA$ and $\epsilon_y =
10^{-11}$ m, i.e. $\hat{\epsilon}_y = 0.6$, a value $\hat{\beta}_y
=6$ corresponds to $\hat{D}_y = 0.1$ and $\hat{N}_y = 3.6$.

Although in these two cases, the weakly quasi-homogeneous
assumption is not fulfilled we will see that the choice $\hat{z}_o
\gg 1$ will allow us to treat these situations in analogy with
respect to some weakly quasi-homogeneous case we already dealt
with.

\subsection{\label{sub:caso1} Case $\hat{N}_x
\sim 1$, $\hat{D}_x \ll 1$.}

Assuming a vertical emittance of the ring much smaller than the
horizontal emittance $\hat{\epsilon}_y \ll \hat{\epsilon}_x$, we
have, automatically, $\hat{N}_y \ll 1$ and $\hat{D}_y \ll 1$. We
start with Eq. (\ref{G2D}), that can be specialized to an equation
dependent on the $x$ coordinates only and, in the case for
$\hat{z}_o \gg 1$, can be written as

\begin{eqnarray}
\hat{G}(\hat{z}_o,\bar{\theta},\Delta \hat{\theta})
=\frac{\exp{\left(i 2 \bar{\theta}\Delta \hat{\theta}
\hat{z}_o\right)}}{\sqrt{2\pi(\hat{N}/\hat{z}_o^2 + \hat{D})}}
\exp{\left[-\frac{ 4 \hat{N} \Delta \hat{\theta}^2 \hat{D} + 4 i
(\hat{N}/\hat{z}_o)\bar{\theta}\Delta \hat{\theta}
}{2(\hat{N}/\hat{z}_o^2+\hat{D})}\right]}
 &&\cr
\times \int_{-\infty}^{\infty} d \hat{\phi}
\exp{\left[-\frac{\left(\hat{\phi}+\bar{\theta}\right)^2+4 i
\hat{\phi} (\hat{N}/\hat{z}_o) \Delta \hat{\theta} }{2
(\hat{N}/\hat{z}_o^2+\hat{D})}\right]} &&\cr\times\mathrm{sinc}
{\left[(\hat{\phi}-\Delta \hat{\theta})^2/4\right]}
 \mathrm{sinc} {\left[(\hat{\phi}+\Delta
\hat{\theta})^2/4\right]}~,\label{Gzlarge9gengen}
\end{eqnarray}
where we systematically omitted $x$ subscripts. Since $\hat{A} +
\hat{D} \ll 1$, the exponential factor in $(\phi+\bar{\theta})^2$
inside the integral sign in Eq. (\ref{Gzlarge9gengen}) behaves
like a $\delta$-Dirac function with respect to the
$\mathrm{sinc}(\cdot)$ functions inside the same integral: as a
result we can substitute $\phi$ with $\bar{\theta}$ in the
$\mathrm{sinc}(\cdot)$ functions, which drop out of the integral
sign. Then, the integral in $d \hat{\phi}$ can be calculated
analytically so that we have:

\begin{eqnarray}
\hat{G}(\hat{z}_o,\bar{\theta},\Delta \hat{\theta}) ={\exp{\left(i
2 \bar{\theta}\Delta \hat{\theta} \hat{z}_o\right)}}
\exp{\left[-\frac{ 2 \hat{N} \Delta \hat{\theta}^2 \hat{D}
}{\hat{N}/\hat{z}_o^2+\hat{D}}\right]} &&\cr\times\mathrm{sinc}
{\left[(\bar{\theta}-\Delta \hat{\theta})^2/4\right]}
\mathrm{sinc} {\left[(\bar{\theta}+\Delta
\hat{\theta})^2/4\right]} &&\cr \times \exp{\left[-\frac{2
(\hat{N}^2/\hat{z}_o^2) \Delta
\hat{\theta}^2}{\hat{N}/\hat{z}_o^2+\hat{D}} \right]}
~.\label{Disomo1}
\end{eqnarray}
Combination of the second and the third exponential function
yields

\begin{eqnarray}
\hat{G}(\hat{z}_o,\bar{\theta},\Delta \hat{\theta}) ={\exp{\left(i
2 \bar{\theta}\Delta \hat{\theta} \hat{z}_o\right)}} \exp{\left[-{
2 \hat{N}  \Delta \hat{\theta}^2  }\right]}
&&\cr\times\mathrm{sinc} {\left[(\bar{\theta}-\Delta
\hat{\theta})^2/4\right]} \mathrm{sinc}
{\left[(\bar{\theta}+\Delta \hat{\theta})^2/4\right]}
~.\label{Disomo2}
\end{eqnarray}
Finally, in order to obtain the degree of spectral coherence $g$
we should normalize $\hat{G}$ according to Eq. (\ref{normfine2})
\footnote{Note that, in this case, we are dealing with non
quasi-homogeneous wavefronts and, as has already been said,
normalizing according to $\hat{G}(\hat{z}_o,\bar{\theta},0)=1$ is
not the same of normalizing according to Eq. (\ref{normfine2}).}
thus obtaining

\begin{eqnarray}
g(\hat{z}_o,\bar{\theta},\Delta \hat{\theta}) ={\exp{\left(i 2
\bar{\theta}\Delta \hat{\theta} \hat{z}_o\right)}} \exp{\left[-{ 2
\hat{N}  \Delta \hat{\theta}^2  }\right]} ~.\label{Disomo3}
\end{eqnarray}
According to Eq. (\ref{Disomo3}) the spectral degree of coherence
$g$ is such that $|g|$ is only function of $\Delta \hat{\theta}$.
Moreover, the dependence of $g$ on the phase $2 \bar{\theta}\Delta
\hat{\theta} \hat{z}_o$ is a feature for radiation from Schell's
model sources in the far field, that we have already encountered
many times in the study of weakly quasi-homogenous cases. As a
result we can conclude that the radiation of the undulator source
at $\hat{N}_x \sim 1$ and $\hat{D}_x \ll 1$ represents the far
field radiation of a Schell's model source. Finally, it should be
noted that in a two-pinhole experiment, for any vertical position
of the pinholes, the fringe visibility, i.e. the modulus of the
spectral degree of coherence depends only on the separation along
the horizontal $x$ direction. As a result, while in Section
\ref{sub:oned} we put $\hat{\theta}_{y1}=\hat{\theta}_{y2}$, thus
selecting from the very beginning a horizontal plane,  the present
case can be fully described by a one-dimensional model,
independently on the choice of transverse coordinates of the
pinholes.

\subsection{\label{sub:caso2} Case $\hat{N}_y \lesssim 1$ and $\hat{D}_y \ll 1$  with $\hat{N}_x \gg 1$,
$\hat{D}_x \gg 1$}

In this situation we go back to a two-dimensional model. When
$\hat{z}_o \gg 1$ we have $\hat{A}_y \ll 1$, which is the limiting
case treated in Paragraph \ref{paragr:2} \textit{(A)}: the
difference is that, now $\hat{N}_{y} \sim 1$. Let us start with
Eq. (\ref{G2Dnewsimplif2}) written as

\begin{eqnarray}
\hat{G}(\hat{z}_o,\bar{\theta}_x,\bar{\theta}_y,\Delta
\hat{\theta}_x,\Delta \hat{\theta}_y) ={\exp{\left[i 2\left(
\bar{\theta}_x\hat{z}_o\Delta
\hat{\theta}_x+\bar{\theta}_y\hat{z}_o\Delta \hat{\theta}_y\right)
\right]}}&&\cr \times \exp{\left[-\frac {\bar{\theta}_x^2 + 4
\hat{A}_x \hat{z}_o^2 \Delta \hat{\theta}_x^2 \hat{D}_x + 4 i
\hat{A}_x \bar{\theta}_x \hat{z}_o \Delta \hat{\theta}_x
}{2(\hat{A}_x+\hat{D}_x)}\right]}&&\cr \exp{\left[ - \frac {  2
\hat{A}_y \hat{z}_o^2 \Delta \hat{\theta}_y^2 \hat{D}_y + 2 i
\hat{A}_y \bar{\theta}_y \hat{z}_o \Delta \hat{\theta}_y
}{\hat{A}_y+\hat{D}_y}\right] } &&\cr \times
\int_{-\infty}^{\infty} d \hat{\phi}_y
\exp{\left[-\frac{(\bar{\theta}_y+\hat{\phi}_y)^2+4i\hat{\phi}_y
\hat{A}_y \hat{z}_o \Delta \hat{\theta}_y }{2
(\hat{A}_y+\hat{D}_y)}\right]}&&\cr \times \int_{-\infty}^{\infty}
d \hat{\phi}_x
S^*{\left[\hat{z}_o,\hat{\phi}_x^2+(\hat{\phi}_y-\Delta
\hat{\theta}_y)^2\right]}
S{\left[\hat{z}_o,\hat{\phi}_x^2+(\hat{\phi}_y+\Delta
\hat{\theta}_y)^2\right]}~,\label{G2Dmenodis2}
\end{eqnarray}
$\hat{A}_y \ll 1$ and $\hat{D}_y \ll 1$ impose a maximal value of
$(\bar{\theta}_y+\hat{\phi}_y)^2 \sim \hat{A}_y + \hat{D}_y \ll
1$. Moreover the $S$ functions can be substituted with
$\mathrm{sinc}$ functions since we are working in the limit for
$\hat{z}_o \gg 1$. Then, from Eq. (\ref{G2Dmenodis2}) we obtain

\begin{eqnarray}
\hat{G}(\hat{z}_o,\bar{\theta}_x,\bar{\theta}_y,\Delta
\hat{\theta}_x,\Delta \hat{\theta}_y) ={\exp{\left[i 2\left(
\bar{\theta}_x\hat{z}_o\Delta
\hat{\theta}_x+\bar{\theta}_y\hat{z}_o\Delta \hat{\theta}_y\right)
\right]}}&&\cr \times \exp{\left[-\frac {\bar{\theta}_x^2 + 4
\hat{A}_x \hat{z}_o^2 \Delta \hat{\theta}_x^2 \hat{D}_x + 4 i
\hat{A}_x \bar{\theta}_x \hat{z}_o \Delta \hat{\theta}_x
}{2(\hat{A}_x+\hat{D}_x)}\right]}&&\cr \exp{\left[ - \frac {  2
\hat{A}_y \hat{z}_o^2 \Delta \hat{\theta}_y^2 \hat{D}_y + 2 i
\hat{A}_y \bar{\theta}_y \hat{z}_o \Delta \hat{\theta}_y
}{\hat{A}_y+\hat{D}_y}\right] } &&\cr \times
\int_{-\infty}^{\infty} d \hat{\phi}_x
\mathrm{sinc}{\left[\frac{\hat{\phi}_x^2+(\bar{\theta}_y-\Delta
\hat{\theta}_y)^2}{4}\right]}
\mathrm{sinc}{\left[\frac{\hat{\phi}_x^2+(\bar{\theta}_y+\Delta
\hat{\theta}_y)^2}{4}\right]} &&\cr \times \int_{-\infty}^{\infty}
d \hat{\phi}_y
\exp{\left[-\frac{(\bar{\theta}_y+\hat{\phi}_y)^2+4i\hat{\phi}_y
\hat{A}_y \hat{z}_o \Delta \hat{\theta}_y }{2
(\hat{A}_y+\hat{D}_y)}\right]}~,\label{G2Dmenodis3}
\end{eqnarray}
As done before, the integral in $d \hat{\phi}_y$ can be performed
giving

\begin{eqnarray}
\hat{G}(\hat{z}_o,\bar{\theta}_x,\bar{\theta}_y,\Delta
\hat{\theta}_x,\Delta \hat{\theta}_y) ={\exp{\left[i 2\left(
\bar{\theta}_x\hat{z}_o\Delta
\hat{\theta}_x+\bar{\theta}_y\hat{z}_o\Delta \hat{\theta}_y\right)
\right]}}&&\cr \times \exp{\left[-\frac {\bar{\theta}_x^2 + 4
\hat{A}_x \hat{z}_o^2 \Delta \hat{\theta}_x^2 \hat{D}_x + 4 i
\hat{A}_x \bar{\theta}_x \hat{z}_o \Delta \hat{\theta}_x
}{2(\hat{A}_x+\hat{D}_x)}\right]}&&\cr \exp{\left[ - \frac {  2
\hat{A}_y \hat{z}_o^2 \Delta \hat{\theta}_y^2 \hat{D}_y + 2 i
\hat{A}_y \bar{\theta}_y \hat{z}_o \Delta \hat{\theta}_y
}{\hat{A}_y+\hat{D}_y}\right] } &&\cr \times \exp{\left[-\frac{2
\hat{A}_y^2 \hat{z}_o^2 \Delta \hat{\theta}_y^2}
{\hat{A}_y+\hat{D}_y} \right]} \exp{\left[\frac{2 i\hat{A}_y
\bar{\theta}_y \hat{z}_o \Delta
\hat{\theta}_y}{\hat{A}_y+\hat{D}_y}\right]} &&\cr \times
\int_{-\infty}^{\infty} d \hat{\phi}_x
\mathrm{sinc}{\left[\frac{\hat{\phi}_x^2+(\bar{\theta}_y-\Delta
\hat{\theta}_y)^2}{4}\right]}
\mathrm{sinc}{\left[\frac{\hat{\phi}_x^2+(\bar{\theta}_y+\Delta
\hat{\theta}_y)^2}{4}\right]} ~,\label{G2Dmenodis4}
\end{eqnarray}
that is

\begin{eqnarray}
\hat{G}(\hat{z}_o,\bar{\theta}_x,\bar{\theta}_y,\Delta
\hat{\theta}_x,\Delta \hat{\theta}_y) ={\exp{\left[i 2\left(
\bar{\theta}_x\hat{z}_o\Delta
\hat{\theta}_x+\bar{\theta}_y\hat{z}_o\Delta \hat{\theta}_y\right)
\right]}}&&\cr \times \exp{\left[-\frac {\bar{\theta}_x^2 + 4
\hat{A}_x \hat{z}_o^2 \Delta \hat{\theta}_x^2 \hat{D}_x + 4 i
\hat{A}_x \bar{\theta}_x \hat{z}_o \Delta \hat{\theta}_x
}{2(\hat{A}_x+\hat{D}_x)}\right]} \exp{\left[ -  {  2 \hat{A}_y
\hat{z}_o^2 \Delta \hat{\theta}_y^2 }\right] }  &&\cr \times
\int_{-\infty}^{\infty} d \hat{\phi}_x
\mathrm{sinc}{\left[\frac{\hat{\phi}_x^2+(\bar{\theta}_y-\Delta
\hat{\theta}_y)^2}{4}\right]}
\mathrm{sinc}{\left[\frac{\hat{\phi}_x^2+(\bar{\theta}_y+\Delta
\hat{\theta}_y)^2}{4}\right]} ~.\label{G2Dmenodis5}
\end{eqnarray}
Finally, normalization according to Eq. (\ref{normfine2}) yields

\begin{eqnarray}
g(\hat{z}_o,\bar{\theta}_x,\bar{\theta}_y,\Delta
\hat{\theta}_x,\Delta \hat{\theta}_y) ={\exp{\left[i 2\left(
\bar{\theta}_x\hat{z}_o\Delta
\hat{\theta}_x+\bar{\theta}_y\hat{z}_o\Delta \hat{\theta}_y\right)
\right]}}&&\cr \times \exp{\left[-\frac { 2 \hat{A}_x \hat{z}_o^2
\Delta \hat{\theta}_x^2 \hat{D}_x + 2 i \hat{A}_x \bar{\theta}_x
\hat{z}_o \Delta \hat{\theta}_x }{\hat{A}_x+\hat{D}_x}\right]}
\exp{\left[ -  {  2 \hat{A}_y \hat{z}_o^2 \Delta \hat{\theta}_y^2
}\right] }  &&\cr \times {\int_{-\infty}^{\infty} d \hat{\phi}_x
\mathrm{sinc}{\left[\frac{\hat{\phi}_x^2+(\bar{\theta}_y-\Delta
\hat{\theta}_y)^2}{4}\right]}
\mathrm{sinc}{\left[\frac{\hat{\phi}_x^2+(\bar{\theta}_y+\Delta
\hat{\theta}_y)^2}{4}\right]}}&&\cr \times
\left[\int_{-\infty}^{\infty} d
\hat{\phi}_x\mathrm{sinc}^2\left\{\frac{\hat{\phi}_x^2+(\bar{\theta}_y-\Delta
\hat{\theta}_y)^2}{4}\right\}\right]^{-1/2} &&\cr\times
{\left[\int_{-\infty}^{\infty} d
\hat{\phi}_x\mathrm{sinc}^2\left\{\frac{\hat{\phi}_x^2+(\bar{\theta}_y+\Delta
\hat{\theta}_y)^2}{4}\right\}\right]^{-1/2}} ~.\label{G2Dmenodis6}
\end{eqnarray}
Obviously $|g|$ is a function of both $\bar{\theta}_y$ and
$\Delta{\theta}$, so that in this case we have neither weak
quasi-homogeneity, neither wavefronts which can be described by
Schell's model. However it should be noted that the integrals in
Eq. (\ref{G2Dmenodis6}) do not contain any parametric dependence.

It is interesting to have some final comment on Eq.
(\ref{G2Dmenodis6}). In the limit for $\hat{N}_y \ll 1$, we
recover the case discussed in Section \ref{sub:oned}, the only
difference being that we did not set $\hat{\theta}_{y1}=
\hat{\theta}_{y2}=0$: in Section \ref{sub:oned} we chose to deal
with a one-dimensional model putting ourselves on the horizontal
plane. Now, Eq. (\ref{G2Dmenodis6}) allows to study the full
two-dimensional situation in the limit for $\hat{N}_y \ll 1$. In
this case we see that the exponential function $\exp{[ -  { 2
\hat{A}_y \hat{z}_o^2 \Delta \hat{\theta}_y^2 }] }$ can be
neglected because we have a maximum value of $\Delta
\hat{\theta}_y \sim 1$. As a result we obtain:

\begin{eqnarray}
g(\hat{z}_o,\bar{\theta}_x,\bar{\theta}_y,\Delta
\hat{\theta}_x,\Delta \hat{\theta}_y) ={\exp{\left[i 2
\bar{\theta}_x\hat{z}_o\Delta \hat{\theta}_x \right]}}&&\cr \times
\exp{\left[-\frac {2 \hat{A}_x \hat{z}_o^2 \Delta \hat{\theta}_x^2
\hat{D}_x + 2 i \hat{A}_x \bar{\theta}_x \hat{z}_o \Delta
\hat{\theta}_x }{\hat{A}_x+\hat{D}_x}\right]} {\exp{\left[i
2\bar{\theta}_y\hat{z}_o\Delta \hat{\theta}_y
\right]}}\chi(\bar{\theta}_y,\Delta \hat{\theta}_y)
 ~,\label{G2Dmenodis6lim}
\end{eqnarray}
where

\begin{eqnarray}
\chi(\bar{\theta}_y,\Delta \hat{\theta}_y)=
{\int_{-\infty}^{\infty} d \hat{\phi}_x
\mathrm{sinc}{\left[\frac{\hat{\phi}_x^2+(\bar{\theta}_y-\Delta
\hat{\theta}_y)^2}{4}\right]}
\mathrm{sinc}{\left[\frac{\hat{\phi}_x^2+(\bar{\theta}_y+\Delta
\hat{\theta}_y)^2}{4}\right]}}&&\cr \times
\left[\int_{-\infty}^{\infty} d
\hat{\phi}_x\mathrm{sinc}^2\left\{\frac{\hat{\phi}_x^2+(\bar{\theta}_y-\Delta
\hat{\theta}_y)^2}{4}\right\}\right]^{-1/2} &&\cr\times
{\left[\int_{-\infty}^{\infty} d
\hat{\phi}_x\mathrm{sinc}^2\left\{\frac{\hat{\phi}_x^2+(\bar{\theta}_y+\Delta
\hat{\theta}_y)^2}{4}\right\}\right]^{-1/2}}~.\label{G2Dmenodis6lim2}
\end{eqnarray}

\begin{figure}
\begin{center}
\includegraphics*[width=140mm]{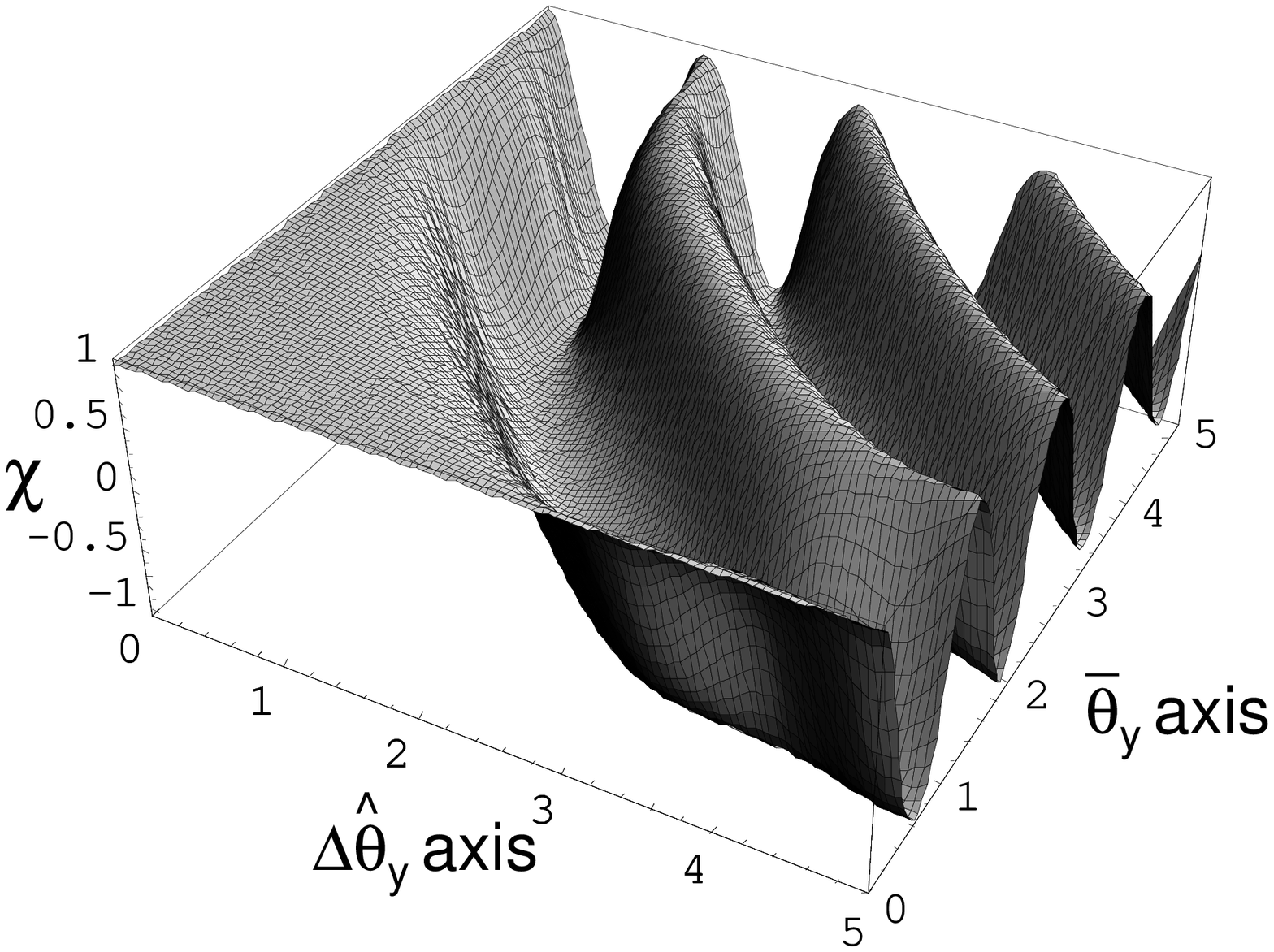}% Here is how to import EPS art
\caption{\label{3DPlot} Three-dimensional representation of $\chi$
as a function of $\bar{\theta}_y$ and $\Delta \hat{\theta}_y$.}
\end{center}
\end{figure}
\begin{figure}
\begin{center}
\includegraphics*[width=140mm]{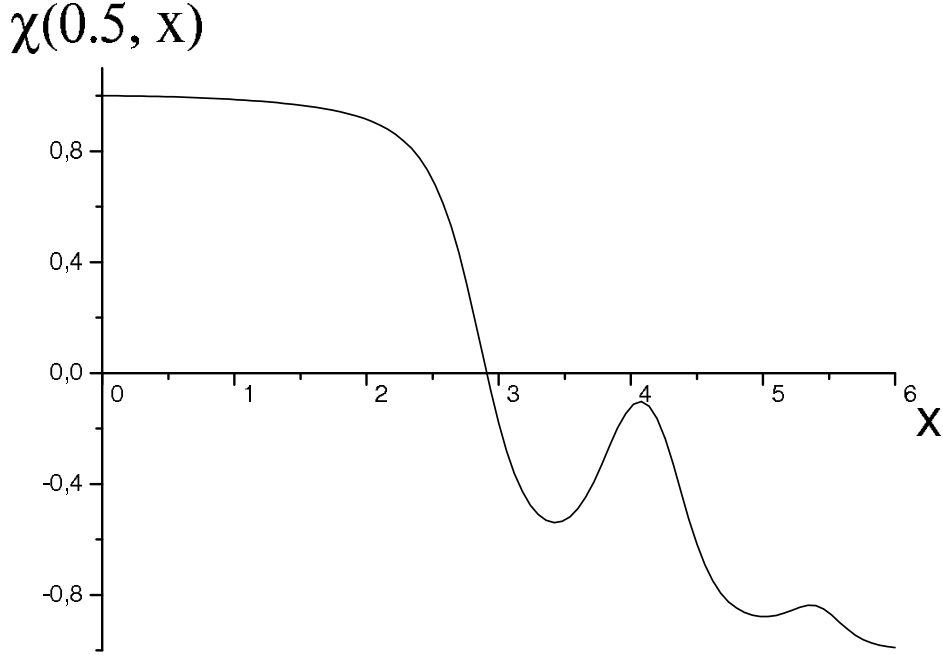}% Here is how to import EPS art
\caption{\label{2DPlotbibis} Plot $\chi(0.5,x)$, illustrating the
cut of Fig. \ref{3DPlot} at $\bar{\theta}_y=0.5$ (or fixed $\Delta
\hat{\theta}_y=0.5$). }
\end{center}
\end{figure}
\begin{figure}
\begin{center}
\includegraphics*[width=140mm]{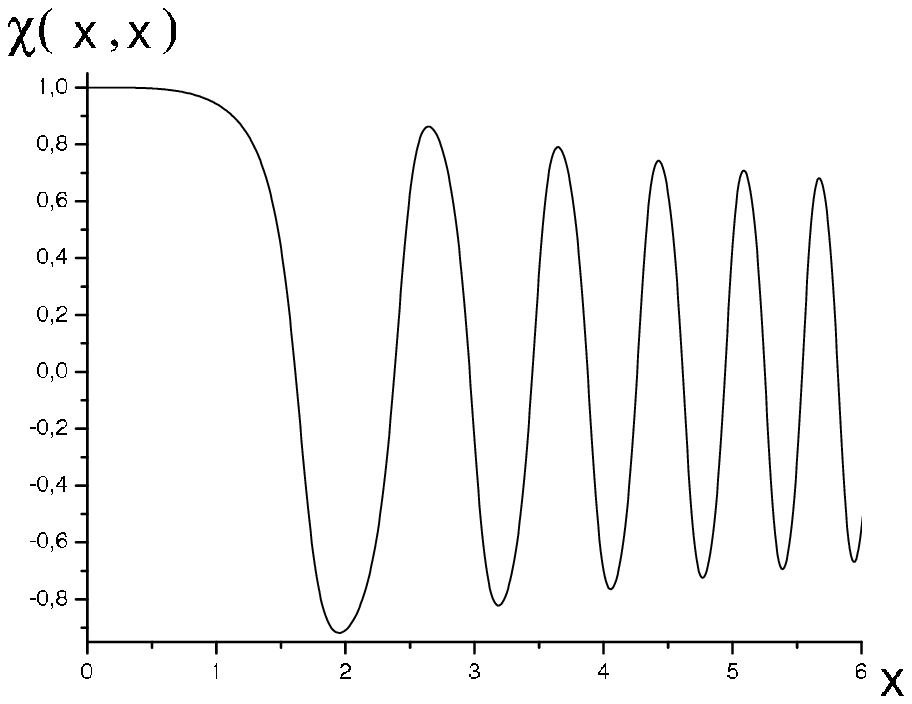}% Here is how to import EPS art
\caption{\label{2DPlottris} Plot $\chi(x,x)$, illustrating the cut
of Fig. \ref{3DPlot} at $\bar{\theta}_y=\Delta \hat{\theta}_y$. }
\end{center}
\end{figure}
\begin{figure}
\begin{center}
\includegraphics*[width=140mm]{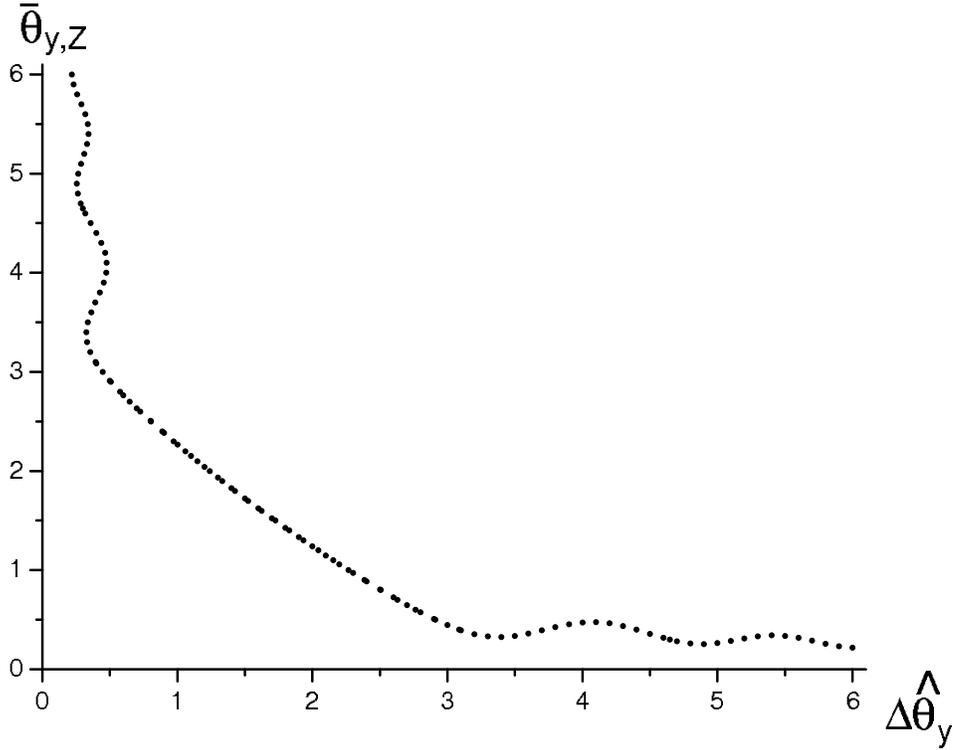}% Here is how to import EPS art
\caption{\label{2DPlot} Plot of $\bar{\theta}_{y,Z}$ as a function
of $\Delta \hat{\theta}_y$. $\bar{\theta}_{y,Z}$ are some zeros of
$\chi$, i.e. some of the values of $\bar{\theta}_{y}$ such that
$\chi(\Delta \hat{\theta}_y,\bar{\theta}_{y,Z}) = 0$.}
\end{center}
\end{figure}
\begin{figure}
\begin{center}
\includegraphics*[width=140mm]{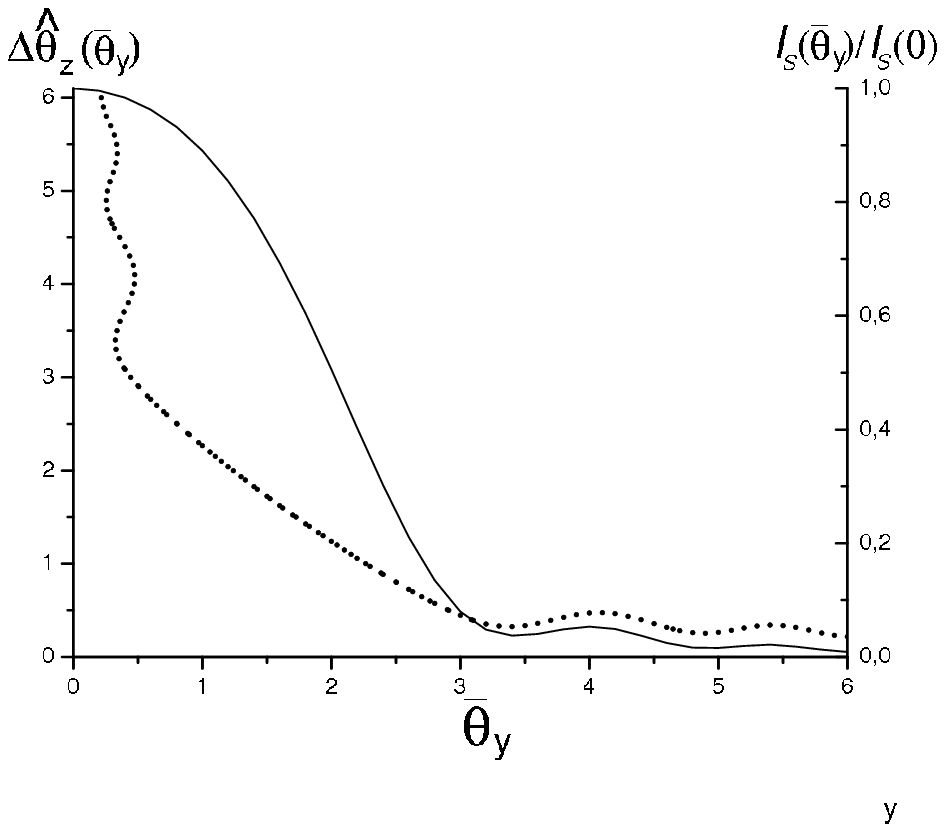}% Here is how to import EPS art
\caption{\label{plot2Dcomp} Comparison between some zeros of
$\chi$, $\Delta \hat{\theta}_{z}(\bar{\theta}_y)$ (black circles),
and the directivity diagram of undulator radiation in the vertical
direction at very large horizontal electron beam divergence
$\hat{D}_x \gg 1$ and negligible vertical divergence $\hat{D}_y
\ll 1$ (solid line).}
\end{center}
\end{figure}
It should be noted that Eq. (\ref{G2Dmenodis6lim2}) does not
depend on parameters and is, in fact, a universal function.
Besides a geometrical factor ${\exp{[i
2\bar{\theta}_y\hat{z}_o\Delta \hat{\theta}_y ]}}$, the function
$\chi$ represents the spectral degree of coherence in the vertical
direction, once the horizontal coordinates are fixed. The fact
that it is a universal function means that even in the case of
zero vertical emittance we never have full coherence in the
vertical direction. On the one hand, this phenomenon can be seen
to be an influence of the presence of horizontal emittance on the
vertical coherence properties of the photon beam, as the integral
in $d\hat{\phi}_x$ in $\chi$ comes from an integration over the
horizontal electron beam distribution. On the other hand, being
$\chi$ a universal function, the influence of the horizontal
emittance on the vertical coherence does not depend, in the limit
for $\hat{N}_x \gg 1$ and $\hat{D}_x \gg 1$, on the actual values
of $\hat{N}_x$ and $\hat{D}_x$. It is straightforward to see that
$\chi$ is symmetric with respect to $\Delta \hat{\theta}_y$ and
with respect to the exchange of $\Delta \hat{\theta}_y$  with
$\bar{\theta}_y$. When $\bar{\theta}_y =0$, i.e.
$\hat{\theta}_{y1}= -\hat{\theta}_{y2}$, we obviously obtain
$\chi(0,\Delta \hat{\theta}_y)=1$ that corresponds to complete
coherence.  In Fig. \ref{3DPlot}  we plot the three-dimensional
representation of $\chi(\bar{\theta}_y,\Delta \hat{\theta}_y)$. In
order to get a feeling for the behavior of $\chi$ we also plot, in
Fig. \ref{2DPlotbibis} and Fig. \ref{2DPlottris}, two cuts of Fig.
\ref{3DPlot} illustrating, respectively, the behavior of $\chi$
for a fixed $\bar{\theta}_y=0.5$ (or fixed $\Delta
\hat{\theta}_y=0.5$) and at $\bar{\theta}_{y}=\Delta
\hat{\theta}_y$.

As it is evident from Fig. \ref{3DPlot}, $\chi$ exhibits, for any
fixed value of $\Delta \hat{\theta}_y$, many different zeros in
$\bar{\theta}_y$. In Fig. \ref{2DPlot} we illustrate some of these
zeros as a function of $\Delta \hat{\theta}_y$,
$\bar{\theta}_{y,Z}(\Delta \hat{\theta}_y)$.  The interest of this
plot is that, once a certain distance $\hat{z}_o \Delta
\hat{\theta}_y$ between two pinholes is fixed, it illustrates at
what position of the pinhole system, $\bar{\theta}_{y,Z}$, the
spectral degree of coherence drops from unity to zero for the
first time.

It is interesting to compare Fig. \ref{2DPlot}, with the
directivity diagram of the radiant intensity
$I_S(\bar{\theta}_y)$. This comparison is shown in Fig.
\ref{plot2Dcomp}. In the limit for $\hat{N}_y \ll 1$ and
$\hat{D}_y \ll 1$, one may increase the degree of coherence of the
beam by spatially filtering the radiation in the far field. If a
vertical slit is used with aperture $d$ much larger than the
horizontal coherence length, i.e. $d \gg \hat{\xi}_{cx}$, one
would have poor coherence. Decreasing the aperture of the slit
will increase the coherence of the X-ray beam up to some value $d$
smaller than $\hat{\xi}_c$. Within our assumption $\hat{N}_x \gg
1$ one has the far field approximation $\hat{\xi}_{cx} =
(\pi/\hat{N}_x)^{1/2} \hat{z}_o$. When $d$ becomes smaller and
smaller with respect to $(\pi/\hat{N}_x)^{1/2} \hat{z}_o$ the
spectral degree of coherence $g$ can be identified with the
universal function $\chi$, as once can see by inspecting Eq.
(\ref{G2Dmenodis6lim}). As a result, as $d$ becomes smaller one
loses photons, but the X-ray beam transverse coherence ceases to
improve because, as is seen in Fig. \ref{plot2Dcomp}, the
transverse degree of coherence $g = \chi$ drops to zero along the
vertical radiation pattern of the filtered X-ray beam: for
instance, from Fig. \ref{plot2Dcomp} one can see that $\chi$ drops
to zero for the first time at $\Delta\hat{\theta}\sim1$
$\bar{\theta}_y \sim 2$, where the X-ray flux is still intense.
This behavior of the degree of coherence should be taken into
account at the stage of planning experiments. To give an example,
after spatial filtering, one may conduct a two-pinhole experiment
(like the one illustrated in Fig. \ref{ed12c}) and find,
surprisingly, that for some vertical position $\bar{\theta}_y$ of
the pinholes (at fixed $\Delta \hat{\theta}_y$) well within the
radiation pattern diagram he will have no fringes, but for some
other vertical position he can find perfect visibility. So,
without the knowledge of the function $\chi$ a user would not even
have the possibility to predict the outcomes of a simple
two-pinhole experiment.

From the definitions of $\chi$, $\beta$, $\gamma$ and $I_S$ it can
be seen that all universal functions introduced in this work are
partial cases of the more generic

\begin{eqnarray}
M(\bar{\theta}_y,\Delta \hat{\theta}_y)= {\int_{-\infty}^{\infty}
d \hat{\phi}_x
\mathrm{sinc}{\left[\frac{\hat{\phi}_x^2+(\bar{\theta}_y-\Delta
\hat{\theta}_y)^2}{4}\right]}
\mathrm{sinc}{\left[\frac{\hat{\phi}_x^2+(\bar{\theta}_y+\Delta
\hat{\theta}_y)^2}{4}\right]}}~.\label{Motherofall}
\end{eqnarray}
In fact

\begin{eqnarray}
\chi(\bar{\theta}_y,\Delta \hat{\theta}_y)=
\frac{M(\bar{\theta}_y,\Delta \hat{\theta}_y)}{\left[M
(\bar{\theta}_y+\Delta\hat{\theta}_y,0)\right]^{1/2}\left[
M(\bar{\theta}_y-\Delta\hat{\theta}_y,0) \right]^{1/2}}&&\cr
I_S(\bar{\theta}_y) = M(\bar{\theta}_y,0) && \cr \beta(\Delta
\hat{\theta}_y) = \frac{1}{2\pi^2} \int_{-\infty}^{\infty} d \xi
M(\xi, \Delta\hat{\theta}_y) && \cr \gamma(x) = \frac{1}{2\pi^2}
\int_{-\infty}^{\infty} d \xi  \exp[- i (2 x) \xi] M(\xi,0)
~.\label{childrenofM}
\end{eqnarray}
\begin{figure}
\begin{center}
\includegraphics*[width=140mm]{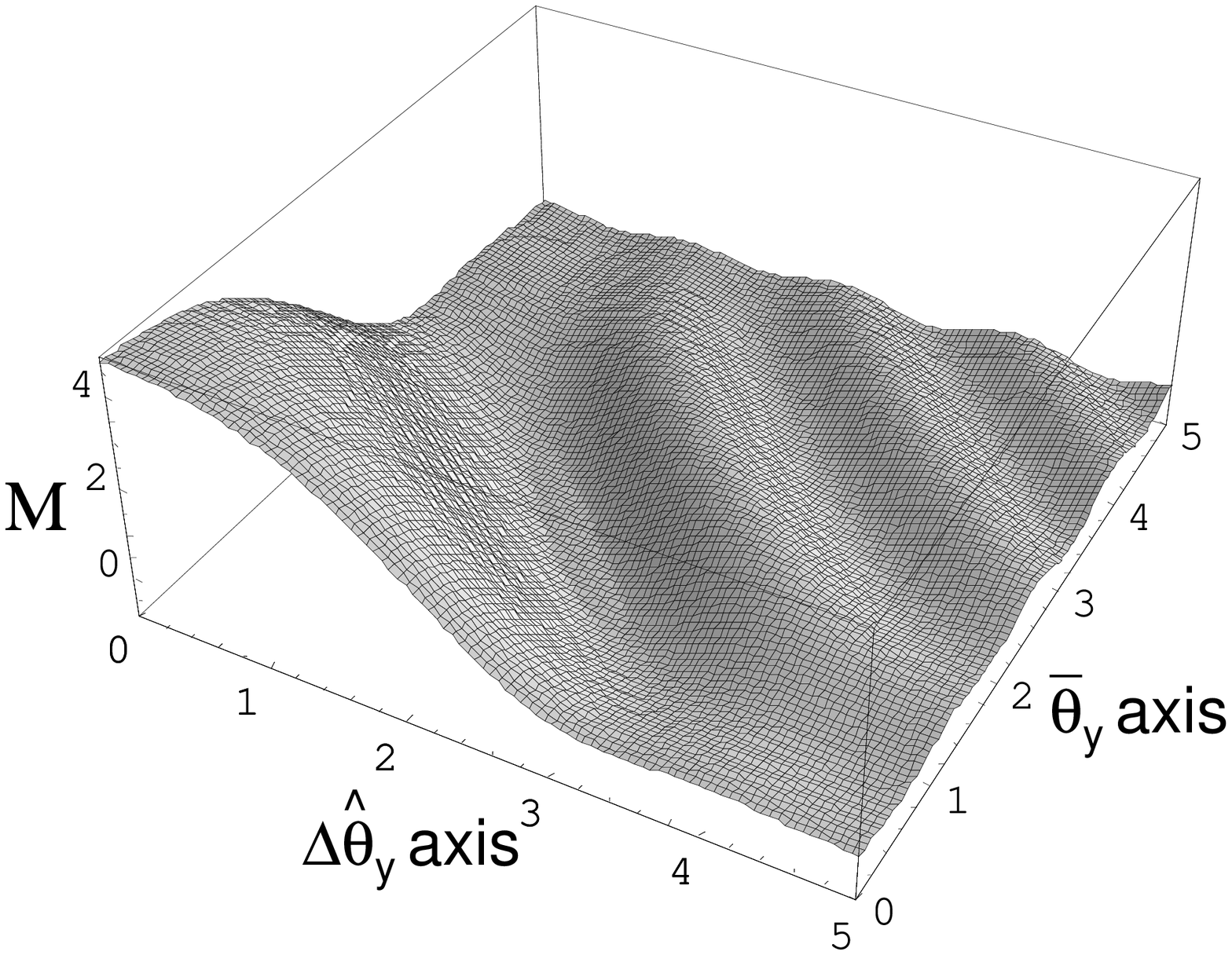}% Here is how to import EPS art
\caption{\label{Mother} Three-dimensional representation of $M$ as
a function of $\bar{\theta}_y$ and $\Delta \hat{\theta}_y$.}
\end{center}
\end{figure}
The knowledge of $M$ is all one needs to calculate coherence
properties out of many experimental setups, in very practical
situations. It is therefore worth to tabulate $M$. We present a 3D
plot of $M$, obtained from such tabulation, in Fig. \ref{Mother}.

Finally, it is interesting to sum up and compare results for the
far field region obtained in this Section (non-homogeneous
undulator source) with results obtained in Section \ref{sec:twod}.
Many users performing coherent experiments with X-ray beams are
interested in the beam coherence properties in the far field. We
have seen that in the most general situation for third generation
light sources one is interested in the case $\hat{N}_x \gg 1$ and
$\hat{D}_x \gg 1$, which guarantees factorization of results in
the $x$ and $y$ direction, with arbitrary $\hat{N}_y$ and
$\hat{D}_y$.  In this case we will not have, in general, weakly
quasi-homogeneous radiation in  the vertical direction. The
spectral degree of coherence can be found by simplifying Eq.
(\ref{G2Dnewsimplif2}) in the mathematical limit $\hat{z}_o
\longrightarrow \infty$:

\begin{eqnarray}
g(\hat{z}_o,\bar{\theta}_x,\bar{\theta}_y,\Delta
\hat{\theta}_x,\Delta \hat{\theta}_y) = {\exp{\left[i 2\left(
\bar{\theta}_x\hat{z}_o\Delta
\hat{\theta}_x+\bar{\theta}_y\hat{z}_o\Delta \hat{\theta}_y\right)
\right]}}&&\cr \times  \exp{\left[-2\hat{N}_x \Delta
\hat{\theta}_x^2 \right]}\exp{\left[-2\hat{N}_y  \Delta
\hat{\theta}_y^2 \right]} \int_{-\infty}^{\infty} d \hat{\phi}_y
\exp{\left[-\frac{(\bar{\theta}_y+\hat{\phi}_y)^2}{2
\hat{D}_y}\right]}M(\hat{\phi}_y,\Delta \hat{\theta}_y) && \cr
\times \left\{\int_{-\infty}^{\infty} d \hat{\phi}_y
\exp{\left[-\frac{(\hat{\phi}_y+\bar{\theta}_y+\Delta
\hat{\theta}_y)^2}{2
\hat{D}_y}\right]}I_S(\hat{\phi}_y)\right\}^{-1/2}&&\cr \times
\left\{\int_{-\infty}^{\infty} d \hat{\phi}_y
\exp{\left[-\frac{(\hat{\phi}_y+\bar{\theta}_y-\Delta
\hat{\theta}_y)^2}{2
\hat{D}_y}\right]}I_S(\hat{\phi}_y)\right\}^{-1/2} ~
.\label{resu1}
\end{eqnarray}
We can see that for any value of $\hat{N}_y$ and $\hat{D}_y$, in
the far field limit we obtain a contribution to the cross-spectral
density for the $x$ and for the $y$ direction. The contribution
for the $y$ direction can be expressed in terms of the product of
an exponential function and  convolutions between the (Gaussian)
electron beam divergence and  universal functions.

On the one hand, as $\hat{D}_y \ll 1$ we have

\begin{eqnarray}
g(\hat{z}_o,\bar{\theta}_x,\bar{\theta}_y,\Delta
\hat{\theta}_x,\Delta \hat{\theta}_y) = {\exp{\left[i 2\left(
\bar{\theta}_x\hat{z}_o\Delta
\hat{\theta}_x+\bar{\theta}_y\hat{z}_o\Delta \hat{\theta}_y\right)
\right]}}&&\cr \times  \exp{\left[-2\hat{N}_x  \Delta
\hat{\theta}_x^2 \right]}\exp{\left[-2\hat{N}_y \Delta
\hat{\theta}_y^2 \right]}\chi(\bar{\theta}_y,\Delta
\hat{\theta}_y)~.\label{resu2}
\end{eqnarray}
On the other hand, as $\hat{D}_y \gg 1$ we have weakly
quasi-homogeneous wavefronts and

\begin{eqnarray}
g(\hat{z}_o,\bar{\theta}_x,\bar{\theta}_y,\Delta
\hat{\theta}_x,\Delta \hat{\theta}_y) = {\exp{\left[i 2\left(
\bar{\theta}_x\hat{z}_o\Delta
\hat{\theta}_x+\bar{\theta}_y\hat{z}_o\Delta \hat{\theta}_y\right)
\right]}}&&\cr \times \exp{\left[-2\hat{N}_x  \Delta
\hat{\theta}_x^2 \right]}\exp{\left[-2\hat{N}_y \Delta
\hat{\theta}_y^2 \right]}\beta(\Delta
\hat{\theta}_y)~\label{resu2bissss}
\end{eqnarray}
where $\beta(\Delta \hat{\theta}_y)$ is given in Eq.
(\ref{gtildedue}).

Moreover, as $\hat{N}_y \gg 1$, and for arbitrary $\hat{D}_y$ we
have:

\begin{eqnarray}
g(\hat{z}_o,\bar{\theta}_x,\bar{\theta}_y,\Delta
\hat{\theta}_x,\Delta \hat{\theta}_y) = {\exp{\left[i 2\left(
\bar{\theta}_x\hat{z}_o\Delta
\hat{\theta}_x+\bar{\theta}_y\hat{z}_o\Delta \hat{\theta}_y\right)
\right]}}&&\cr \times\exp{\left[-2\hat{N}_x  \Delta
\hat{\theta}_x^2 \right]}\exp{\left[-2\hat{N}_y \Delta
\hat{\theta}_y^2 \right]}~.\label{resu3}
\end{eqnarray}
It should be noted that Eq. (\ref{resu3}) is simply a consequence
of the application of the VCZ theorem in both horizontal and
vertical directions.

\section{\label{sec:spot} Application: Coherent X-ray beam expander scheme}

In this Section we show how transverse coherence properties of an
X-ray beam can be manipulated to obtain a larger coherent
spot-size on a sample.

\begin{figure}
\begin{center}
\includegraphics*[width=140mm]{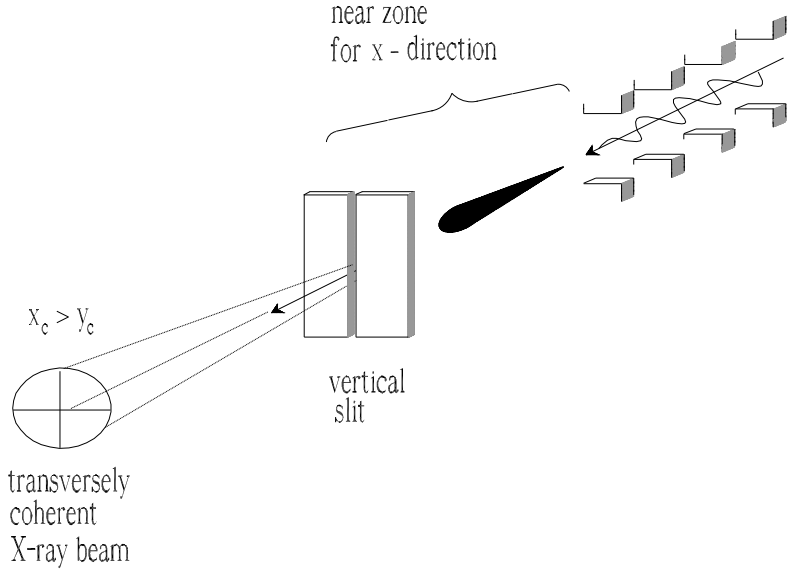}% Here is how to import EPS art
\caption{\label{ed1} Undulator radiation with a coherent X-ray
beam expander.}
\end{center}
\end{figure}

The idea of increasing the horizontal width of the coherence spot
is based on the use of a downstream slit for selection of the
transversely coherent fraction of undulator radiation. Imagine a
slit very close to the exit of the undulator with an aperture $d$
comparable with the coherent length of the radiation at the exit
of the undulator $\hat{\xi}_{c x} = \sqrt{\pi/\hat{D}_x}$, as
illustrated in Fig. \ref{ed1}. The new radiation source after the
slit is now coherent and characterized by a horizontal dimension
of the light spot equal to $\sqrt{\pi/\hat{D}_x}$. In the far
field one can  take advantage of the reciprocal width relations of
Fourier transform pairs or, equivalently, the expression for the
Fraunhoffer diffraction pattern from a slit, i.e. a
$\mathrm{sinc}$ function, to calculate the magnitude of the
coherence spot. There is, of course, some arbitrary convention to
agree upon when it comes to the definition of the width of the
$\mathrm{sinc}$ function but, numerical factors aside, this
reasoning shows qualitatively that the coherence spot is of order
$\sqrt{\hat{D}_x}\hat{z}_o$ which is $\hat{\epsilon}_x$ times
larger than the spot size dimension in the case of free-space
propagation, of order $\hat{z}_o/\sqrt{\hat{N}_x}$. The radiation
beyond the slit must be then spectrally filtered by a
monochromator (not shown in Fig. \ref{ed1}) to further narrow the
spectral bandwidth. Here we assume that the radiation frequency
$\omega$ is equal to the fundamental frequency $\omega_o$. The
radiation beyond the slit is transversely coherent when the
aperture $d$ is equal (at most) to the coherence length $\xi_{c
x}$.

\begin{figure}
\begin{center}
\includegraphics*[width=140mm]{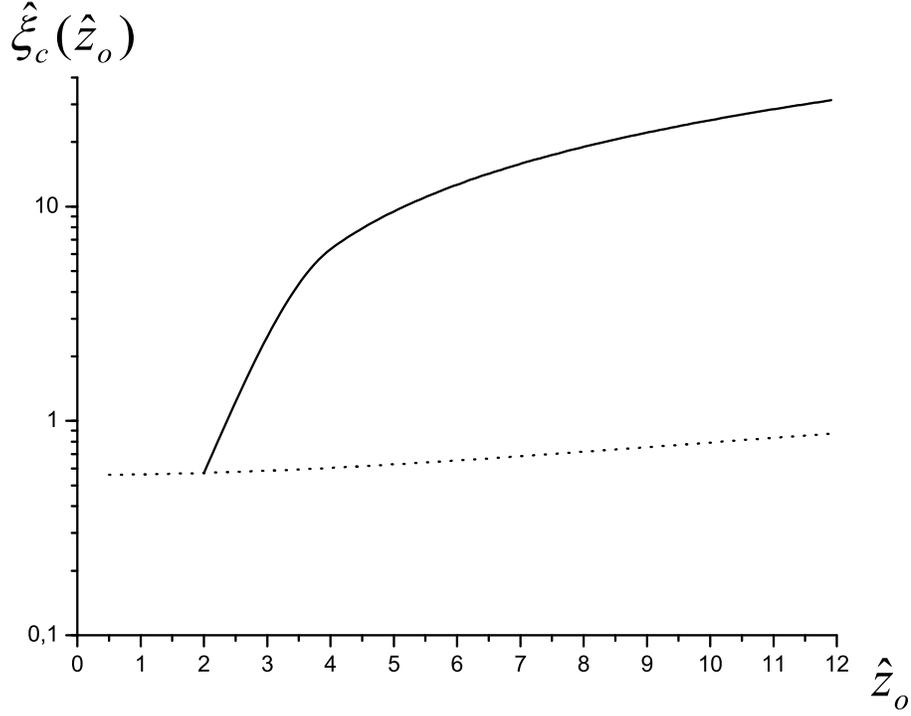}% Here is how to import EPS art
\caption{\label{blog} Behavior of the horizontal coherence length
as a function of distance for free space (dashed line) and
comparison with asymptotic behaviors near the slit and at large
distances (solid lines). }
\end{center}
\end{figure}

Let us present a numerical example illustrating the improvement of
the horizontal coherence length obtained by slit application. Let
us consider the case when the electron horizontal emittance is
large $\hat{\epsilon}_x \gg 1$ and the vertical emittance is small
$\hat{\epsilon}_y \ll 1$, with $\hat{N}_x \gg \hat{D}_x$. Since
$\hat{N}_x = \hat{\epsilon}_x \hat{\beta}_x$ and $\hat{D}_x =
\hat{\epsilon}_x / \hat{\beta}_x$, this is the case, for instance
when $\hat{\beta}_x \simeq 10$. Then, for $\hat{\epsilon}_x = 100$
we have $\hat{N}_x = 1000$ and $\hat{D}_x = 10$. This particular
numerical example has been considered in Section \ref{sub:oned} to
illustrate, in free space, the behavior of the coherence length as
a function of the position along the beamline, given in Eq.
(\ref{cohlen2}). Suppose we install the slit at $\hat{z}_o = 2$.
From Eq. (\ref{cohlen2}) we have $\hat{\xi}_{cx \mid \hat{z}_o =
2} \sim \sqrt{\pi/\hat{D}_x} \sim 0.6$. At $\hat{z}_o = 12$ we
have a situation close to the asymptotic behavior where the field
diffracted by the slit can be treated in the Fraunhofer
approximation and $\hat{\xi}_{cx \mid \hat{z}_o  = 12} \sim
\sqrt{\hat{D}_x} \hat{z}_o \sim 30$. Comparison of the asymptotic
behaviors after spatial filtering with respect to the free-space
propagation case is given in Fig. \ref{blog}. Note that only the
asymptotic behaviors near the slit and at large distance
$\hat{z}_o \gg 1$ are plotted for the spatially filtered
radiation. What is important is, in fact, the comparison between
the coherent distance at large values of $\hat{z}_o$ with and
without spatial filtering in the near zone.

This is an example in which evolution of transverse coherence
through the beam line plays an important role. In fact, the
ability of spatially filter radiation by a slit requires the
knowledge of the transverse coherence length variation along the
beamline.

Let us calculate the number of coherent photons observed beyond
the aperture. In the region of parameters where $\hat{N}_x \gg 1$
and $\hat{D}_x \gg 1$, the number of transversely coherent photons
into the slit aperture $d = \xi_{c x}$ can be calculated as

\begin{equation}
(N_{\mathrm{ph}})_\mathrm{coh} = \frac{d N_\mathrm{ph}}{d x}
\xi_{c x}~.\label{num1}
\end{equation}
In the near-zone limit the slit is positioned at a position down
the beamline $z_s \simeq \beta_x$ so that, from Eq.
(\ref{cohlen2}) and Eq. (\ref{Gzlarge9gen}) we have:

\begin{equation}
\frac{d N_\mathrm{ph}}{d x}=\frac{N_\mathrm{ph}}{\sqrt{2\pi
\sigma_{x}^2}}~\label{num2}
\end{equation}
and

\begin{equation}
\xi_{c x}={\sqrt{\pi}}\frac{\lambda}{2\pi \sigma_{x'}}
~.\label{num22}
\end{equation}

Therefore we can write

\begin{equation}
\frac{d N_\mathrm{ph}}{d x} \xi_{c x}={\sqrt{\pi}}\frac{\lambda
N_\mathrm{ph} }{\sqrt{(2\pi)^3 \sigma_{x}^2
\sigma_{x'}^2}}~.\label{num1b}
\end{equation}
At the opposite extreme, with the distance $z_s$ much larger than
$\beta_x$ we find that, always from Eq. (\ref{cohlen2}) and Eq.
(\ref{Gzlarge9gen})

\begin{equation}
\frac{d N_\mathrm{ph}}{d x}=\frac{N_\mathrm{ph}}{\sqrt{2\pi z_s^2
\sigma_{x'}^2}} \label{num2b}
\end{equation}
and

\begin{equation}
\xi_{c x} = {\sqrt{\pi}} \frac{\lambda z_s}{2\pi \sigma_{x}}~.
\label{num2bb}
\end{equation}
Therefore we can write

\begin{equation}
\frac{d N_\mathrm{ph}}{d x} \xi_{c x}={\sqrt{\pi}}\frac{\lambda
N_\mathrm{ph} }{\sqrt{(2\pi)^3 \sigma_{x}^2
\sigma_{x'}^2}}~.\label{num3b}
\end{equation}
Thus, the number of transversely coherent photons into the slit
aperture $d=\xi_{c x}$ will be independent of the distance $z_s$.
This means that operation of spatial filtering in the near field
region, as proposed by us, will not diminish the  number of
coherent photons with respect to the usual practice in which
spatial filtering to obtain coherent radiation is performed in the
far field: its effect will only be that of increasing the
horizontal dimension of the coherent spot size.  This possibility
to create transversely coherent radiation with large divergence in
the horizontal direction is important for many experiments. The
distance of the slit from the exit of the undulator sets a limit
to achievable linear dimension of coherence area. If we build a
coherent X-ray beam line, we want to have large linear dimension
of coherence area at the specimen position.  Therefore, in order
to have a largest linear dimension of coherence area at the
specimen position $\xi_{c x}(z_o)\simeq\sigma_{x'} z_o$ we must
have a slit aperture of at most of the size $d \simeq
\lambda/(2\pi \sigma_{x'})$ installed in the near zone at $z_s
\simeq \beta_x$. In order not to loose coherent photons instead,
the slit aperture must be at least of the size $d \simeq
\lambda/(2\pi \sigma_{x'})$. The right compromise is thus a slit
aperture of the size $d \simeq \lambda/(2\pi \sigma_{x'})$
installed in the near zone at $z_s \simeq \beta_x$.

\section{\label{sec:concl} Conclusions}

Before this work, no satisfactory theory describing spatial
coherence from undulator radiation sources has been built. In this
paper we developed such a theory of transverse coherence dealing
with X-ray beams, with particular attention to third generation
light sources.

First we studied Synchrotron Radiation as a random statistical
process using the language of Statistical Optics. Statistical
Optics developed around Gaussian, stationary processes
characterized by quasi-homogeneous sources; under these
assumptions, the characterization of statistical properties of the
process are greatly simplified and the van Cittert-Zernike theorem
(or its generalized version) can be used in order to describe the
X-ray beam partial coherence properties in the far field region.
However, for Synchrotron Radiation, there is no \textit{a priori}
reason to hold these assumptions satisfied.

We showed that Synchrotron Radiation is a Gaussian random process.
As a result, statistical properties of Synchrotron Radiation are
described satisfactory by second-order field correlation
functions. We used a frequency domain analysis to describe them
from a mathematical viewpoint. This choice is very natural. In
fact, up-to-date detectors are limited to about $100$ ps time
resolution: therefore, in real-life experiments with third
generation light sources, detectors are by no means able to
resolve a single X-ray pulse in time domain and work, instead, by
counting the number of photons at a certain frequency over an
integration time longer than the radiation pulse. As a consequence
of the frequency domain analysis we could study the spatial
correlation for a given frequency content using the cross-spectral
density of the system which, independently of the spectral
correlation function, can be used to extract useful information
even if the process is not stationary.

We gave an expression for the process cross-spectral density
dependent on six dimensionless parameters. Subsequently we tuned
parameters at perfect resonance, thus obtaining a simplified
expression.

First we studied the limit of applicability of the
quasi-homogeneous model from an analytical viewpoint, within the
framework of simplifying assumptions, namely in the limit of small
electron beam divergence and Fresnel number in the vertical
direction, of large electron beam divergence and Fresnel number in
the horizontal direction, and performing calculations for the
cross-spectral density on the horizontal plane only. This
simplified study allowed us to introduce the concept of weakly
quasi-homogeneous radiation and virtual quasi-homogeneous source
while discussing the applicability region of the VCZ theorem.

Second, we studied the effect of the vertical emittance on the
cross-spectral density. This study led us to analyze both cases in
which the sources are weakly quasi-homogeneous and cases when they
are not quasi-homogeneous at all. In the limit for large
horizontal beam divergence and Fresnel number, which is always
satisfied for third generation light sources, we found that the
spectral degree of coherence factorizes in the product of factors
depending separately on the horizontal and on the vertical
coordinates. In the far field limit the vertical part of the
spectral degree of coherence can be expressed in terms of the
product of an exponential function (which, alone, would simply
satisfy the VCZ theorem)  and convolutions between the electron
beam divergence in the vertical direction and a universal
functions, that we introduced in our work. The universality of
such a function implied that even for zero vertical emittance we
never have full coherence in the vertical direction. This
unexpected result is due to the influence of the horizontal
emittance on the vertical coherence properties of the photon beam.
Because of this, the degree of coherence changes between zero and
unity within the diffraction angle. We also studied the near field
zone. When one is interested in the evolution of the degree of
coherence along the beamline back up to the exit of the undulator
the situation becomes much more complicated with respect to the
far zone case, as the observation distance is one of the problem
parameters. There are many more asymptotic situations which can be
studied, and a large part of our paper is devoted to the
calculation of these asymptotic situations. We provided
approximate estimations for the vertical coherence length that are
valid from the far zone and back, up to the exit of the undulator
in the case when either the vertical Fresnel number or the
vertical electron beam divergence are much larger than unity.
These can be used at the stage of planning experiments.

It should be noted that, throughout this work, we did not discuss
the accuracy of the approximation of small and large parameters.
In order to do so, one needs to develop a perturbation theory for
each asymptotic case studied here, which would considerably
increase the size of this paper. As a result we leave this issue
for future work.

%It should be noted, though that when the only vertical Fresnel
%number is much larger than unity is arbitrary, our approximate
%formula is valid for any average observation position only in the
%near and in the far zone, because there is a transition region
%between these two regions where the radiation is not weakly
%quasi-homogeneous. On the contrary, when the vertical electron
%beam divergence is much larger than unity, our approximated
%relation is always valid because, in this case, radiation is
%weakly quasi-homogeneous along all the beamline .

Finally, we selected an application to show the power of our
approach. We discussed how the transverse coherence properties of
an X-ray beam can be manipulated to obtain a convenient coherent
spot-size on the sample with the help of a simple vertical slit;
this invention was predicted almost entirely on the basis of
theoretical ideas of rather complex and abstract nature discussed
in the previous parts of the paper.

\newpage

\section*{Appendix A: Random phasor sum}

The field of thermal light can be regarded as a sum of a great
many independent contributions. The complex envelope of polarized
thermal light at fixed time and a fixed point in space is a sum of
a very large number of complex phasors

\begin{equation}
E(\vec{r},t) = \sum_{k=1}^N \alpha_k e^{i \psi_k}
~,\label{sumrand}
\end{equation}
where $N$ is the number of radiating atoms. Statistical properties
of elementary phasors that are generally satisfied in thermal
light problems of interest are as follows:

a) The amplitudes $\alpha_k = Re(\alpha_k)$ and the phases
$\psi_k$ are statistically independent of each other and of the
amplitudes and phases of all other elementary phases for different
values of $k$.

b) The random variables $\alpha_k$ are identically distributed for
all $k$ with mean value $<\alpha>$ and second moment $<\alpha^2>$.

c) The phases $\psi_k$ are uniformly distributed over the interval
$(0, 2\pi)$.

The reader can find in \cite{GOOD} that when assumptions from a)
to c) are satisfied, the real ($Re(E)$) and imaginary ($Im(E)$)
parts of the field are distributed in accordance with the Gaussian
law in the limit for $N \longrightarrow \infty$, so that

\begin{eqnarray}
\left\langle Re(E) \right\rangle =\left\langle Im(E) \right\rangle
= 0~, && \cr \left\langle [Re(E)]^2 \right\rangle =\left\langle
[Im(E)]^2 \right\rangle = \frac{\left\langle \alpha^2
\right\rangle}{2}N = \sigma^2~,&& \cr \left\langle Re(E) Im(E)
\right\rangle = 0~, && \cr p(Re(E),Im(E))=\frac{1}{2\pi \sigma^2}
\exp{\left[-\frac{[Re(E)]^2+[Im(E)]^2}{2\sigma^2}\right]}
\label{Gausslaw}~,
\end{eqnarray}
where $p(Re(E),Im(E))$ is the joint probability density function.

In Section \ref{sub:def} we discussed statistical properties of
Synchrotron Radiation and we were led to assumptions 1), 2) and 3)
which are weaker than a), b) and c). Here we will demonstrate that
assumptions from a) to c) can be relaxed to assumptions from 1) to
3) without changes in results. We will derive results valid when
the amplitudes $\alpha_k$ are complex $\alpha_k = |\alpha_k|
\exp{(i \phi_k)}$.
%and phases $\phi_k$ take an arbitrary
%probability density function $P_\phi(\phi)$, while remaining
%identically distributed and independent.

After denoting with $r$ and $i$ the real and imaginary parts of
the fields and after substituting notation $\langle Q \rangle$
with $\bar{Q}$ we first demonstrate that $\bar{r}=\bar{i}=0$. We
have straightforwardly

\begin{eqnarray}
\bar{r} = \frac{1}{N}\sum_{k=1}^{N}\Bigg(
\overline{\mid\alpha_k\mid \cos \phi_k}~ {\overline{\cos \psi_k}}
- \overline{\mid\alpha_k\mid \sin \phi_k} ~\overline{\sin
\psi_k}\Bigg)  = 0~, && \cr \bar{i} =
\frac{1}{N}\sum_{k=1}^{N}\Bigg( \overline{\mid\alpha_k\mid \cos
\phi_k}~ \overline{\sin \psi_k} + \overline{\mid\alpha_k\mid \sin
\phi_k} ~\overline{\cos \psi_k} \Bigg) = 0~, \label{mediaRI}
\end{eqnarray}
because all averages over trigonometric functions are zero.

Second, we demonstrate that $\overline{r^2}=\overline{i^2} =
\overline{\alpha^2}/2$. Again, direct calculation shows

\begin{eqnarray}
\bar{r^2} = \frac{1}{N}\sum_{k,n=1}^{N} \Bigg(
{\overline{\mid\alpha_k\mid \mid\alpha_n \mid \cos \phi_k \cos
\phi_n}} ~{\overline{\cos \psi_k \cos \psi_n}}  &&\cr +
{\overline{\mid\alpha_k\mid \mid\alpha_n \mid \sin \phi_k \sin
\phi_n}} ~{\overline{\sin \psi_k \sin \psi_n}} &&\cr -
{\overline{\mid\alpha_k\mid \mid\alpha_n \mid \cos \phi_k \sin
\phi_n}} ~{\overline{\cos \psi_k \sin \psi_n}} && \cr -
{\overline{\mid\alpha_k\mid \mid\alpha_n \mid \sin \phi_k \cos
\phi_n}} ~{\overline{\sin \psi_k \sin \psi_n}}\Bigg)&& \cr =
\frac{1}{2 N}\sum_{k=1}^{N} \overline{\mid\alpha_k\mid^2
\left(\cos^2 \psi_k+\sin^2
\psi_k\right)}=\frac{\overline{\alpha^2}}{2}~. \label{mediaR2}
\end{eqnarray}
Moreover it is easy to see that $\overline{i^2} = \overline{r^2}$.

Finally, we show that $\overline{ri} = 0$. In fact

\begin{eqnarray}
\overline{r i} = \frac{1}{N} \sum_{k,n=1}^{N} \Bigg(
\overline{\mid\alpha_k\mid\mid\alpha_n\mid \cos \phi_k \cos
\phi_n} ~ {\overline{\cos \psi_k \sin \psi_n}} &&\cr +
\overline{\mid\alpha_k\mid\mid\alpha_n\mid \cos \phi_k \sin
\phi_n} ~ {\overline{\cos \psi_k \cos \psi_n}} &&\cr
-\overline{\mid\alpha_k\mid\mid\alpha_n\mid \sin \phi_k \cos
\phi_n} ~ {\overline{\sin \psi_k \sin \psi_n}} &&\cr
-\overline{\mid\alpha_k\mid\mid\alpha_n\mid \sin \phi_k \sin
\phi_n} ~ {\overline{\sin \psi_k \cos \psi_n}} \Bigg)
 &&\cr  = \frac{1}{2 N} \sum_{k}
\overline{{|\alpha_k|^2} \left({{\cos \phi_k \sin\phi_k}}-{{\sin
\phi_k \cos\phi_k}}\right)} = 0~.\label{mediafinale}
\end{eqnarray}
As a result we have that real and imaginary parts have zero means,
equal variances and are uncorrelated. Use of the central limit
theorem allows to conclude that the resulting phasor sum is a
circular complex Gaussian random variable.

\newpage

\section*{Appendix B: A useful transformation of the expression for the undulator radiation field}

We start reporting here, for convenience, Eq. (\ref{undunormfin}),
that represents the field (in normalized units) produced by a
particle with offset and deflection at any distance $\hat{z}_o
\geqslant 1/2$ from the exit of the undulator, where the undulator
center is taken at $\hat{z}_o=0$:

\begin{eqnarray}
\hat{E}_{s\bot}=  \hat{z}_o  \int_{-1/2}^{1/2} d\hat{z}'
\frac{1}{\hat{z}_o-\hat{z}'} \exp \left\{i
\left[\left(\hat{C}+\frac{\left.\vec{\hat{\eta}}\right.^2}{2}\right)\hat{z}'
+ \frac{\left(\vec{\hat{{r}}}_{\bot
o}-\vec{{\hat{l}}}-\vec{\hat{\eta}}\hat{z}' \right)^2 }{2
(\hat{z}_o-\hat{z}')}\right] \right\} .~ \label{undunormfinapp}
\end{eqnarray}
In this Appendix we show that $\hat{E}_{s\bot}$ as reported in Eq.
(\ref{undunormfinapp}) may be described as

\begin{eqnarray}
\hat{E}_{s\bot} =   \int_{-1/2}^{1/2} \frac{ \hat{z}_o  d\hat{z}'
}{\hat{z}_o-\hat{z}'} \exp \left\{i \left[\Phi_U +\hat{C} \hat{z}'
+ \frac{\hat{z}_o \hat{z}'}{2 (\hat{z}_o-\hat{z}')}
\left(\vec{\hat{\theta}}- \frac{\vec{\hat{l}}}{\hat{z}_o}-
\vec{\hat{\eta}}\right)^2 \right] \right\}
\label{undunormfinappultini}
\end{eqnarray}
where $\Phi_U$ is given by

\begin{equation}
\Phi_U =
\left[\left(\hat{\theta}_x-\frac{\hat{l}_x}{\hat{z}_o}\right)^2
+\left(\hat{\theta}_y-\frac{\hat{l}_y}{\hat{z}_o}\right)^2\right]
\frac{\hat{z}_o}{2} ~.\label{phisnormapp}
\end{equation}
This means that $\hat{E}_{s\bot}$ is of the form

\begin{equation}
\hat{E}_{s\bot}\left(\hat{C},\hat{z}_o,\vec{\hat{\theta}}-
(\vec{\hat{l}}/\hat{z}_o)- \vec{\hat{\eta}}\right) =
\exp({i\Phi_U}) S\left(\hat{C},\hat{z}_o,\vec{\hat{\theta}}-
(\vec{\hat{l}}/\hat{z}_o)- \vec{\hat{\eta}}\right)~.
\label{Esumapp}
\end{equation}
Let us introduce in this Appendix, for simplicity of notation,
$\vec{\hat{\xi}} = \vec{\hat{{r}}}_{\bot o}-\vec{{\hat{l}}}$ and
$\vec{\hat{\phi}} =\vec{\xi}/\hat{z}_o$. It is easy to rewrite the
phase in the integrand of Eq. (\ref{undunormfinapp}), which we
denote with $\Phi_T$ as

\begin{eqnarray}
\Phi_T =  \hat{C} \hat{z}'
+\frac{\left.\vec{\hat{\eta}}\right.^2}{2}\hat{z}' + \frac{1}{2}
\left\{\left[ \vec{\hat{\phi}}^2 \hat{z}_o - 2
\vec{\hat{\phi}}\cdot \vec{\hat{\eta}} \hat{z}' +
\frac{\vec{\hat{\eta}}^2 \hat{z}'^2}{\hat{z}_o}\right]\left[1+
\sum_{n=1}^{\infty} \left(\frac{\hat{z}'}{\hat{z}_o}\right)^n
\right] \right\}~.\label{phaseapp}
\end{eqnarray}
Further algebraic manipulation of Eq. (\ref{phaseapp}) yields

\begin{eqnarray}
\Phi_T =  \hat{C} \hat{z}' + \frac{\vec{\hat{\phi}}^2 \hat{z}_o
}{2}  +  \frac{\hat{z}'}{2}
\left(\vec{\hat{\phi}}-\vec{\hat{\eta}}\right)^2 + \frac{1}{2}
\left\{\frac{\vec{\hat{\eta}}^2 \hat{z}'^2}{\hat{z}_o} +
\frac{\vec{\hat{\eta}}^2 \hat{z}'^2 }{\hat{z}_o}
\sum_{n=1}^{\infty} \left(\frac{\hat{z}'}{\hat{z}_o}\right)^n
\right.&&\cr\left. + \vec{\hat{\phi}}^2 \hat{z}_o
\sum_{n=2}^{\infty} \left(\frac{\hat{z}'}{\hat{z}_o}\right)^n - 2
\vec{\hat{\phi}}\cdot \vec{\hat{\eta}} \hat{z}'\sum_{n=1}^{\infty}
\left(\frac{\hat{z}'}{\hat{z}_o}\right)^n
\right\}~.\label{phaseapp2}
\end{eqnarray}
It is easy to recognize $\Phi_U$ in the second term on the right
hand side of Eq. (\ref{phaseapp2}). Furthermore, the fourth term
on the right hand side of Eq. (\ref{phaseapp2}) can be further
manipulated, leading to

\begin{eqnarray}
\Phi_T =  \hat{C} \hat{z}' + \Phi_U  +  \frac{\hat{z}'}{2}
\left(\vec{\hat{\phi}}-\vec{\hat{\eta}}\right)^2 &&\cr +
\frac{1}{2} \left\{\hat{z}_o \left(\vec{\hat{\phi}}^2
+\vec{\hat{\eta}}^2\right) \sum_{n=2}^{\infty}
\left(\frac{\hat{z}'}{\hat{z}_o}\right)^n - 2
\vec{\hat{\phi}}\cdot \vec{\hat{\eta}}
\hat{z}_o\sum_{n=2}^{\infty}
\left(\frac{\hat{z}'}{\hat{z}_o}\right)^n
\right\}~,\label{phaseapp3}
\end{eqnarray}
that is

\begin{eqnarray}
\Phi_T =  \hat{C} \hat{z}' + \Phi_U  +  \frac{\hat{z}'}{2}
\left(\vec{\hat{\phi}}-\vec{\hat{\eta}}\right)^2  +
\frac{\hat{z}_o}{2} \left\{ \left(\vec{\hat{\phi}}
-\vec{\hat{\eta}}\right)^2 \sum_{n=2}^{\infty}
\left(\frac{\hat{z}'}{\hat{z}_o}\right)^n
\right\}~\label{phaseapp4}
\end{eqnarray}
or

\begin{eqnarray}
\Phi_T =  \hat{C} \hat{z}' + \Phi_U  +  \frac{\hat{z}_o
\hat{z}'}{2 (\hat{z}_o-\hat{z}')}
\left(\vec{\hat{\phi}}-\vec{\hat{\eta}}\right)^2
~\label{phaseapp5}
\end{eqnarray}

Therefore, since

\begin{equation}
\vec{\hat{\phi}}-\vec{\hat{\eta}} = \vec{\hat{\theta}}-
\frac{\vec{\hat{l}}}{\hat{z}_o}- \vec{\hat{\eta}}~\label{diffapp}
\end{equation}
we have

\begin{eqnarray}
\hat{E}_{s\bot} =    \int_{-1/2}^{1/2} \frac{\hat{z}_o  d\hat{z}'
}{\hat{z}_o-\hat{z}'} \exp \left\{i \left[\Phi_U +\hat{C} \hat{z}'
+ \frac{\hat{z}_o \hat{z}'}{2 (\hat{z}_o-\hat{z}')}
\left(\vec{\hat{\theta}}- \frac{\vec{\hat{l}}}{\hat{z}_o}-
\vec{\hat{\eta}}\right)^2 \right] \right\}
\label{undunormfinappult}
\end{eqnarray}
that is Eq. (\ref{undunormfinappultini}), \textit{quantum erat
demonstrandum}.

\newpage

\section*{Appendix C: Autocorrelation function for undulator sources}

In this Appendix we demonstrate the validity of Eq.
(\ref{impores}). Having defined

\begin{eqnarray}
\tilde{f}(\hat{z}_o,\Delta\hat{\theta}_y)=
\frac{1}{2\pi^2}\int_{-\infty}^{\infty} d \hat{\phi}_y
\int_{-\infty}^{\infty} d \hat{\phi}_x
S^*{\left[\hat{z}_o,\hat{\phi}_x^2+(\hat{\phi}_y-\Delta
\hat{\theta}_y)^2\right]}
S{\left[\hat{z}_o,\hat{\phi}_x^2+(\hat{\phi}_y+\Delta
\hat{\theta}_y)^2\right]}~,~~~~~~\label{G2D2perDFApp}
\end{eqnarray}
and

\begin{eqnarray}
\beta(\Delta\hat{\theta}_y)=
\frac{1}{2\pi^2}\int_{-\infty}^{\infty} d \hat{\phi}_y
\int_{-\infty}^{\infty} d \hat{\phi}_x
\mathrm{sinc}{\left[\frac{\hat{\phi}_x^2+(\hat{\phi}_y-\Delta
\hat{\theta}_y)^2}{4}\right]}
\mathrm{sinc}{\left[\frac{\hat{\phi}_x^2+(\hat{\phi}_y+\Delta
\hat{\theta}_y)^2}{4}\right]}~,~~~~~~\label{G2D2perDGApp}
\end{eqnarray}
we have want to demonstrate that Eq. (\ref{impores}) holds, that
is

\begin{equation}
\tilde{f}(\hat{z}_o,\Delta\hat{\theta}_y) =
\beta(\Delta\hat{\theta}_y)~.\label{imporesApp}
\end{equation}
The proof is based on the autocorrelation theorem, which states
that if the (two-dimensional) Fourier Transform of a function
$w(x,y)$ with respect to variables $\alpha_x$ and $\alpha_y$ is
indicated by $\bar{w}(\alpha_x,\alpha_y)$, then the Fourier
transform of the two-dimensional autocorrelation function of
$w(x,y)$ with respect to the same variables $\alpha_x$ and
$\alpha_y$ is given by $|\bar{w}(\alpha_x,\alpha_y)|^2$. In
formulas, after definition of the autocorrelation function

\begin{equation}
\mathcal{A}[w](x,y) = \int_{-\infty}^{\infty} d\eta
\int_{-\infty}^{\infty} d\xi w(\eta+x,\xi+y)
w^*(\eta,\xi)~,\label{autoapp}
\end{equation}
which is equivalent to

\begin{equation}
\mathcal{A}[w](x,y) = \int_{-\infty}^{\infty} d\eta
\int_{-\infty}^{\infty} d\xi w(\eta+x/2,\xi+y/2)
w^*(\eta-x/2,\xi-y/2)~,\label{auto2app}
\end{equation}
the autocorrelation theorem states that

\begin{equation}
\int_{-\infty}^{\infty} dx \int_{-\infty}^{\infty} dy  \exp{[i
(\alpha_x x+ \alpha_y y)]} \mathcal{A}[w](x,y) =
|\bar{w}(\alpha_x,\alpha_y)|^2~. \label{fourapp}
\end{equation}
First we extend the definition of $\tilde{f}$

\begin{eqnarray}
\tilde{f}(\hat{z}_o,\Delta\hat{\theta}_x',\Delta\hat{\theta}_y')=\frac{1}{2\pi^2}
\int_{-\infty}^{\infty} d \hat{\phi}_y \int_{-\infty}^{\infty} d
\hat{\phi}_x S^*{\left[\hat{z}_o,(\hat{\phi}_x-\Delta
\hat{\theta}_x'/2)^2+(\hat{\phi}_y-\Delta
\hat{\theta}_y'/2)^2\right]} &&\cr \times
S{\left[\hat{z}_o,(\hat{\phi}_x+\Delta
\hat{\theta}_x'/2)^2+(\hat{\phi}_y+\Delta
\hat{\theta}_y'/2)^2\right]}~,~~~~~~\label{G2D2perDFgenApp}
\end{eqnarray}
where we changed variables from $\Delta \hat{\theta}_{x,y}$ to
$\Delta \hat{\theta}_{x,y}'= 2 \Delta \hat{\theta}_{x,y}$ Then we
can apply the autocorrelation theorem in Eq. (\ref{fourapp}) to
the function $\tilde{f}$ thus obtaining the following relation:

\begin{eqnarray}
\int_{-\infty}^{\infty} d \Delta\hat{\theta}_x'
\int_{-\infty}^{\infty} d\Delta\hat{\theta}_y'  \exp{[i (\alpha_x
\Delta\hat{\theta}_x'+ \alpha_y \Delta\hat{\theta}_y')]}
\tilde{f}(\hat{z}_o,\Delta\hat{\theta}_x',\Delta\hat{\theta}_y')
&& \cr  = \frac{1}{2\pi^2}\left|\int_{-\infty}^{\infty} d
\hat{\phi}_x \int_{-\infty}^{\infty} d \hat{\phi}_y \exp{[i
(\alpha_x \hat{\phi}_x+ \alpha_y \hat{\phi}_y)]}
S{\left[\hat{z}_o,\hat{\phi}_x^2+\hat{\phi}_y^2\right]}
\right|^2~, \label{appliedapp}
\end{eqnarray}
where $\alpha_{x,y}$ are now conjugated variables with respect to
the angles $\hat{\phi}_{x,y}$ on which $S$ depends. We will denote
with $\bar{S}$ the two-dimensional Fourier Transform of $S$, that
is:

\begin{eqnarray}
\bar{S}({\alpha}_x,{\alpha}_y)=\int_{-\infty}^{\infty} d
\hat{\phi}_x \int_{-\infty}^{\infty} d \hat{\phi}_y \exp{[i
(\alpha_x \hat{\phi}_x+ \alpha_y \hat{\phi}_y)]}
S{\left[\hat{z}_o,\hat{\phi}_x^2+\hat{\phi}_y^2\right]}~.
\label{appliedFTapp}
\end{eqnarray}
The relation between the function $S$ and the undulator field is
given by Eq. (\ref{Esum}), and one has

\begin{eqnarray}
\bar{S}({\alpha}_x,{\alpha}_y)=\int_{-\infty}^{\infty} d
\hat{\phi}_x \int_{-\infty}^{\infty} d \hat{\phi}_y \exp{[i
(\alpha_x \hat{\phi}_x+ \alpha_y \hat{\phi}_y)]} \exp{[-i
(\hat{\phi}_x^2+\hat{\phi}_y^2)\hat{z}_o/2]}&&\cr \times
\hat{E}_{s\bot}
{\left[\hat{z}_o,\hat{\phi}_x^2+\hat{\phi}_y^2\right]}~.
\label{appliedFTapp-1}
\end{eqnarray}
After definition of $\bar{\alpha}_{x,y} = \alpha_{x,y}/\hat{z}_o$
one can write  $\bar{S}$, as a function of $\bar{\alpha}_{x,y}$
instead of $\alpha_{x,y}$. Then, one can switch to the new
integration variables  $\hat{x} = \hat{\phi}_x\hat{z}_o$ and
$\hat{y} = \hat{\phi}_y\hat{z}_o$ to obtain:

\begin{eqnarray}
\bar{S}(\bar{\alpha}_x,\bar{\alpha}_y)=\frac{1}{\hat{z}_o^2}\int_{-\infty}^{\infty}
d \hat{x} \int_{-\infty}^{\infty} d \hat{y} \exp{[i
(\bar{\alpha}_x \hat{x}+ \bar{\alpha}_y \hat{y})]}\exp{[-i
(\hat{x}^2+\hat{y}^2)/(2\hat{z}_o)]}&&\cr \times \hat{E}_{s\bot}
{\left[\hat{z}_o,\hat{x}^2+\hat{y}^2\right]} ~.&&\cr
\label{appliedFT2app}
\end{eqnarray}
where the expression for $\hat{E}_{s\bot}
{[\hat{z}_o,\hat{x}^2+\hat{y}^2]}$ is given in Eq.
(\ref{undunormfin}). Now we have to calculate the Fourier
transform of the product of two factors: $\exp{[-i
(\hat{x}^2+\hat{y}^2)/(2\hat{z}_o)]}$ and $\hat{E}_{\bot s}$. Let
us look for the Fourier transform of each factor.

A direct calculation shows that

\begin{eqnarray}
\int_{-\infty}^{\infty} d \hat{x} \int_{-\infty}^{\infty} d
\hat{y} \exp{[i (\bar{\alpha}_x \hat{x}+ \bar{\alpha}_y \hat{y})]}
\hat{E}_{s\bot }\left(\hat{z}_o,\hat{x}^2+\hat{y}^2\right)&&\cr= 2
i \pi \hat{z}_o \exp{\left[-i
\frac{(\bar{\alpha}_x^2+\bar{\alpha}_y^2) \hat{z}_o}{2}\right]}
\mathrm{sinc}\left[\frac{\bar{\alpha}_x^2+\bar{\alpha}_y^2}{4}\right]~.
\label{appliedFT3app}
\end{eqnarray}
Second, let us deal with the Fourier transform of $\exp{[-i
(\hat{x}^2+\hat{y}^2)/(2\hat{z}_o)]}$:

\begin{eqnarray}
\int_{-\infty}^{\infty} d \hat{x} \int_{-\infty}^{\infty} d
\hat{y} \exp{[i (\bar{\alpha}_x \hat{x}+ \bar{\alpha}_y
\hat{y})]}\exp{[-i (\hat{x}^2+\hat{y}^2)/(2\hat{z}_o)]} &&\cr= -4
i \hat{z}_o \exp{[i(\bar{\alpha}_x^2+\bar{\alpha}_y^2)
\hat{z}_o/2]} ~. \label{appliedFT4app}
\end{eqnarray}
Since the Fourier transform of a product is equal to the
convolution of the Fourier transforms of the factors we have

\begin{eqnarray}
\bar{S}(\bar{\alpha}_x,\bar{\alpha}_y) = 8 \pi
\int_{-\infty}^{\infty} d u \int_{-\infty}^{\infty} d w
\exp{\{i[(\bar{\alpha}_x-u)^2+(\bar{\alpha}_y-w)^2]
\hat{z}_o/2\}}&&\cr \times \exp{[-i (u^2+w^2) \hat{z}_o/2]}
\mathrm{sinc}[(u^2+w^2)/4]~. \label{appliedFT5app}
\end{eqnarray}
and therefore we have

\begin{eqnarray}
|\bar{S}(\bar{\alpha}_x,\bar{\alpha}_y)| = 8 \pi
\left|\int_{-\infty}^{\infty} d u \int_{-\infty}^{\infty} d w
\exp{[i(\bar{\alpha}_x u+\bar{\alpha}_y w) \hat{z}_o]}
\mathrm{sinc}[(u^2+w^2)/4]\right|~. \label{appliedFT6app}
\end{eqnarray}
Going back to old variables $\alpha_{x,y}$  we obtain

\begin{eqnarray}
|\bar{S}({\alpha}_x,{\alpha}_y)| = 8 \pi
\left|\int_{-\infty}^{\infty} d u \int_{-\infty}^{\infty} d w
\exp{[i({\alpha}_x u+{\alpha}_y w) ]}
\mathrm{sinc}[(u^2+w^2)/4]\right|~. \label{appliedFT6appbissss}
\end{eqnarray}
which is independent of $\hat{z}_o$. As a result Eq.
(\ref{appliedapp}) is also independent on $\hat{z}_o$, i.e. the
Fourier transform of $\tilde{f}$ is independent of $\hat{z}_o$.

Now, on the one hand in the limit for $\hat{z}_o \longrightarrow
\infty$ the function $\tilde{f}$ transforms into $\beta$, because
the $S$ functions in $\tilde{f}$ tend asymptotically to the
$\mathrm{sinc}$ functions in $\beta$. On the other hand, if the
Fourier transform of $\tilde{f}$ is independent of $\hat{z}_o$,
also $\tilde{f}$ is independent of $\hat{z}_o$.  As a result it
can only be $\tilde{f}(\Delta \hat{\theta}_y') = \beta(\Delta
\hat{\theta}_y')$, and $\tilde{f}(\Delta \hat{\theta}_y) =
\beta(\Delta \hat{\theta}_y)$ that is Eq. (\ref{imporesApp})
holds, \textit{quantum erat demonstrandum}.

Note that, based on the autocorrelation theorem, it is also
possible to give an analytic expression for the Fourier transform
of $\beta$. After definition of
$\beta=\beta(\hat{z}_o,\Delta\hat{\theta}_x',\Delta\hat{\theta}_y')$
as in Eq. (\ref{G2D2perDFgenApp}), application of the
autocorrelation theorem simply states that the two-dimensional
Fourier Transform of $\beta$, that will be indicated with
$\bar{\beta}$ can be written as

\begin{equation}
\bar{\beta}(\alpha_x,\alpha_y) =\frac{1}{2\pi^2}\left|
\int_{-\infty}^{\infty} d\hat{\phi}_x\int_{-\infty}^{\infty}
d\hat{\phi}_y \exp{[i (\alpha_x \hat{\phi}_x+\alpha_y
\hat{\phi}_y)]}
\mathrm{sinc}\left(\frac{\hat{\phi}_x^2+\hat{\phi}_y^2}{4}\right)
\right|^2~. \label{tildegtrapp}
\end{equation}
Introducing $\alpha^2 = \alpha_x^2+\alpha_y^2$ and representing
the two-dimensional Fourier Transform of $\beta$ in terms of
Fourier-Bessel transform we obtain

\begin{equation}
\bar{\beta}(\alpha) = \frac{1}{2\pi^2}\left| 2\pi
\int_{0}^{\infty} d r ~r J_0(r \alpha)
\mathrm{sinc}\left(\frac{r^2}{4}\right) \right |^2 = 2 \left[\pi -
2\mathrm{Si}(\alpha^2)\right]^2~, \label{tildegtrapp2}
\end{equation}
where $\mathrm{Si}$ indicates the sine integral function.

Finally, one can get back a simpler representation of the function
$\beta$ in terms of a one-dimensional integration simply
performing an anti Fourier-Bessel transform:

\begin{equation}
{\beta}(\Delta \hat{\theta}') =  \frac{1}{2\pi} \int_{0}^{\infty}
d \alpha ~ \alpha J_0(\alpha \Delta \hat{\theta}')
\bar{\beta}(\alpha)~, \label{tildegtrapp3}
\end{equation}
where $\Delta \hat{\theta}^{'2}=\Delta \hat{\theta}_x^{'2}+\Delta
\hat{\theta}_y^{'2}$. For $\Delta \hat{\theta}_x=0$ we obtain

\begin{equation}
{\beta}(\Delta \hat{\theta}_y) = \frac{1}{\pi} \int_{0}^{\infty} d
\alpha ~ \alpha J_0\left(\alpha \frac{\Delta
\hat{\theta}_y}{2}\right) \left[\pi -
2\mathrm{Si}(\alpha^2)\right]^2~. \label{tildegtrapp4}
\end{equation}

\newpage

\section*{Acknowledgements}

The authors wish to thank Hermann Franz, Petr Ilinski and Ivan
Vartanyants for many useful discussions,  Jochen Schneider and
Edgar Weckert for their interest in this work.

\end{document}